\documentclass{jfm}

\usepackage{graphicx}
\usepackage{natbib}
\usepackage{hyperref}
\usepackage{amsmath}
\usepackage{ar}
\usepackage{tikz}
\usepackage{pgfplots}
\usepackage{tikz-3dplot}
\usepackage{bm}
\usepackage{multirow}

\newcommand{\aver}[1]{ \! \left\langle {#1} \right \rangle \!}

\title[Kolmogorov-size particles in homogeneous and isotropic turbulence]{Kolmogorov-size particles in homogeneous and isotropic turbulence}

\author[A.~Chiarini, S.~Tandurella, M.E.~Rosti]{Alessandro Chiarini\corresp{Present address: Dipartimento di Scienze e Tecnologie Aerospaziali, Politecnico di Milano, via La Masa 34, 20156 Milano, Italy.}\corresp{\email{alessandro.chiarini@polimi.it}}, 
Simone Tandurella  
and Marco Edoardo Rosti\corresp{\email{marco.rosti@oist.jp}}}
\affiliation{Complex Fluids and Flows Unit, Okinawa Institute of Science and Technology Graduate University, 1919-1 Tancha, Onna-son, Okinawa 904-0495, Japan}

\begin{document}
\maketitle

\begin{abstract}

We investigate the fluid-solid interaction of suspensions of Kolmogorov-size spherical particles moving in homogeneous isotropic turbulence at a microscale Reynolds number of $Re_\lambda \approx 140$. Two volume fractions are considered, $10^{-5}$ and $10^{-3}$, and the solid-to-fluid density ratio is set to $5$ and $100$. 
We present a comparison between interface-resolved (PR-DNS) and one-way-coupled point-particle (PP-DNS) direct numerical simulations. 
We find that the modulated energy spectrum shows the classical $-5/3$ Kolmogorov scaling in the inertial range of scales and a $-4$ scaling at smaller scales, with the latter resulting from a balance between the energy injected by the particles and the viscous dissipation, in an otherwise smooth flow. An analysis of the small-scale flow topology shows that the particles mainly favour events with axial strain and vortex compression. The dynamics of the particles and their collective motion studied for PR-DNS are used to assess the validity of the PP-DNS. 
We find that the PP-DNS predicts fairly well both the Lagrangian and Eulerian statistics of the particles motion for the low-density case, while some discrepancies are observed for the high-density case. Also, the PP-DNS is found to underpredict the level of clustering of the suspension compared to the PR-DNS, with a larger difference for the high-density case. 

\end{abstract}


\section{Introduction}
\label{sec:intro}

Particle-laden turbulent flows have been extensively investigated over the years, because of their relevance from both the fundamental and applicative viewpoints. They are indeed ubiquitous in several natural and engineering scenarios \citep{delillo-etal-2014,sengupta-etal-2017}, such as in volcanic ash and cloud droplets in atmospheric turbulence, dust particles in protoplanetary disks, sandstorms, ocean microplastics, and fuel droplets in spray combustion.

\subsection{The Maxey-Riley-Gatignol (MRG) Equation}

In particle-laden turbulent flows the turbulent scales of the fluid phase are coupled in a non trivial manner with the solid phase. Properly resolving the flow around each particle is thus crucial to capture the fluid-solid interaction and describe the dynamics of the particles. Due to the prohibitive computational cost, however, most of the theoretical and numerical studies rely on approximations and models. 
As an example, in the context of particle clustering and preferential sampling --- i.e. the tendency of particles to explore flow regions with specific properties --- we mention the recent theoretical works by \cite{goto-vassilicos-2008}, \cite{coleman-vassilicos-2009}, \cite{bragg-ireland-collins-2015} and \cite{matsuda-etal-2024}. Most of the models used are based on the seminal works of \cite{maxey-riley-1983} and \cite{gatignol-1983}, where the equation for small rigid spheres in a non-uniform flow (hereafter referred to as MRG equation) has been derived by exploiting linear perturbation theory. In these models, each particle is treated as a mathematical point source of mass, momentum and energy. In point-particle models, particles are indeed assumed to be much smaller than any structure of the flow, as the MRG equation holds when the fluid velocity field does not show a turbulent behaviour at the particle scale. In other words, the Reynolds number $Re_p$ based on the particle diameter and the particle-fluid relative velocity has to be small, i.e. $Re_p \ll 1$.

The models based on the MRG equation do not resolve the flow around the particles, and the influence of the solid phase on the fluid phase has to be modelled; see for example \cite{ferrante-elghobashi-2003}, \cite{gualtieri-etal-2015} and \cite{vreman-2016}. However, the back-reaction of the particles on the carrier flow and the inter-particle collisions are usually negligible in the limit of very dilute regimes with $\Phi_V = V_s/(V_s + V_f) \le 10^{-5}$ (where $\Phi_V$ is the volume fraction, and $V_s$ and $V_f$ are the volumes of the solid and fluid phases, respectively), small particles $D_p < \eta$ (where $\eta$ is the turbulent Kolmogorov scale) and small solid-to-fluid density ratios $\rho_p/\rho_f$ \citep{brandt-coletti-2022}. In this case, the influence of the solid phase on the carrier fluid can be neglected and the fluid-particle interaction is often modelled with one-way coupling models, in which the particles move under the action of the flow, but they do not modulate it. Starting from the MRG equation, several corrections have been proposed over the years to account for several effects, and extend its range of validity to a wider range of parameters. For example, \cite{saffman-1965} introduced a lift force which is crucial to properly model the particle dynamics in the presence of a linear shear. This force has been later extended by \cite{mei-1992} to fit the numerical data of \cite{mclaughlin-1991} at large Reynolds numbers. Other corrections have been introduced to account for finite values of the particle Reynolds number. For example, we refer to the corrections to the drag term reported in \cite{balachandar-2009}, and to the different convolutional kernels for the Basset time-history force contribution proposed by \cite{mei-adrian-1992}. 

The point particle approximation coupled with Direct Numerical Simulations of the Navier--Stokes equations (PP-DNS) has been widely used to investigate turbulent particle-laden flow in the one-way coupling regime in several scenarios \citep[see][and references therein]{balachandar-eaton-2010}. Despite the large number of studies based on PP-DNS, however, clear understanding of the range of validity of the underlying assumptions of the point-particle model is not yet present, and studies that investigate its limitations are needed. In this respect, a step forward has been done in the last years thanks to the introduction of several numerical methods (PR-DNS) which couple the Direct Numerical Simulations of the Navier--Stokes equations with techniques that resolve the flow around each particle and capture the effect of the no-slip boundary condition at the particles' surface on the flow. We mention the arbitrary Lagrangian-Eulerian (ALE) technique \citep{hu-patankar-zhu-2001}, the overset-grid approaches \citep{burton-eaton-2005,koblitz-etal-2017,vreman-2017,horne-mahesh-2019}, the Physalis technique introduced by \cite{prosperetti-oguz-2001} and the immersed boundary method (IBM) \citep{kajishima-etal-2001,uhlmann-2005,huang-etal-2007,breugem-2012,kempe-frolich-2012,hori-rosti-takagi-2022}. Unlike experiments, PR-DNS allows to explain the underlying physical mechanisms and investigate the limitations of the point-particle approximation.


In the context of forced homogeneous isotropic turbulence (HIT) with fixed particles, \cite{burton-eaton-2005} and \cite{vreman-2016} found that, when compared with PR-DNS, the two-way coupled PP-DNS based on the Schiller-Naumann drag correlation captures fairly well the turbulence attenuation and the fraction of the turbulence dissipation rate due to the particles. 
Surprisingly, they found a good agreement also for $D_p/\eta = 1.5$ \citep{burton-eaton-2005} and $D_p/\eta \approx 2.2$ \citep{vreman-2016}, although the model is expected to largely under-perform for these particle sizes. \cite{mehrabadi-etal-2018} used PR-DNS to assess the validity of PP-DNS in a decaying isotropic turbulent particle-laden flow, focusing on the particle acceleration model. They found that the predictions of the PP-DNS models they considered are in excellent agreement compared with PR-DNS for small Stokes numbers. For large Stokes numbers, however, they found that PP-DNS under-predicts the true particle acceleration and that second moment quantities are not properly captured. They showed that the predictions improve once considering finite Reynolds number corrections to the model. \cite{costa-brandt-picano-2020} tested the one-way point-particle approximation in a turbulent channel flow laden with small inertial particles, with high particle-to-fluid density ratios. They considered a volume fraction of $\Phi_V \approx 10^{-5}$ to ensure that the feedback of the particles on the fluid phase was negligible. They found that in the bulk of the channel the model predicts fairly well the statistics of the particles velocity. Close to the wall, however, they observed that the model fails, as it is not able to capture the shear-induced lift force acting on the particles, which instead is well predicted by the lift force model introduced by \cite{saffman-1965}. 

In this work, we further address the limit of the one-way point-particle approximation, and we use PR-DNS to investigate the reliability of one-way-coupled PP-DNS in the context of forced homogeneous isotropic turbulence laden with Kolmogorov-size particles.

\subsection{Particles in homogeneous isotropic turbulence}

The fluid-solid interaction of suspension of spherical particles moving in homogeneous isotropic turbulence has been widely investigated over the years by means of both simulations and experiments. In the following, we list some of the main contributions, and we refer the interested reader to \cite{balachandar-eaton-2010} and \cite{brandt-coletti-2022} for a comprehensive review. 

The majority of the numerical studies dealing with small particles $D_p \ll \eta$ are based on the point-particle approximation, as particle-resolved methods are prohibitively expensive in this case. Although the related model error is not known, the available numerical studies have shown that small particles may either amplify or damp the turbulence of the carrier flow, in agreement with previous experimental studies \citep{gore-crowe-1989}. \cite{squires-eaton-1990} used two-way-coupled PP-DNS to investigate homogeneous isotropic turbulence laden with small heavy spherical particles at $Re_\lambda = u' \lambda / \nu \approx 38$ (where $u'$ is the average velocity fluctuation, $\lambda$ is the Taylor length scale, and $\nu$ is the kinematic fluid viscosity). 
Compared to the particle-free case, they reported a significant attenuation of the fluid kinetic energy and of the dissipation rate. By looking at the energy spectrum, they found that the addition of the particles results in a relative enhancement of the energy at the small scales compared to the energy content at the large scales. They also showed that heavier particles cause a less selective modification of the turbulence properties. Heavy particles are indeed more uniformly dispersed by the turbulence, and cause a more homogeneous modification of the flow properties compared to lighter particles, that instead show a stronger preferential collection in regions of low vorticity and high strain \citep{maxey-1987}. \cite{elghobashi-truesdell-1993} used two-way coupled PP-DNS to investigate decaying homogeneous isotropic turbulence laden with small spherical particles. Besides confirming the non-uniform modulation of the energy spectrum, they observed that the energy enhancement at the small scales is accompanied by an increase of the viscous dissipation rate and, thus, by an enhancement of the rate of energy transfer from larger to smaller scales. \cite{boivin-simonin-squires-1998} made use of two-way-coupled PP-DNS to study the influence of small and heavy particles on forced homogeneous isotropic turbulence at $Re_\lambda = 62$. 
They reported that the influence of the particles changes with their inertia, with the small-scale energy content being attenuated/enhanced by large/small particles. By investigating the spectrum of the fluid-particle energy exchange rate, they observed that particles act as a sink of kinetic energy at large scales, while they add kinetic energy to turbulence at the smallest scales. The large scale motions of the fluid drag the particles, while the small-scale fluctuations are driven by the presence of the solid phase.
\cite{druzhinin-2001} investigated the influence of small and heavy particles in decaying isotropic homogeneous turbulence at the initial Reynolds number of $Re_\lambda = 30$ and $Re_\lambda = 50$, with a focus on 
particles with very small inertia and small relaxation time. Unlike the previous works, they found that the turbulent kinetic energy and the viscous dissipation rate increase at all times compared to the particle-free case. The presence of the particles, indeed, largely enhances the small-scale energy content, while slightly reduces the large-scale energy content, with positive integral variation. 
This was later confirmed by \cite{ferrante-elghobashi-2003}, that investigated by PP-DNS the influence of particles on decaying homogeneous isotropic turbulence with an initial Reynolds number of $Re_\lambda = 75$. 

For large particles with $D_p>\eta$ the numerical studies are based on PR-DNS, as in this case the point-particle approximation does not hold. \cite{lucci-etal-2010} and \cite{lucci-etal-2011} investigated the influence of particles with size $D_p \approx \lambda$ 
 in decaying homogeneous turbulence. They observed that in contrast to what happens when $D_p<\eta$, the presence of the particles damps the turbulent kinetic energy of the fluid compared to the particle-free case at all times, and that the two-way coupling rate of change is always positive. \cite{tenCate-etal-2004} investigated the influence of particles with a solid-to-fluid density ratio of $\rho_p/\rho_f = 1.15$ and $1.73$ on forced homogeneous isotropic turbulence at $Re_\lambda = u' \lambda / \nu = 61$, varying the volume fraction between $\Phi_V = 0.02$ and $\Phi_V = 0.1$. They found that 
the energy spectrum is enhanced for wavenumbers $\kappa > \kappa_p \approx 0.75 \kappa_D$, where $\kappa_D = 2 \pi /D_p$, while it is attenuated for $\kappa < \kappa_p$. These results were later confirmed by \cite{yeo-etal-2010} at $Re_\lambda \approx 60$, $\rho_p/\rho_f = 1.4$ and $\Phi_V = 0.06$. \cite{uhlmann-chouippe-2017} considered particles with $D_p/\eta \approx 5-8 $ and $\rho_p/\rho_f = 1.5$ at the larger Reynolds number of $Re_\lambda \approx 130$. They focused on the dynamics of the particles, and observed that finite-size inertial particles exhibit a moderate level of clustering, as later confirmed also by \cite{chiarini-rosti-2024}. \cite{olivieri-cannon-rosti-2022} considered the effect of particles on homogeneous isotropic turbulence at the larger Reynolds number of $Re_\lambda \approx 400$, which ensures a well developed inertial range of scales. They set the volume fraction at $\Phi_V = 0.079$, and investigated the turbulence modulation by particles with size $D_p/\eta=123$ and solid-to-fluid density ratio between $1.3 \le \rho_p/\rho_f \le 100$. They showed that the solid phase modifies the energy cascade described by Richardson and Kolmogorov; the fluid-solid coupling drives the energy cascade at large scales, while the classical energy cascade is restored at scales smaller than the particle size. \cite{oka-goto-2022} studied the turbulence modulation due to spherical particles with $7.8 \le D_p/\eta \le 64$ by setting the volume fraction at $\Phi_V = 8.1 \times 10^{-3}$ and the reference Reynolds number at $Re_\lambda \le 100$. They found that the turbulent kinetic energy content monotonically decreases with $D_p$, due to the increase of the energy dissipation rate in the wake of the particles. More recently, \cite{chiarini-etal-2023} and \cite{chiarini-rosti-2024} investigated by PR-DNS how the flow modulation changes with $D_p$ and $\rho_p$. They set the Reynolds number to $Re_\lambda \approx 400$ and the volume fraction to $\Phi_V = 0.079$, and varied the particles size and the solid-to-fluid density ratio in the $16 \le D_p/\eta \le 123$ and $1.3 \le \rho_p/\rho_f \le 100$ range. 
 \cite{chiarini-rosti-2024} observed that interface-resolved particles enhance flow intermittency favouring events with large localised velocity gradients. For the smallest and heaviest particles, they found that the classical energy cascade is subdominant at all scales, and that the energy transfer is completely driven by the fluid-solid coupling term. \cite{cannon-olivieri-rosti-2024} investigated the effect of the Reynolds number on the flow modulation by finte-size particles in homogeneous isotropic turbulence. Notably, they observed that the modulation of the turbulent kinetic energy has little dependence on $Re_\lambda$, and that particles modulate turbulence also at the smallest Reynolds numbers.

While a relatively larger body of literature has investigated the dynamics of particle of size larger and smaller than Kolmogorov size, few works have considered Kolmogorov-size particles with $D_p \approx \eta$, which are the focus of the present work. From an experimental point of view the $D_p \approx \eta$ case is complex, as it requires a resolution of sub-Kolmogorov scales when measuring the velocity perturbations near the particles \citep{tanaka-eaton-2010}. Numerical schemes based on the MRG equation, which are commonly used for $D_p \ll \eta$, are generally thought of as not valid when $D_p \approx \eta$ \citep{balachandar-eaton-2010}. PR-DNS, on the other hand, becomes prohibitively expensive as $D_p$ decreases when $Re$ is sufficiently large, due to the extra resolution required to properly resolve both the flow perturbations induced by the particles and all the turbulence scales. Among the few works available, we mention \cite{hwang-eaton-2006} that experimentally investigated the influence of a dilute dispersion of particles with $D_p \approx \eta$ in forced homogeneous isotropic turbulence at $Re_\lambda \approx 230$. They observed that Kolmogorov-size particles attenuate the turbulent global kinetic energy and the viscous dissipation rate up to $40\%$ and $50\%$ for a mass loading $\Phi_M = \rho_p V_p/(\rho_f V_f) $ of $ \Phi_M = 0.3$. \cite{yang-shy-2005} experimentally investigated the settling of small solid particles in homogeneous isotropic turbulence, with $\rho_p \gg \rho_f$ and $D_p \le \eta$ such that $Re_p<1$ and $0 \le St \le 2$. They observed that at these parameters the tendency of particles to form clusters is maximum for $St \approx 1$, and that the clusters are distributed along the periphery of  intense vortical structures. \cite{poelma-etal-2007}, instead, investigated the influence of $D/\eta = O(1)$ and $\rho_p/\rho_f = O(1)$ particles on the decay rate of grid generated turbulence. According to their experiments, the presence of the particles moves the onset of the turbulence decay upstream, and promotes the flow anisotropy. For recent experimental works concerning sub-Kolmogorov $D_p < \eta$ particles, we refer the reader to \cite{sumbekova-etal-2017,petersen-etal-2019,hassaini-etal-2023} (clustering) and to \cite{hassaini-coletti-2022} (turbulence modification). \cite{schneiders-etal-2017} studied by PR-DNS the interaction of decaying isotropic turbulence with finite-size $D_p \approx \eta$ particles. They varied the solid-to-fluid density ratio between $40 \le \rho_p/\rho_f \le 5000$ and the mass loading between $0.01 \le \Phi_M \le 1$. They set the Reynolds number of the flow at $Re_\lambda(t_0) = 79$, a starting value for which the flow lacks a well-defined inertial range. 
They observed that in the vicinity of the particles the viscous dissipation rate of the fluid is amplified due to the large velocity gradients that are generated by the boundary conditions at the surface of the particles \citep[see also][]{tanaka-eaton-2010,brandle-etal-2016,chiarini-rosti-2024}. Particles also release kinetic energy to the fluid by locally accelerating the surrounding flow, similarly to what seen by \cite{chiarini-rosti-2024} for larger particles. From a global viewpoint, \cite{schneiders-etal-2017} observed that, for large $\rho_p/\rho_f$, particles with $D/\eta \approx 1$ induce local velocity disturbances that significantly modulate the distribution and the decay of the fluid kinetic energy at all scales. Overall, despite the interest an exhaustive characterisation of how Kolmogorov-size particles modulate turbulence at a Reynolds number that is large enough to ensure a proper separation of scales is still lacking.

\subsection{Present study}

In this study, we investigate the fluid-solid interaction of suspension of Kolmogorov-size spherical particles moving in homogeneous isotropic turbulence at the relatively large microscale Reynolds number of $Re_\lambda \approx 140 $ by use of direct numerical simulations. The study is based on both PR-DNS and one-way-coupled PP-DNS. The specific objective of the present study is twofold. We aim \textit{(i)} to investigate (for the first time) the modulation of forced homogeneous isotropic turbulence by finite Kolmogorov-size particles at a Reynolds number which is large enough to ensure a proper separation of scales, and \textit{(ii)} to address the limits and the range of validity of the one-way-coupled PP-DNS in the simplest configuration of homogeneous isotropic turbulence. To do this, we consider a portion of the parameter space which is on the edge of the range of validity of the one-way-coupled PP-DNS \citep{brandt-coletti-2022}, and compare the results in terms of both Eulerian and Lagrangian particles' statistics. 
Two volume fractions are considered, $\Phi_V = 10^{-3}$ and $\Phi_V = 10^{-5}$, and the solid-to-fluid density ratio is set equal to $\rho_p/\rho_f \approx 5$ and $\rho_p/\rho_f \approx 100$. 

The structure of the work is as follows. After this introduction, the computational set up and the numerical methods are described in \S\ref{sec:method}. Then, section \S\ref{sec:flow} is devoted to the assessment of the flow modulation, and discusses the results of the PR-DNS. The influence of the particles on the energy spectrum, on the scale-by-scale energy budget and on the local structure of the flow are discussed. Sections \S\ref{sec:particles} and \S\ref{sec:collective} deal respectively with the dynamics of the particles and with the inhomogeneity of their distribution in the flow. In these sections, we assess the validity of the one-way-coupled PP-DNS. Eventually, concluding remarks are provided in \S\ref{sec:conclusions}.

\section{Mathematical formulations and numerical method}
\label{sec:method}

We consider a turbulent flow in a triperiodic box of size $L=2\pi$ laden with $N$ spherical particles; see figure \ref{fig:setup}. The carrier flow is governed by the incompressible Navier--Stokes equations
\begin{figure}
\end{figure}
\begin{figure}
  \centering
  \vspace{0.0cm}
  \includegraphics[width=1\textwidth]{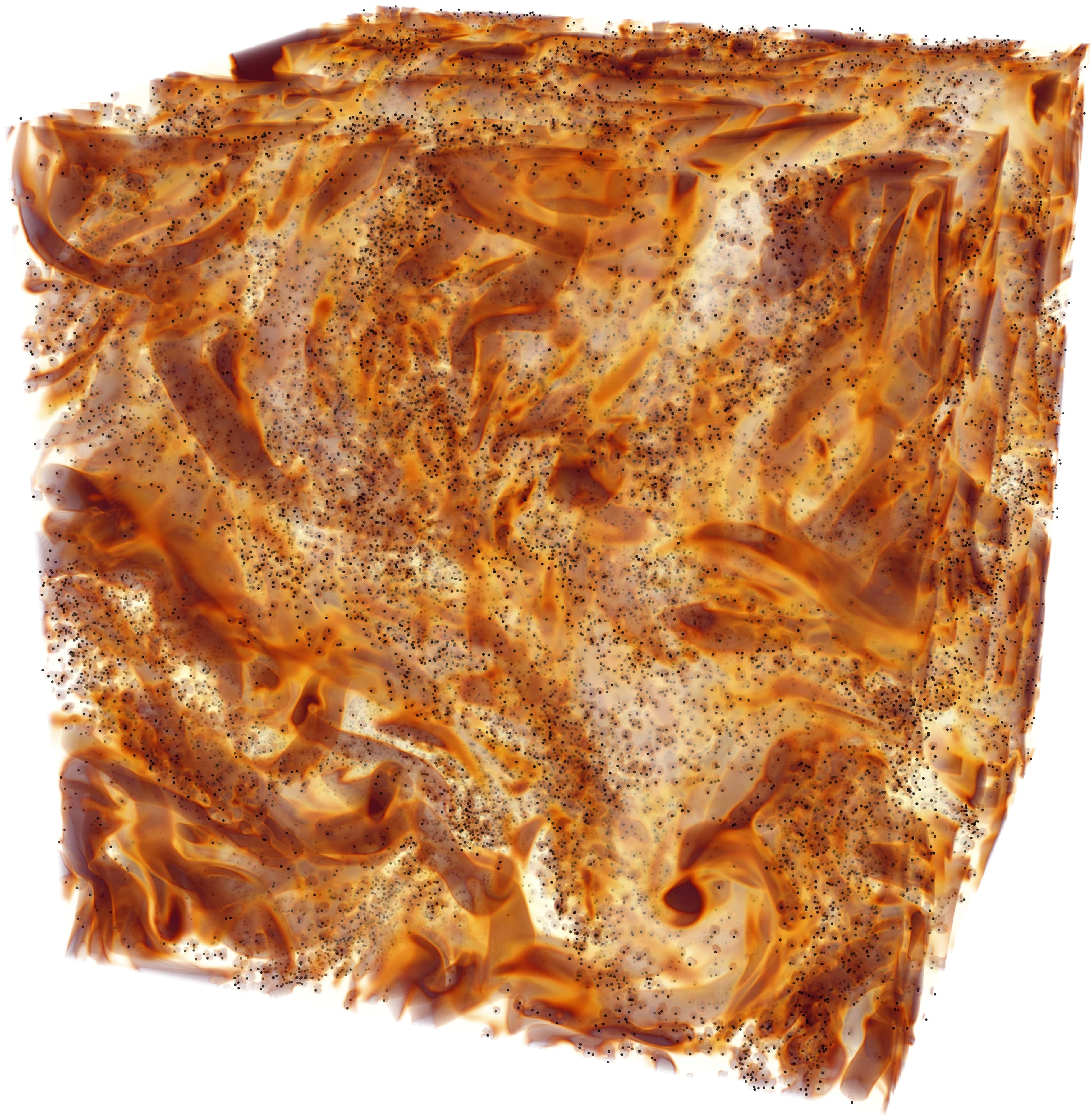} 
  \caption{Volumetric rendering of a snapshot of the PR-DNS with $\Phi_V = 10^{-3}$ and $\rho_p/\rho_f=100$. Darker coloured areas correspond to higher enstrophy $\omega^2$ regions of the flow. Particles, shown here in black, appear to preferentially sample regions of low $\omega^2$ (see \S\ref{sec:sampling}).}
  \label{fig:setup}
\end{figure}
\begin{equation}
\frac{\partial \bm{u}}{\partial t}  + \bm{\nabla} \cdot \bm{u} \bm{u} = 
- \frac{1}{\rho_f} \bm{\nabla} p + \nu \bm{\nabla}^2 \bm{u} + \bm{f} + \bm{f}^{\leftarrow p}, \
\bm{\nabla} \cdot \bm{u} = 0,
\end{equation}
where $\bm{u}=(u,v,w)$ is the fluid velocity, $p$ is the reduced pressure, and $\rho_f$ and $\nu$ are the fluid density and kinematic viscosity. At the right-hand-side of the momentum equation $\bm{f}$ is the Arnold-Beltrami-Childress (ABC) cellular forcing \citep{podvigina-pouquet-1994} used to inject energy at the largest scales and sustain turbulence, while $\bm{f}^{\leftarrow p}$ is the force the particles exert on the fluid phase. In this work $\bm{f}^{\leftarrow p}$ is non null for PR-DNS only.

\subsection{Particle-resolved simulations (PR-DNS)}

The motion of a rigid particle can be described using the translational velocity $\bm{u}_p$ and the rotational velocity $\bm{\omega}_p$ of its centre of mass, that obey the classical Newton-Euler equations for rigid body dynamics,
\begin{equation}
m_p \frac{\text{d} \bm{u}_p}{\text{d} t} = \bm{f}^{\leftarrow f}  + \bm{f}_p^{\leftrightarrow p}, \ 
I_p \frac{\text{d} \bm{\omega}_p}{\text{d} t} = \bm{L}_p^{\leftarrow f},
\label{eq:NEeq}
\end{equation}
where $m_p = \pi \rho_p D_p^3/6$ and $I_p = m_p D_p^2/10$ are the mass and inertial moment of the particle, with $\rho_p$ being the particle density and $D_p$ the particle diameter. Here $\bm{f}^{\leftrightarrow p}$ is the force due to particle collisions, while $\bm{f}^{\leftarrow f}$ and $\bm{L}_p^{\leftarrow f}$ are the force and momentum due to the fluid-solid interaction, namely
\begin{equation}
\bm{f}^{\leftarrow f}   = \oint_{\partial V_p} \bm{\sigma} \cdot \bm{n} \text{d}A \ \text{and} \
\bm{L}_p^{\leftarrow f} = \oint_{\partial V_p} \bm{r} \times \left( \bm{\sigma} \cdot \bm{n} \right) \text{d} A,
\end{equation}
where $\bm{\sigma}= -p \mathcal{I} + 2 \mu \mathcal{D}$ is the Cauchy stress tensor, with $\mathcal{I}$ being the identity tensor, $\mu$ the fluid kinematic viscosity, $\mathcal{D}$ the strain rate tensor, and $\bm{n}$ the unit vector normal to the surface of the particle.

\subsection{One-way-coupled point particle simulations (PP-DNS)}

The interface-resolved simulations are complemented with one-way-coupled point particle simulations (PP-DNS). Here the particles move under the action of the fluid flow, but they do not modulate it; the particle-particle interactions are also neglected. In this work, we consider the complete governing equation for a point particle as introduced by \cite{maxey-riley-1983} and \cite{gatignol-1983}, i.e.
\begin{equation}
\label{eq: PP-momentum-setup}
\begin{aligned}[t]
    \rho_p V_p \frac{\text{d}\bm{u}_p}{\text{d}t} &= \underbrace{3 \pi D_p \rho_f \nu \left(\bm{u} - \bm{u_p} + \frac{1}{6} \left( \frac{D_p}{2} \right)^2 \bm{\nabla}^2 \bm{u} \right) }_{\text{Stokes drag}} \\ 
    &+ \underbrace{\frac{\rho_f V_p}{2} \left( 3\frac{D\bm{u}}{Dt} - \frac{\text{d}\bm{u}_p}{\text{d}t} + \frac{1}{10} \left( \frac{D_p}{2} \right)^2 \frac{\text{d} }{\text{d} t} \bm{\nabla}^2 \bm{u}  \right)}_{\text{Added mass}}  \\ 
    &\quad + \underbrace{\frac{3}{2} D_p^2 \rho_f \sqrt{\pi \nu} \int_{-\infty}^t K_B(t-\tau) \left( \frac{\text{d}\bm{u}}{\text{d}\tau} - \frac{\text{d}\bm{u_p}}{\text{d}\tau}+ 
       \frac{1}{6} \left( \frac{D_p}{2} \right)^2 \frac{\text{d}}{\text{d} t} \bm{\nabla}^2 \bm{u} \right) \text{d} \tau}_{\text{Basset force}}. 
\end{aligned}
\end{equation}
Here, $V_p = \pi \rho_p D_p^3/6$ is the volume of the particle, and $D/Dt$ denotes the material derivative. Note that the Fax\'{e}n correction \citep{faxen-1922} proportional to $\bm{\nabla}^2 \bm{u}$ has been included in the Stokes drag, added mass and Basset forces. According to \cite{homann-bec-2010}, this correction reproduces dominant finite-size effects on velocity and acceleration fluctuations for neutrally buoyant particles with diameter up to $D_p/\eta \approx 4$. For the added mass, we have used the form described by \cite{auton-hunt-1988}. The computation of the Basset history force presents some challenges, and its evaluation can become extremely time consuming and memory demanding; indeed, this term requires at each time step the computation of an integral over the complete time history of the particle. Over the years, several attempts have been proposed to approximate this term; see for example \cite{michaelides-1992}, \cite{dorgan-loth-2007} and \cite{prasath-etal-2019}. In this work, we resort on the second-order and memory-efficient algorithm developed by \citet{hinsberg-boonkkamp-clercx-2011}, whose details are briefly reported in appendix \S\ref{sec:basset} for completeness.

\subsection{Computational details}

We consider a single-phase micro-scale Reynolds number of $Re_\lambda = u' \lambda / \nu \approx 140$ to ensure a relatively large inertial range of scales; here $u'$ is the root mean square of the velocity fluctuations and $\lambda$ is the Taylor length scale. The particle diameter is set to $D_p/\eta \approx 0.9$, where $\eta$ is the Kolmogorov length scale for the single-phase case. Two volume fractions are considered, i.e. $\Phi_V = V_s/(V_s + V_f)= 10^{-5}$ and $\Phi_V = 10^{-3}$ for a total number of particles of $N=742$ and $N=74208$, respectively. For each volume fraction two values of the particle density are considered, $\rho_p/\rho_f = 5$ and $\rho_p/\rho_f=100$ to consider both light and heavy particles. This leads to a total of four PR-DNS. In PP-DNS the particles do not modulate the flow and do not interact, therefore, the volume fraction is not a parameter. Because of this, only two PP-DNS simulations have been carried out for the different density ratios.

The governing equations are numerically integrated in time using the in-house solver Fujin (\url{https://groups.oist.jp/cffu/code}). It solves the Navier--Stokes equations using an incremental pressure-correction scheme. The governing equations are written in primitive variables on a staggered grid, and second-order finite differences are used in all the directions. The Adams-Bashforth time scheme is used for advancing the momentum equation in time. The Poisson equation for the pressure enforcing the incompressibility constraint is solved using a fast and efficient approach based on the Fast Fourier Transform.

For the PR-DNS the governing equations for the particles are dealt with by the immersed boundary method introduced by \cite{hori-rosti-takagi-2022}. The fluid-solid coupling is achieved in an Eulerian framework, and accounts for the inertia of the fictitious fluid inside the solid phase, so as to properly reproduce the particles' behaviour in both the neutrally-buoyant case and in the presence of density difference between the fluid and solid phases. The soft sphere collision model \citep{cundall-strack-1979,tsuji-etal-1993} is used to prevent the interpenetration between particles and compute the $\bm{f}^{\leftrightarrow p}$ term in equation \ref{eq:NEeq}. In this model the particles are allowed to slightly penetrate. The collision is modelled as a spring and dashpot dynamical system, with the collision force being proportional to the penetration depth between the particles and to their relative normal velocity. A fixed-radius near neighbours algorithm \citep[see][and references therein]{monti-etal-2021} is used for the particle interaction to avoid an otherwise prohibitive increase of the computational cost when the number of particles becomes large. The PR-DNS simulations are performed without any additional lubrication correction, i.e. we consider only the lubrication that naturally arises from the method.
For the one-way-coupled PP-DNS the governing equation for the particle velocity is advanced in time using the second-order Adams-Bashforth time scheme. At each time step, the fluid velocity is evaluated at the position of the particle using a second-order linear interpolation.

For the PR-DNS the fluid domain is discretised on a cubic domain using $N_p = 2048$ points along each direction, leading to $\eta / \Delta x \approx D_p /\Delta x \approx 6-7$, where $\Delta x$ denotes the grid spacing. The accuracy of the results has been verified by running an additional simulation on a coarser grid with $N_p=1440$ for the case with $\Phi_V = 10^{-3}$ and $\rho_p/\rho_f = 100$, resulting in a negligible difference in the scale-by-scale fluid energy spectrum and budget in figures \ref{fig:spectrum} and \ref{fig:budget}, and in the Lagrangian and Eulerian particles' statistics in figures \ref{fig:kurt} and \ref{fig:Pstrfun}. For the single-phase case and the PP-DNS the fluid domain is discretised using $N_p = 1024$ points in the three directions, leading to $\eta/\Delta x \approx 3-4$. In this case the flow around the particles does not need to be solved. Excluding the initial transient period, all simulations are advanced in time for approximately $50 \tau_f$, where $\tau_f = \mathcal{L}/\sqrt{2\aver{E}/3}$ is the average turnover time of the largest eddies; $\mathcal{L} =\pi/(4\aver{E}/3)\int_0^\infty \mathcal{E}(\kappa)/\kappa \text{d} \kappa$ is the fluid integral scale with $\mathcal{E}(\kappa)$ being the energy spectrum, $E(\bm{x},t)$ is the local and instantaneous fluid kinetic energy, and the $\aver{\cdot}$ operator denotes averages in space and time. 

Details of the PR-DNS are reported in table \ref{tab:simulations}. Note that, when looking at the bulk quantities, the flow modulation due to the solid phase is rather low. For comparison, we mention that at a reference Reynolds number of $Re_\lambda \approx 240$ \cite{hwang-eaton-2006} experimentally report an energy attenuation of $\aver{E}/\aver{E_0} = 0.78$ for a suspension of spherical particles with $D_p/\eta = 0.9$, $\rho_p/\rho_f = 2048$ and $\Phi_V \approx 10^{-4}$, which corresponds to a mass loading of $\Phi_M = \rho_p V_p/(\rho_f V_f) = 0.1$ (the $\cdot_0$ subscript refers to the single phase case). Although those parameters differ from ours, the value of the mass loading matches that of the $\rho_p/\rho_f = 100$ and $\Phi_V = 10^{-3}$ case ($\Phi_M \approx 0.098$), for which we however found a lower energy attenuation, i.e. $\aver{E}/\aver{E_0} = 0.9 \pm 0.26$. This suggests that the value of the mass loading alone is not sufficient to predict the level of the turbulence attenuation by Kolmogorov-size particles. For completeness, we mention that our predictions are close to those obtained by two-way-coupled PP-DNS at similar parameters, although at lower Reynolds numbers. For example, \cite{boivin-simonin-squires-1998} predict an attenuation of $\aver{E}/\aver{E_0} \approx 0.88$ for $D_p/\eta = 0.11$, $Re_\lambda = 62$ and $\Phi_M = 0.2$. At $Re_\lambda = 38$, $D_p/\eta <1$, $\rho_p/\rho_f \gg 1$ and $\Phi_M = 0.1$, instead, \cite{squires-eaton-1990} report an attenuation of $\aver{E}/\aver{E_0} = 0.93$.

\begin{table}
\caption{Details of the PR-DNS considered in the present parametric study. 
$\epsilon$ is the dissipation rate. $St$ is the Stokes number defined as $St=\tau_p/\tau_f$, where $\tau_p=(\rho_p/\rho_f)D_p^2/(18 \nu)$ is the relaxation time of the particle and $\tau_f=\mathcal{L}/\sqrt{2\aver{E}/3}$ is the turnover time of the largest eddies. 
 $Re_p$ is the particle Reynolds number defined as $Re_p = |\Delta \bm{u}| D /\nu$, where $\Delta \bm{u} = \bm{u}_p - \bm{u}_f$ is the fluid-particle relative velocity. Here, $\bm{u}_f$ is the fluid velocity seen by the particle evaluated as the average of the fluid velocity within a shell centred with the particle and with radius $R_{sh} = 3(D_p/2)$ \citep[see][]{uhlmann-chouippe-2017,chiarini-rosti-2024}.} 
\label{tab:simulations}
\centering
\begin{tabular}{cccccccc}
$\Phi_V$ & $\rho_p/\rho_f$ & $N$ & $\aver{\eta}$ & $\aver{E}$ & $\aver{\epsilon}$ & $St$ & $Re_p$ \\
\hline
$-$   &  $-$         & $-$  & $0.0190 \pm 0.0012$ & $60.69 \pm 12.38$ & $63.34 \pm 15.98$ &  $-$  & $-$    \\
      &                     &                     &                       &               &          &  & \\
$10^{-5}$ & $5$ & $742$    & $0.0189 \pm 0.001$ & $63.15 \pm 08.64$ & $64.69 \pm 12.49$  &   $0.0043$ & $0.274\pm0.032$  \\
$10^{-5}$ & $100$ & $742$   & $0.0187 \pm 0.001$ & $61.70 \pm 09.81$ & $58.55 \pm 9.74$ &  $0.1287$ & $2.48\pm0.32$  \\
      &       &                    &         &             & &  \\
$10^{-3}$ & $5$ & $74208$    & $ 0.0190 \pm 0.0004$ & $60.32 \pm 2.75$ & $61.81 \pm 6.40$ &  $0.0038$ & $0.265\pm0.018$  \\
$10^{-3}$ & $100$ & $74208$   & $ 0.0192 \pm 0.0008$ & $54.09 \pm 6.74$ & $ 57.25 \pm 9.12$ &  $0.0737$ & $1.89\pm0.18$  \\
\end{tabular}
\end{table}

\section{Flow modulation}
\label{sec:flow}

In this section we discuss the PR-DNS results and focus on the influence of the particles on the carrier flow. First, we show the influence of the solid phase on the energy spectrum, on the structure functions and on the the scale-by-scale energy budget. Next, the influence of the particles on the local structure of the small scales of the flow is addressed.

\subsection{Energy Spectrum}
\label{sec:spectrum}

\begin{figure}
  \centering
  \includegraphics[width=0.85\textwidth]{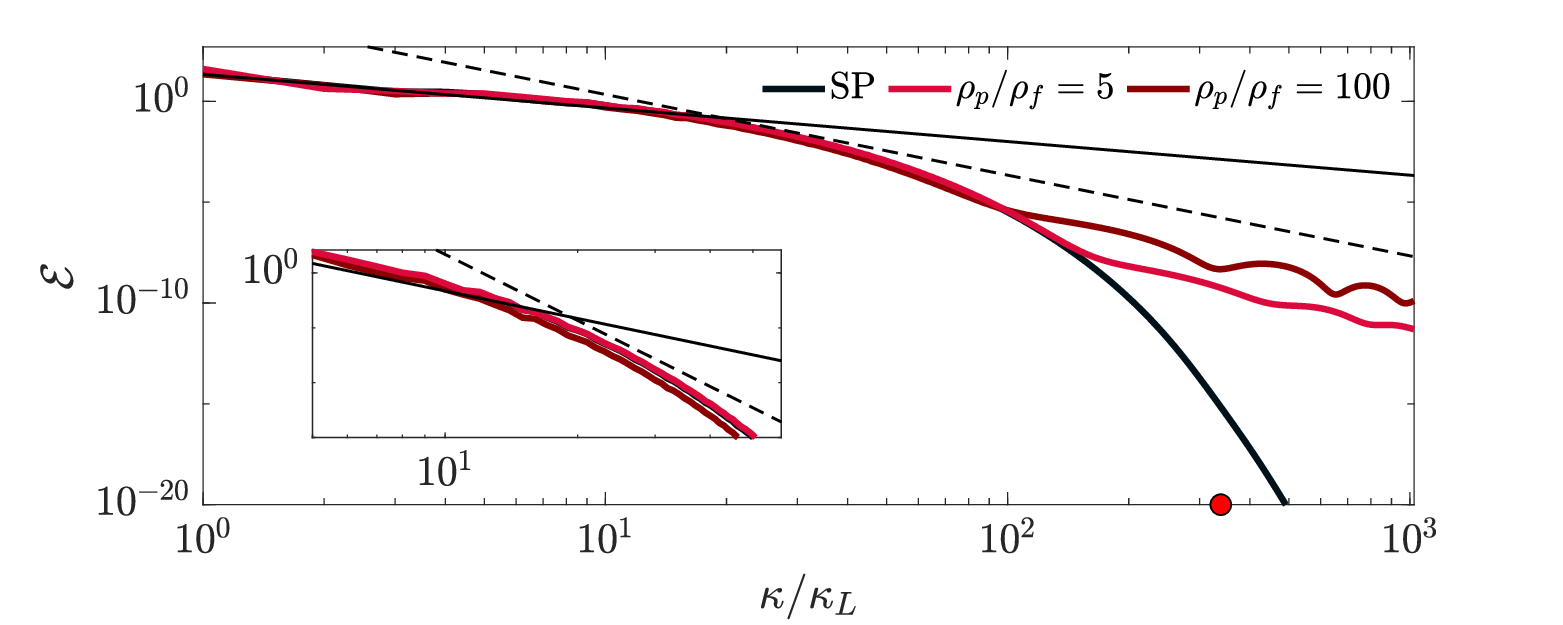}
  \includegraphics[width=0.85\textwidth]{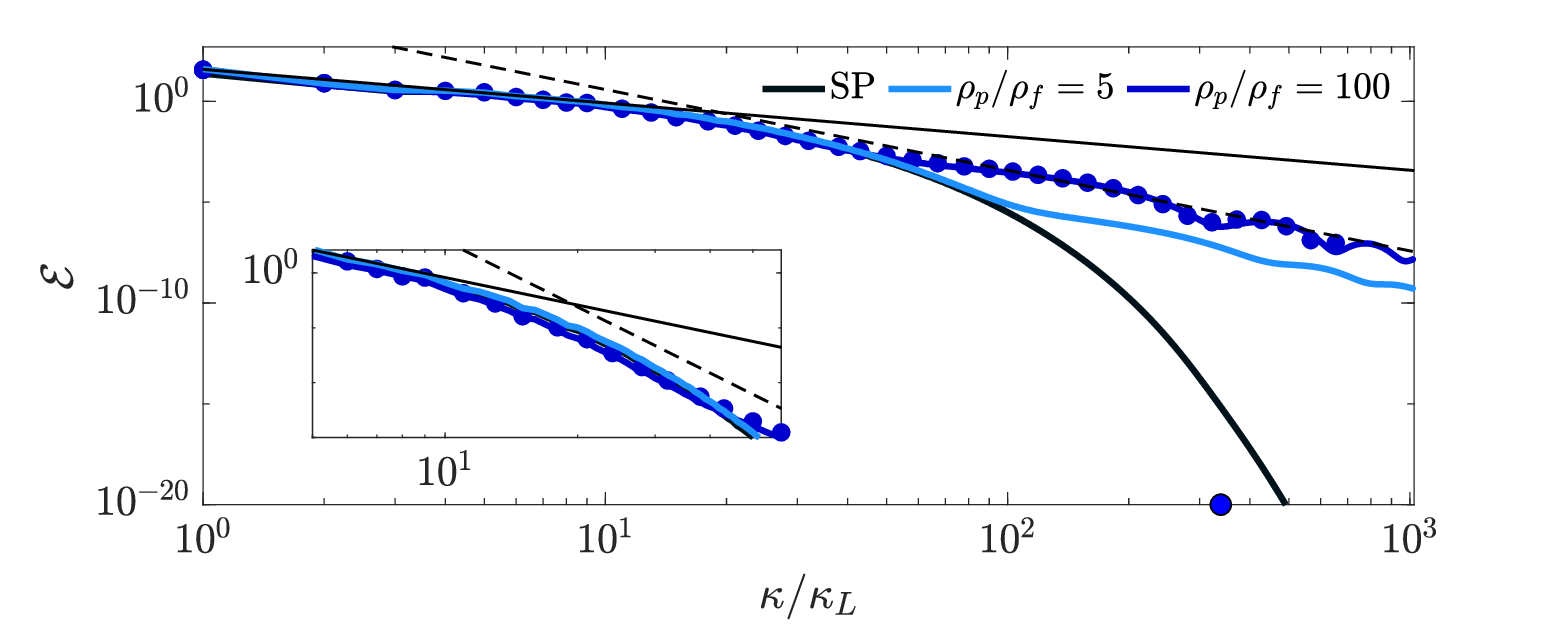}
  \caption{Energy spectrum for (top) $\Phi_V = 10^{-5}$ and (bottom) $\Phi_V = 10^{-3}$. The black line refers to the single-phase case, the red/blue light line is for $\rho_p/\rho_f = 5$, an the red/blue dark line is for $\rho_p/\rho_f = 100$. The symbols in the bottom panel are from the simulation carried out with the coarser grid. The filled circle on the $\kappa$ axis denotes the wavenumber associated with the particle size. The solid line is for the Kolmogorov $\kappa^{-5/3}$ scaling. The dashed line is for $\kappa^{-4}$. Here $\kappa_L = 2 \pi/L$.}
  \label{fig:spectrum}
\end{figure}
Figure \ref{fig:spectrum} shows the influence of the solid phase on the energy spectrum $\mathcal{E}(\kappa)$, and highlights how Kolmogorov-size particles modulate the turbulent fluctuations scale-by-scale. The top and bottom panels are for $\Phi_V = 10^{-5}$ and $\Phi_V=10^{-3}$, respectively. The black solid line refers to the reference unladen case; note the inertial range of scales where $\mathcal{E}(\kappa) \sim \kappa^{-5/3}$ extends for more than one decade of wavenumbers, confirming that the present Reynolds number is large enough to ensure a proper separation of scales. For validation purposes, in the bottom panel we also plot with symbols the energy spectrum obtained for the $\Phi_V=10^{-3}$ and $\rho_p/\rho_f=100$ case with the coarser grid, showing good agreement with that obtained with the standard grid, thus ensuring the suitability of the chosen grid resolution (see \S\ref{sec:method}). We compute the energy spectra (and the structure functions in the next section) using all the grid points of the computational domain including those that are inside the particles. However, as mentioned in \cite{chiarini-rosti-2024}, with the present IBM the results do not change when the points within the particles are neglected.

At large scales, the spectra of the particle-laden cases substantially overlap with the unladen spectrum. We indeed observe only a weak depletion of the energy content at the intermediate scales; see the insets in the two panels. At scales smaller than a certain wavenumber $\kappa_p$, the energy spectra of the particle-laden cases deviate above the reference spectrum. Solid particles enhance the energy content of the small scales by locally deforming the fluid flow around them. Notably, this mechanism is amplified as $\Phi_V$ and/or $\rho_p/\rho_f$ increase, as conveniently visualised by the larger values of $\mathcal{E}(\kappa)$ for $\kappa>\kappa_p$ and by the shift of $\kappa_p$ towards smaller wavenumbers. However, notice that due to the low values of $\Phi_V$ considered, the flow modulation is rather low for all cases, being substantially negligible for $\Phi_V=10^{-5}$ and/or $\rho_p/\rho_f=5$.

Figure \ref{fig:spectrum} shows that the modulated energy spectrum exhibits multiscaling behaviour. The classical $\mathcal{E}(\kappa) \sim \kappa^{-5/3}$ decay in the inertial range of scales is indeed followed by a steeper decay $\mathcal{E}(\kappa) \sim \kappa^{-4}$ starting at wave numbers close to $\kappa_p$.
A similar steep decay has been observed in bubbly flows at scales smaller than the bubble diameter, by means of both experiments \citep{mercado-etal-2010,riboux-risso-legendre-2010,prakash-etal-2016,dung-etal-2022} and simulations \citep{pandey-ramadugu-perlekar-2020,pandey-mitra-perlekar-2022}. 
Additionally, a similar multiscaling behaviour has also been observed in a turbulent planar Couette flow laden with small particles \citep{wang-etal-2023}, and in homogeneous isotropic turbulence laden with slender fibres \citep{olivieri-cannon-rosti-2022,olivieri-mazzino-rosti-2022}. 
In the context of bubbly flows, the emergence of the $\kappa^{-\alpha}$ decay with $\alpha \ge 3$ has been attributed to the wakes the bubbles generate in an otherwise smooth flow \citep[see for example][]{almeras-etal-2017,pandey-ramadugu-perlekar-2020}; here the velocity fluctuations produced by the bubbles are directly dissipated by viscosity \citep{lance-betaille-1991,risso-2018}. Accordingly, this scaling has been indeed observed only when the bubble Reynolds number is large enough and is in the $10 \le Re_{bub} \le 1000$ range \citep{pandey-ramadugu-perlekar-2020}. 
To compare, we computed the local particle Reynolds number using the relative velocity between the particle and the surrounding flow, and found that it is in the $0.25 \lessapprox Re_p \lessapprox 2.5$ range (see table \ref{tab:simulations}). More recently, \cite{zamansky-etal-2024} observed that in bubbly flows the $\kappa^{-3}$ range constitutes a transition between the production-dominated (anisotropic) and the inertial/dissipative (isotropic) ranges of scales, and proposed that the specific $\alpha=3$ scaling results from the mean shear rate imposed by the bubbles which controls the rate of return to isotropy. In view of this, we mention that unlike what is commonly found in bubbly-induced agitation where the $\kappa^{-3}$ subrange is between the energetic and the inertial range of scales \citep[see also figure 5 of][]{risso-2018}, in our case the $\kappa^{-4}$ scaling is observed at the small dissipative scales.
It is also worth mentioning that in a recent work \cite{ramirez-etal-2024} investigated the effect of singularities (e.g. induced by the particles) of various orders on the energy spectrum. They showed that singularities may affect the spectrum beyond the imposition of simple oscillations \citep{lucci-etal-2010}, and actually cause power law scalings at wave numbers $\kappa\geq\kappa_p$, with the exact slope being dependent on the order of the singularity. However, in most of the cases investigated by us, the start of the $\kappa^{-4}$ scaling region appears at wave numbers sensibly lower than $\kappa_p$.  
Further investigation on the link between the fluctuations induced by particles and the $\mathcal{E}(\kappa) \sim \kappa^{-4}$ decay is provided in \S\ref{sec:budget} by looking at the scale-by-scale energy budget. 

Overall, figure \ref{fig:spectrum} shows that at the present parameters particles almost do not modulate the inertial range of scales, where the classical energy cascade described by Richardson and Kolmogorov is preserved, but mainly affect the (otherwise smooth) smallest scales of the flow. 

\subsection{Structure function and intermittency}
\label{sec:strfun}

\begin{figure}
  \centering
  \includegraphics[width=0.85\textwidth]{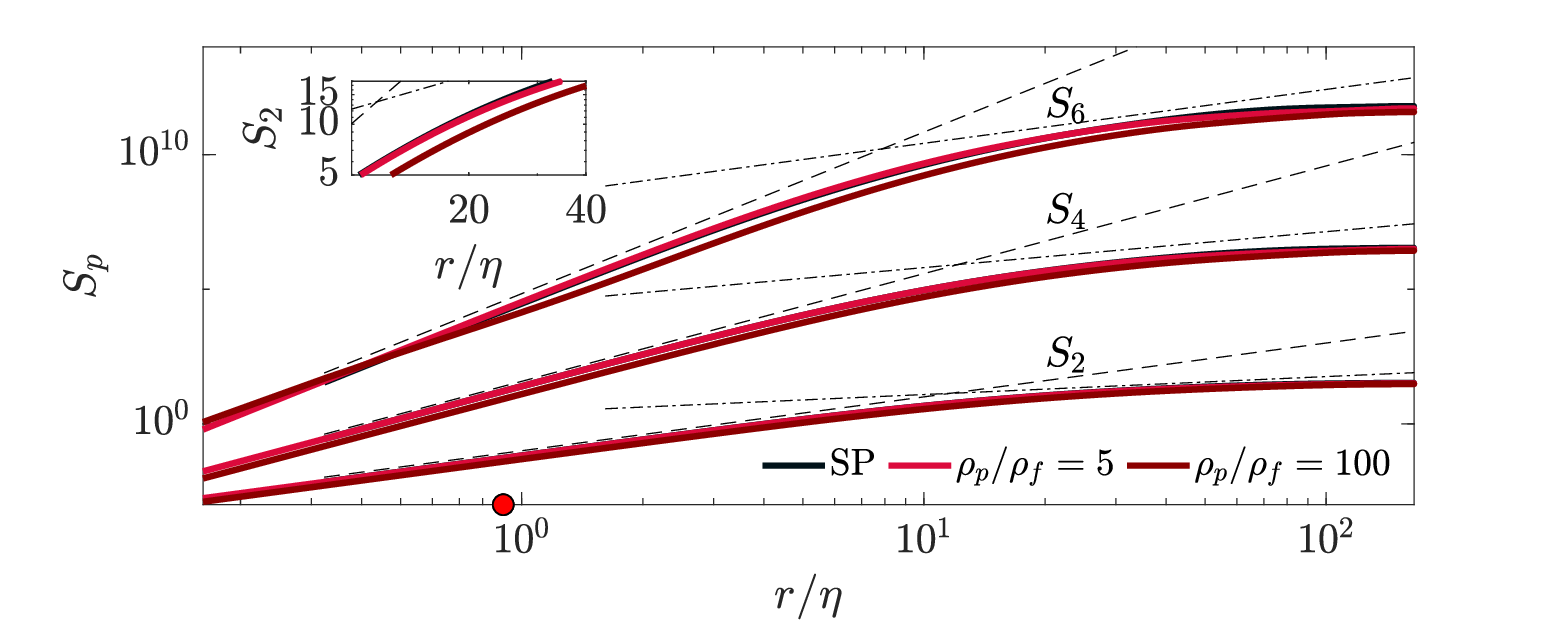}
  \includegraphics[width=0.85\textwidth]{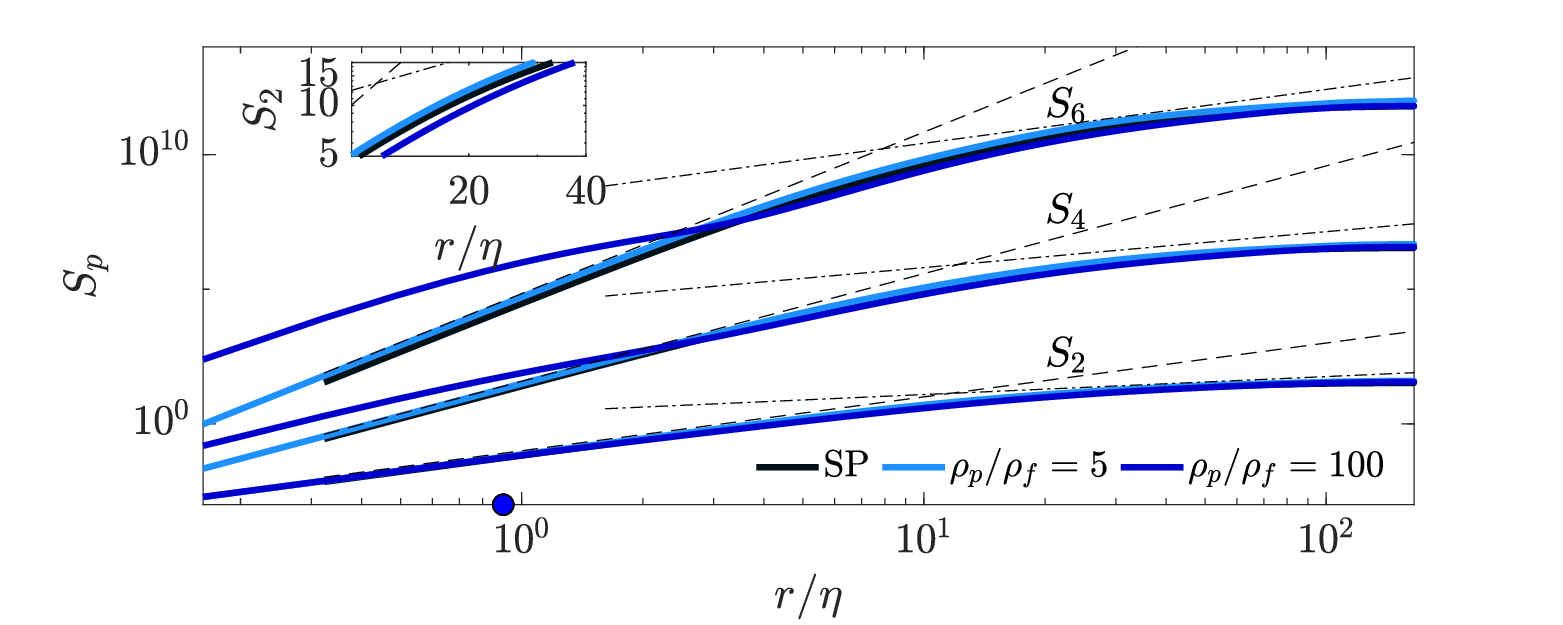}
  \caption{Structure functions for (top) $\Phi_V = 10^{-5}$ and (bottom) $\Phi_V = 10^{-3}$. For each panel, from bottom to top the plots are for $S_2$, $S_4$ and $S_6$. The dashed lines represent $r^p$, while the dash-dotted ones $r^{p/3}$. The insets show a zoom of $S_2$ for $10 \le r/\eta \le 40$.}
  \label{fig:strfun}
\end{figure}
We extend the analysis done in the spectral domain by computing the longitudinal structure functions defined as $S_p(r) = \aver{\delta u(r)^p}$ where $\delta u(r) = (\bm{u}(\bm{x}+\bm{r}) - \bm{u}(\bm{x}) )\cdot \bm{r}/r$ and $r = |\bm{r}|$. In particular, figure \ref{fig:strfun} plots $S_2$, $S_4$ and $S_6$ as a function of $r$ for (top) $\Phi_V=10^{-5}$ and (bottom) $\Phi_V = 10^{-3}$. In the single-phase case, we observe that $S_p \sim r^{p/3}$ in the inertial range of scales is in agreement with the Kolmogorov prediction \citep{kolmogorov-1941}, and that $S_p \sim r^p$ at the small scales, as a result of the differentiability of the fluid velocity field \citep{schumacher-etal-2007}. Recall that although $S_2(r)$ is commonly referred to as scale energy \citep{davidson-2004,davidson-pearson-2005}, its meaning slightly differs from $\mathcal{E}(\kappa)$. Indeed while $\mathcal{E}(\kappa) \text{d}\kappa$ refers to the amount of energy associated with the scale $r = 2\pi/\kappa$, $S_2(r)$ can be interpreted as the amount of energy associated with scales up to $r$.
In agreement with the modulation of the energy spectrum, figure \ref{fig:strfun} shows that particles enhance the energy content at small scales compared to the unladen case. The energy enhancement is more intense for larger $\Phi_V$ and $\rho_p/\rho_f$, and becomes more evident when considering higher order structure functions. At the same time, for $\rho_p/\rho_f=100$ the presence of the particles decreases the amount of energy stored at the larger scales (as seen by the dark blue and dark red curves laying below the black one in the insets of figure \ref{fig:strfun}). By interacting with the vortical structures of the flow, the (heavy) particles drain energy from scales larger than $D$ and reinject it back at smaller scales. 


\begin{figure}
  \centering
  \includegraphics[width=0.85\textwidth]{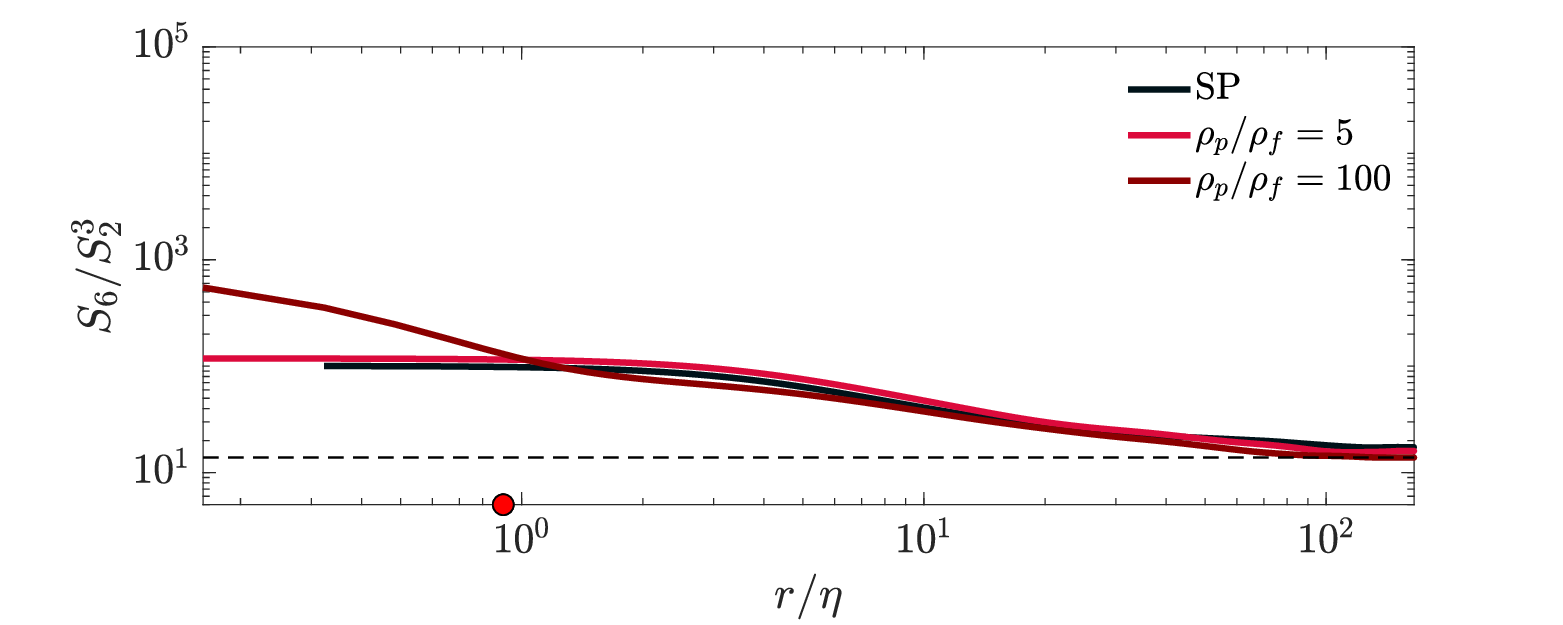}
  \includegraphics[width=0.85\textwidth]{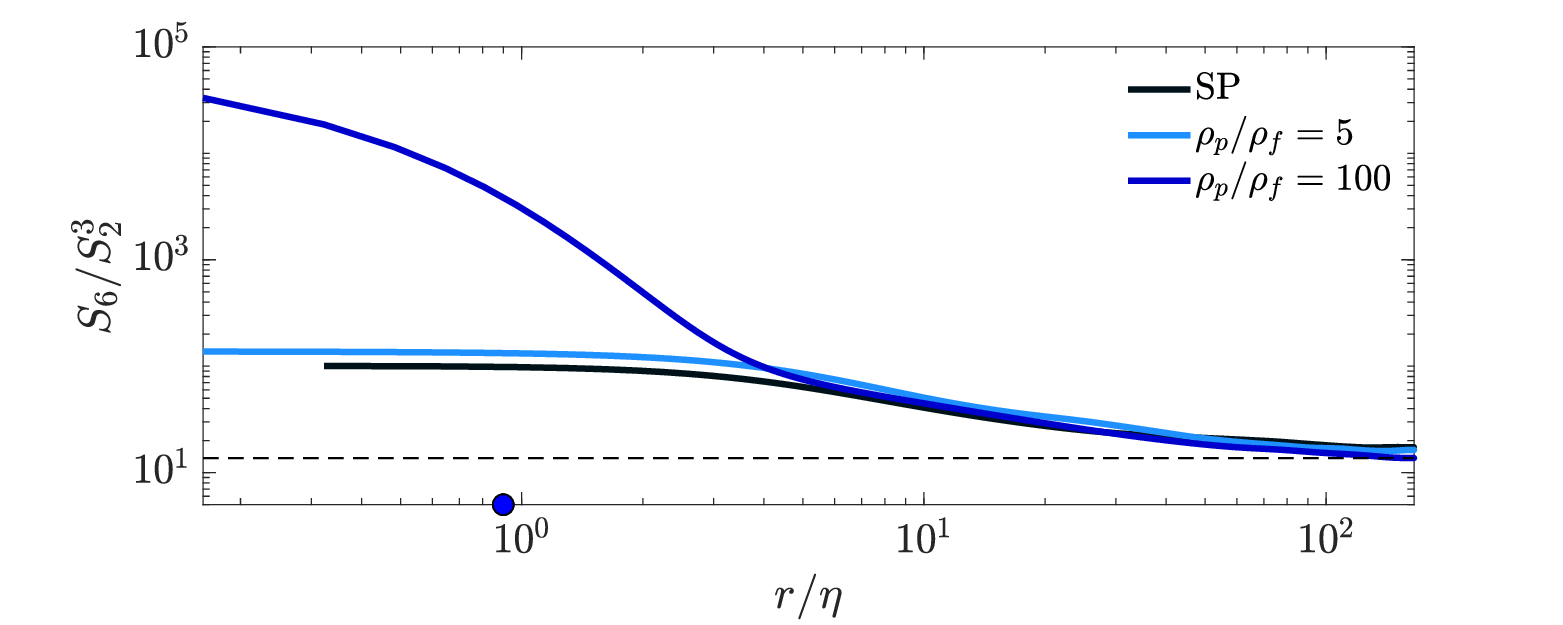}
  \caption{Extendend self-similarity for (top) $\Phi_V = 10^{-5}$ and (bottom) $\Phi_V = 10^{-3}$. $S_6/S_2^3$ is plotted against $r$.}
  \label{fig:ess}
\end{figure}
Structure functions are commonly employed to quantify the flow intermittency, i.e. the relevance of extreme events that are localised in space and time and break the Kolmogorov similarity hypothesis \citep{frisch-1995,pope-2000}. In figure \ref{fig:ess}, we use the extended self similarity introduced by \cite{benzi-1993}, and plot $S_6/S_2^3$ as a function of $r$. In the limit case where extreme events do not occur, the $S_6 \sim S_2^3$ power law holds, i.e. $S_6/S_2^3 \sim constant$, and deviations from this behaviour are a measure of the flow intermittency. Accordingly with the intrinsic intermittent nature of turbulent flows, figure \ref{fig:ess} shows that $S_6/S_2^3$ deviates from the Kolmogorov prediction also in the single-phase case, and this deviation increases in the particle-laden cases, similarly to what is found for suspension of particles with size in the inertial range of scales \citep{chiarini-rosti-2024}. This is due to the no-slip and no-penetration boundary conditions at the particles surface that give origin to localised and intense velocity gradients. Figure \ref{fig:ess} shows that for $\rho_p/\rho_f = 5$ the deviation from the single-phase case is rather small for both $\Phi_V = 10^{-5}$ and $\Phi_V=10^{-3}$, in agreement with the low level of flow modulation discussed above. For $\rho_p/\rho_f = 100$, instead, the $S_6/S_2^3$ curve significantly deviates from the single-phase case at small scales. For the considered parameters, heavy Kolmogorov-size particles increase the intermittency of the small scales of the flow. 

\subsection{Scale-by-scale energy budget}
\label{sec:budget}

We now detail the influence of the particles on the organisation of the fluctuations, by studying the scale-by-scale energy budget equation and characterising thus the dominant energetic mechanisms in the different regimes identified with the energy spectra (see figure \ref{fig:spectrum}). For the present case with three homogeneous directions, the energy balance reads
\begin{equation}
P(\kappa) + \Pi(\kappa) + \Pi_{fs}(\kappa) - D_v(\kappa) = 0,
\label{eq:bud}
\end{equation}
where $P(\kappa)$ is the scale-by-scale turbulent energy production due to the external forcing, $\Pi(\kappa)$ is the energy flux associated with the nonlinear convective term, $\Pi_{fs}(\kappa)$ is the fluid-solid coupling term, and $D_v(\kappa)$ is the scale-by-scale viscous dissipation. Specifically, these terms are defined as

\begin{align}
P(  \kappa) = & \int_\kappa^\infty  \frac{1}{2} \left( \hat{\bm{f}} \cdot \hat{\bm{u}}^* + \hat{\bm{f}}^* \cdot \hat{\bm{u}} \right) \text{d} k, \\
\Pi(\kappa) = & \int_\kappa^\infty -\frac{1}{2} \left( \hat{\bm{G}} \cdot \hat{\bm{u}}^* + \hat{\bm{G}}^* \cdot \hat{\bm{u}} \right) \text{d} k, \\
\Pi_{fs}(\kappa) = & \int_\kappa^\infty  \frac{1}{2} \left( \hat{\bm{f}}^{\leftrightarrow p} \cdot \hat{\bm{u}}^* + \hat{\bm{f}}^{\leftrightarrow p,*} \cdot \hat{\bm{u}} \right) \text{d} k, \\
D_v(\kappa) = & \int_\kappa^\infty \left( 2 \nu k^2 \mathcal{E} \right) \text{d} k,
\end{align}
where $\hat{\cdot}$ denotes the Fourier transfrom operator, and the superscript $\cdot^*$ denotes complex conjugate. The term $\hat{\bm{G}}$ is the Fourier transform of the nonlinear term $\bm{\nabla} \cdot ( \bm{u} \bm{u} )$. Note that here we integrate all the terms from $\kappa$ to $\infty$. $\Pi(\kappa)$ and $\Pi_{fs}(\kappa)$ do not produce nor dissipate energy at any scale, but redistribute it among scales by means of the classical energy cascade and of the fluid-solid interaction. Also, note that since we integrate the viscous term from $\kappa$ to $\infty$, $D_v(0) = \epsilon$. For the complete derivation we refer the reader to \cite{pope-2000}.

\begin{figure}
  \centering
  \includegraphics[width=0.49\textwidth]{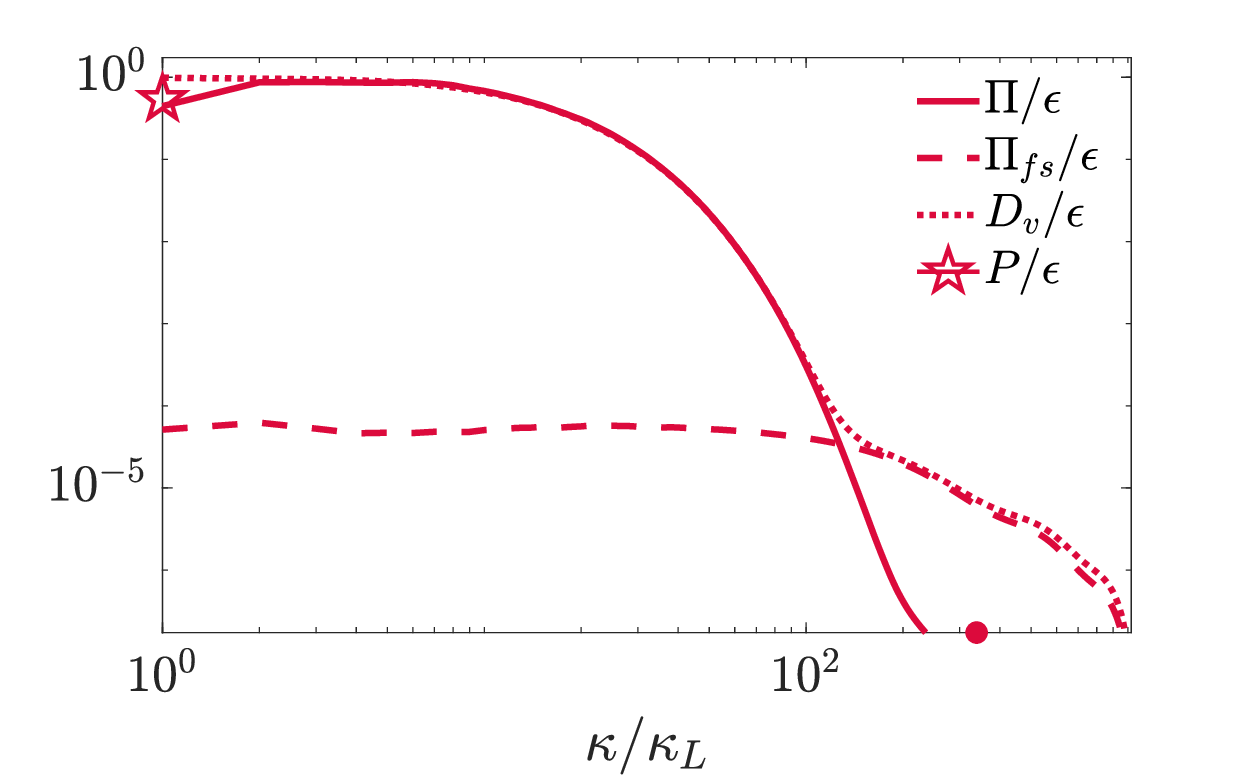}
  \includegraphics[width=0.49\textwidth]{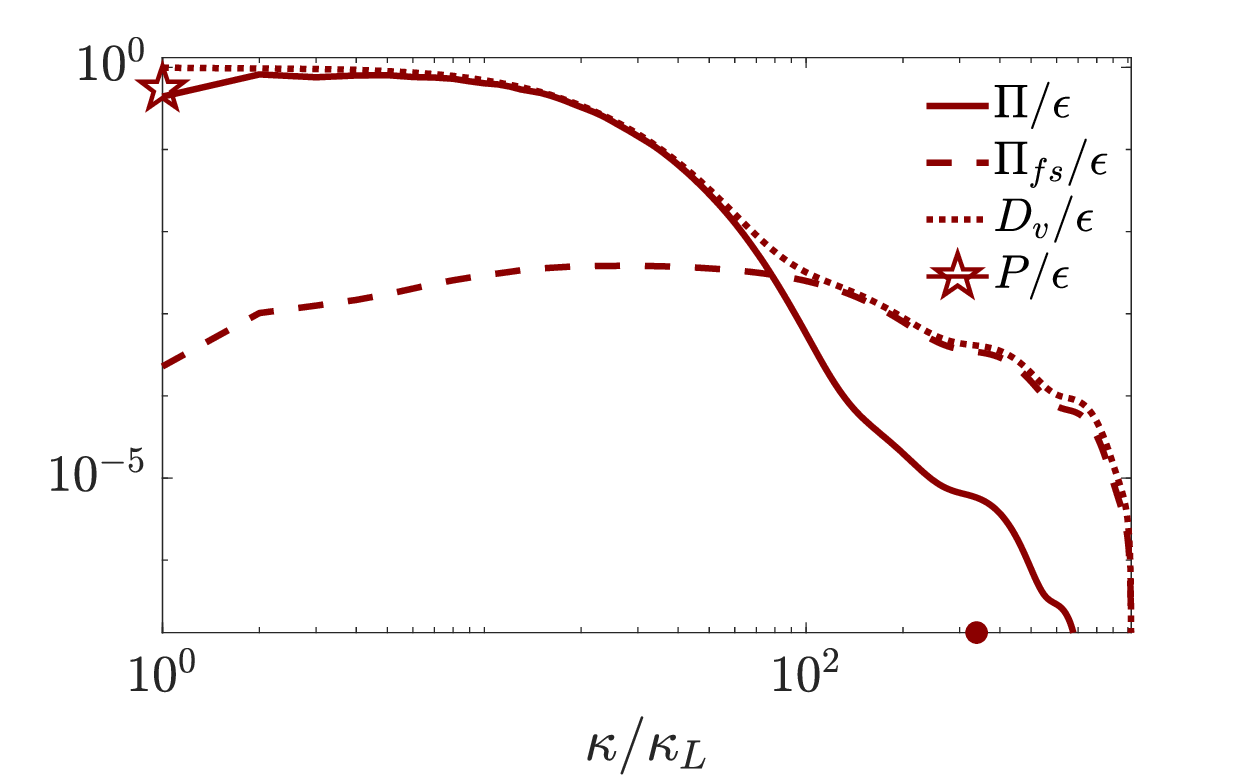}
  \includegraphics[width=0.49\textwidth]{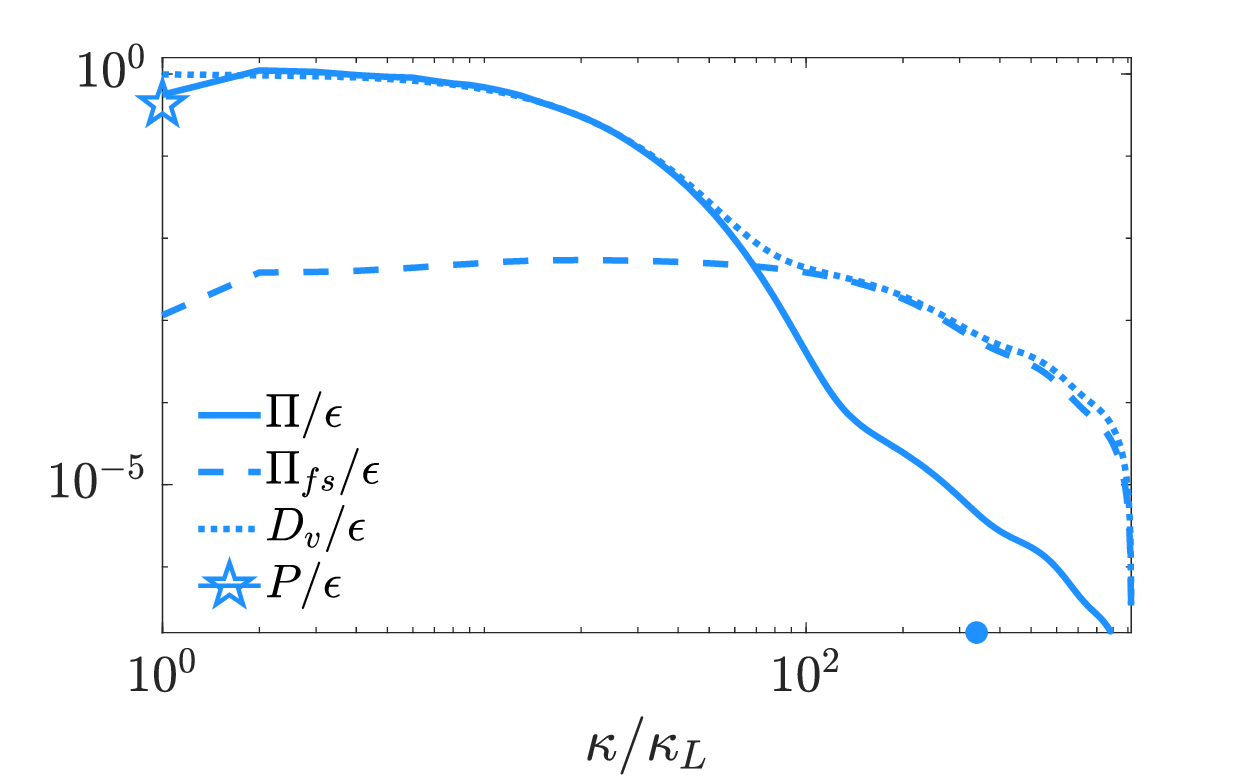}
  \includegraphics[width=0.49\textwidth]{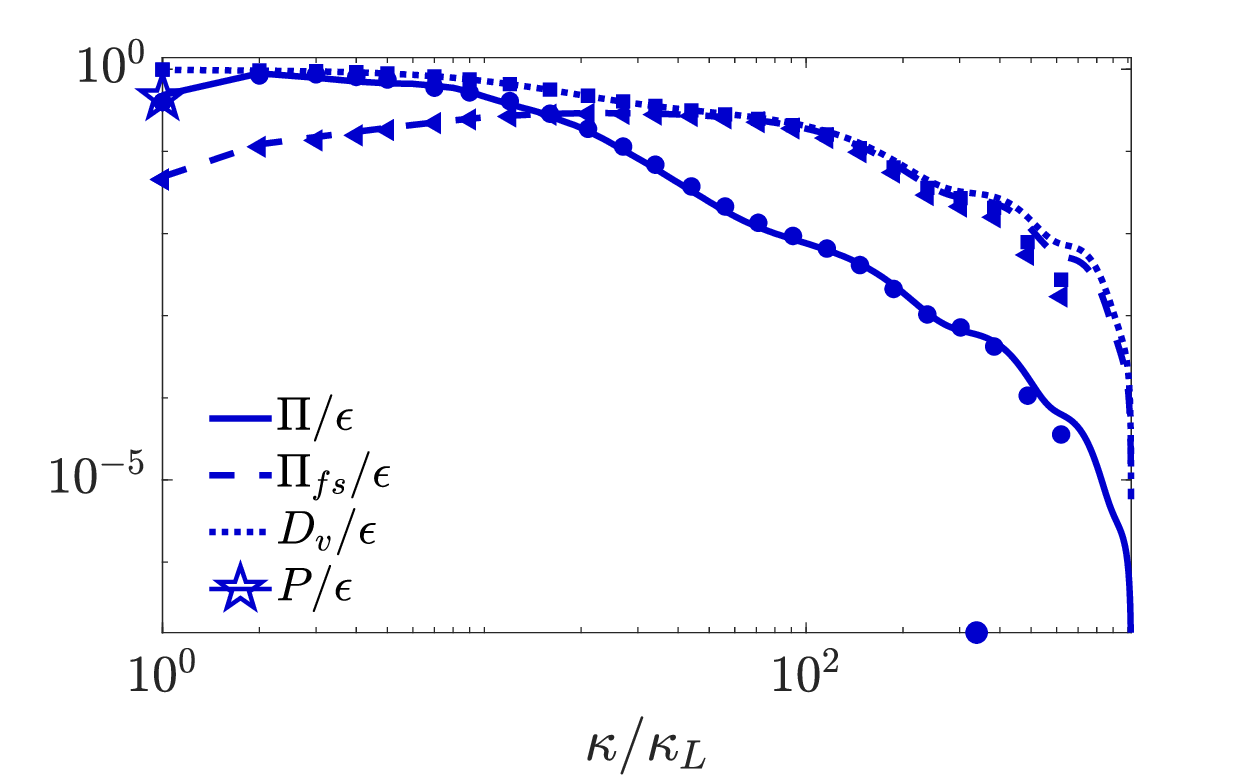}
  \caption{Scale-by-scale energy budget for (top) $\Phi_V = 10^{-5}$ and (bottom) $\Phi_V = 10^{-3}$. Plots in the left column are for $\rho_p/\rho_f = 5$, and those in the right for $\rho/\rho_f = 100$. The production term $P$ acts at the largest scales $\kappa/\kappa_L \le 1$ only, being $P=0$ for $\kappa/\kappa_L \ge 2$ (not visible as the $y$ axes is in log scale). The filled symbols in the bottom right panel are from the simulation carried out with the coarser grid; circles are for $\Pi/\epsilon$, triangles for $\Pi_{fs}/\epsilon$ and squares for $D_v/\epsilon$.}
  \label{fig:budget}  
\end{figure}
Figure \ref{fig:budget} plots the terms of equation \ref{eq:bud} as a function of $\kappa$ for the four particle-laden cases investigated. For validation purposes, we also plot with filled symbols the terms obtained with the coarser grid for $\Phi_V=10^{-3}$ and $\rho_p/\rho_f=100$. In agreement with the multiscaling behaviour shown in the energy spectra (see figure \ref{fig:spectrum}), the energy budgets exhibit two distinct behaviours. Energy is injected at the largest scales at a rate that equals the dissipation rate $P(0) = \epsilon$. In the inertial range of scales $\kappa_L < \kappa< \kappa_p$, the fluid-solid coupling term is subdominant and
\begin{equation}
  \Pi(\kappa) \approx D_v(\kappa) \approx \epsilon.
\end{equation}
Thus, $\Pi(\kappa) \approx \epsilon$ is constant with $\kappa$ at these scales, exhibiting a plateau. In agreement with the Kolmogorov theory, the viscous effects are negligible $2 \nu \kappa^2 \mathcal{E}(\kappa) \approx 0$ and energy is transferred from larger to smaller scales at a rate that matches the energy injection rate $\epsilon$. This corresponds to the range of scales where $\mathcal{E}(\kappa) \sim \kappa^{-5/3}$. Similarly to what observed in the energy spectrum, the range of scales where this relation holds shrinks as $\Phi_V$ and $\rho_p/\rho_f$ increase. For the small scales with $\kappa > \kappa_p$ where the spectrum shows the $\mathcal{E}(\kappa) \sim \kappa^{-4}$ decay, instead, the nonlinear flux is negligible $\Pi(\kappa) \approx 0$. In this range of scales, the viscous effects and the fluid-solid coupling term are not negligible, and the energy budget reduces to
\begin{equation}
  \Pi_{fs}(\kappa) \approx D_v(\kappa).
\end{equation}
Here the fluctuations that are produced by the fluid-solid interaction are (on average) directly dissipated by viscosity and are not transferred among scales. 
In agreement with the energy spectrum, the range of scales where this regime holds widens as $\Phi_V$ and/or $\rho_p/\rho_f$ are increased. 

\subsection{The local structure of the flow}
\label{sec:qr}

As shown in \S\ref{sec:spectrum}, Kolmogorov-size particles mainly modify the organisation of the velocity fluctuations at the smallest scales. Particles indeed modulate the energy spectrum and the structure functions for $\kappa \gtrapprox \kappa_p$ only. Here we characterise the smallest scales of the flow to provide new insights of the influence of the dispersed phase on the structure of the velocity fluctuations. For the sake of brevity, in this section we consider the $\Phi_V = 10^{-3}$ cases only, 
and we investigate how particles modify the velocity gradient field $A_{ij} = \partial u_j/\partial x_i$. In the neighbourhood of a given point $(\bm{x}_0,t)$, indeed, the velocity field can be approximated as $u_i(\bm{x},t) = u_i(\bm{x}_0,t) + A_{ij}(\bm{x}_0,t)(x_j-x_{0,j}) + \mathcal{O}(|\bm{x}-\bm{x}_0|^2)$.
This linear expansion is valid in the region around $\bm{x}_0$, where the fluid is sufficiently smooth and the variations of $A_{ij}$ are small \citep{meneveau-2011}; for a turbulent flow the extent of this region is of the order of the Kolmogorov scale $\eta$. Based on these arguments, we study the influence of the particles on the smallest scales of the flow, by inspecting their effect on the $A_{ij}$ tensor.

We decompose $A_{ij}$ into its symmetric and antisymmetric parts, namely the strain-rate tensor $S_{ij} = (A_{ij} + A_{ji})/2$ and the rotation rate tensor $W_{ij} = (A_{ij} - A_{ji})/2$. The field of the velocity gradient is completely addressed when knowing: \textit{(i)} the three principal rates of strain $\alpha \ge \beta \ge \gamma$, i.e. the three eigenvalues of $S_{ij}$, \textit{(ii)} the magnitude of the vorticity $\omega^2 = \bm{\omega} \cdot \bm{\omega}$, i.e. the enstrophy, and \textit{(iii)} the orientation of $\bm{\omega}$ relative to the three principal axes of strain, i.e. the eigenvectors of $S_{ij}$ \citep{davidson-2004}. Note that, due to the incompressibility constraint $\alpha + \beta + \gamma = 0$, meaning that $\alpha$ is always nonnegative, $\gamma$ is always nonpositive while $\beta$ can have any sign depending on the local straining state.

\begin{figure}
\centering
\includegraphics[width=0.85\textwidth]{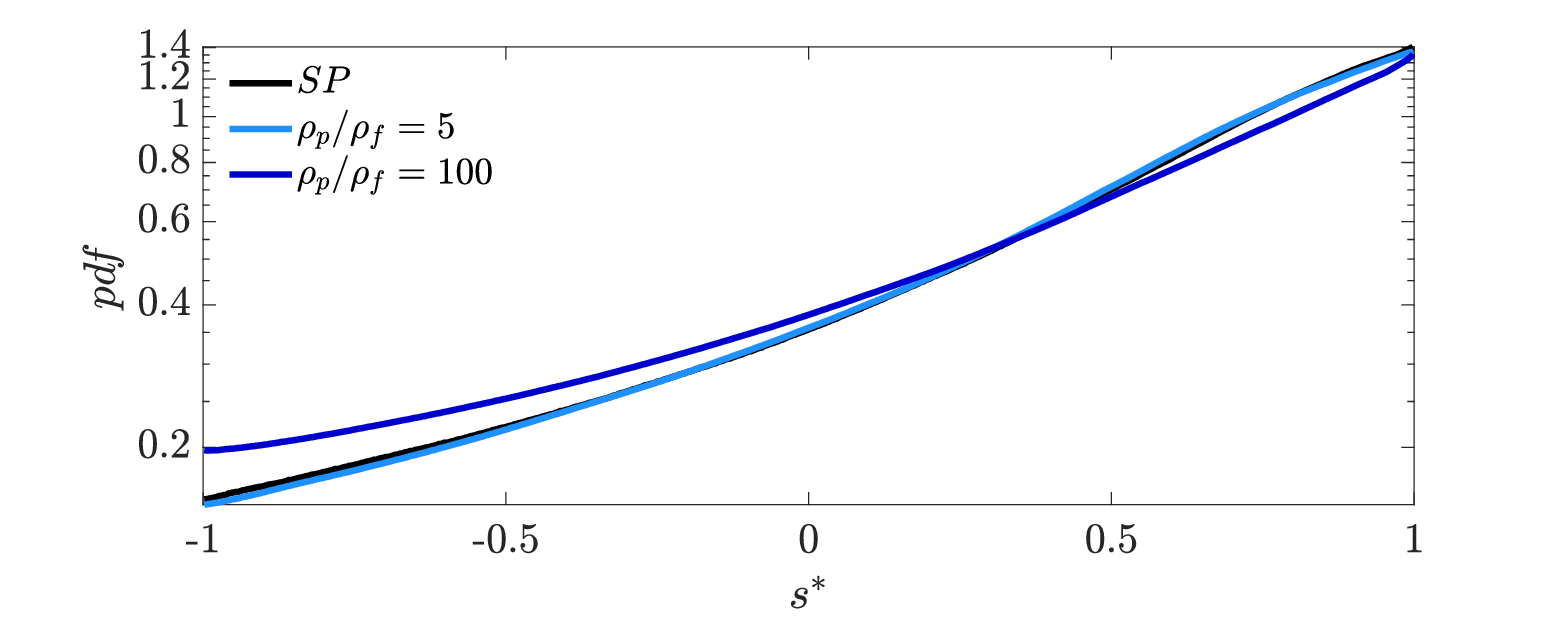}
\caption{Probability density function of $s^*$ for $\Phi_V = 10^{-3}$, with $\rho_p/\rho_f=5$ and $\rho_p/\rho_f=100$.}
\label{fig:sstar}
\end{figure}
We start by characterising $\alpha$, $\beta$ and $\gamma$ in figure \ref{fig:sstar}. Following \cite{lund-rogers-1994}, we use $s^*$ which is defined as
\begin{equation}
  s^* = - \frac{3 \sqrt{6} \alpha \beta \gamma }
               {(\alpha^2 + \beta^2 + \gamma^2)^{3/2}}.
\end{equation}
For a random velocity gradient field with no preferred structure, the distribution of $s^*$ is uniform. When $s^*=1$, $\alpha = \beta= - \gamma/2$, meaning that the state of straining is an axisymmetric extension, in which a small spherical fluid element moving in the flow extends symmetrically in two directions and contracts in the third one, forming thus a disk-like structure. When $s^*=-1$, instead, $\gamma = \beta = - \alpha/2 <0$, and the state of straining is an axisymmetric compression, in which a small fluid element contracts in two directions and extends in the third one, forming thus a vortex tube. Finally, when $s^*=0$ we have $\beta=0$, meaning that the straining state is two-dimensional, as typical for shear dominated regions.

In the absence of particles, the distribution peaks at $s^*=1$, in agreement with the fact that for purely Newtonian turbulence the most likely state of straining is an axisymmetric extension \citep{davidson-2004,meneveau-2011}. Figure \ref{fig:sstar} shows that the addition of Kolmogorov-size particles with $\Phi_V \le 10^{-3}$ leads to a rather small variation of the distribution of $s^*$ for the lighter particles with $\rho_p/\rho_f = 5$. We observe instead that particles with $\rho_p/\rho_f = 100$ decrease the probability of events with large positive $s^*$ and increase the probability of events with $s^* \le 0$. This agrees with the observation of \cite{cannon-olivieri-rosti-2024} who found that particles with size in the inertial range favour the occurrence of events with two-dimensional straining states and with axisymmetric compression. These events indeed are associated with the shear layers that separate from the surface of the particles, that strengthen as $\rho_p/\rho_f$ increases. 

\begin{figure}
\centering
\includegraphics[width=0.85\textwidth]{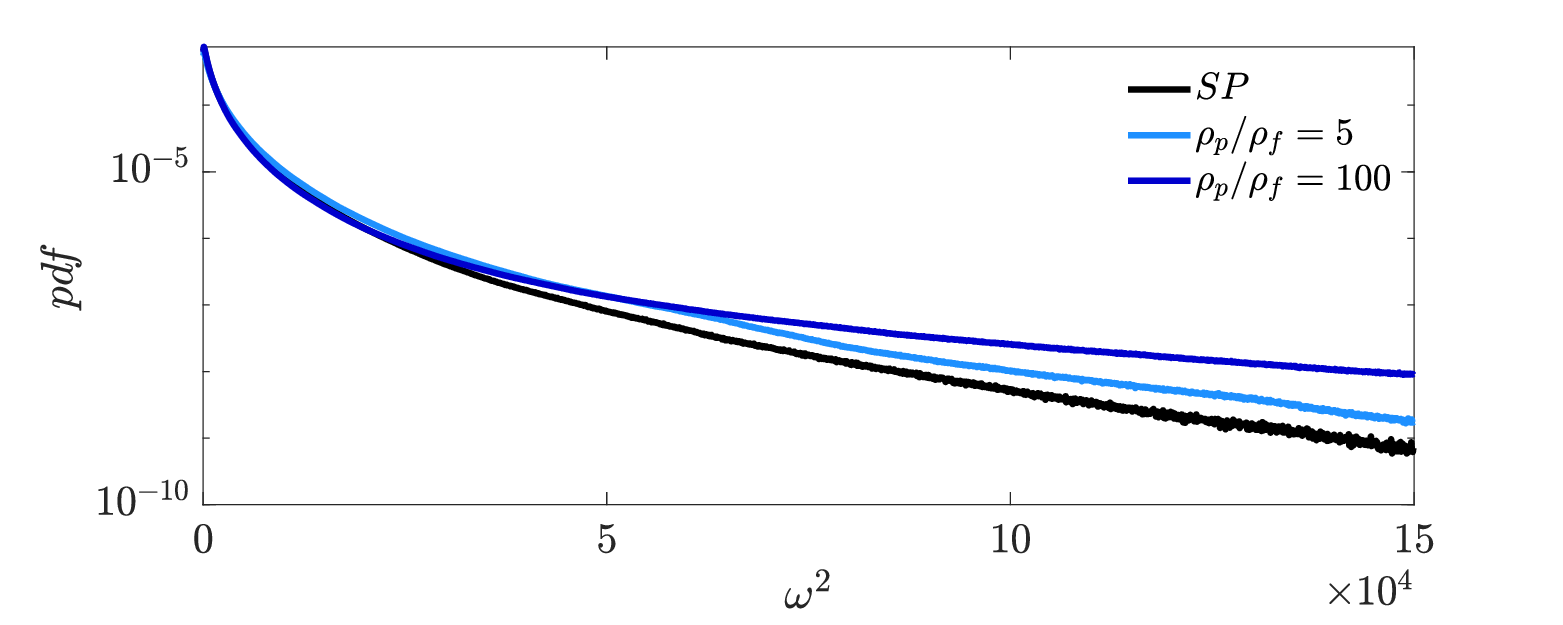}
\includegraphics[width=0.85\textwidth]{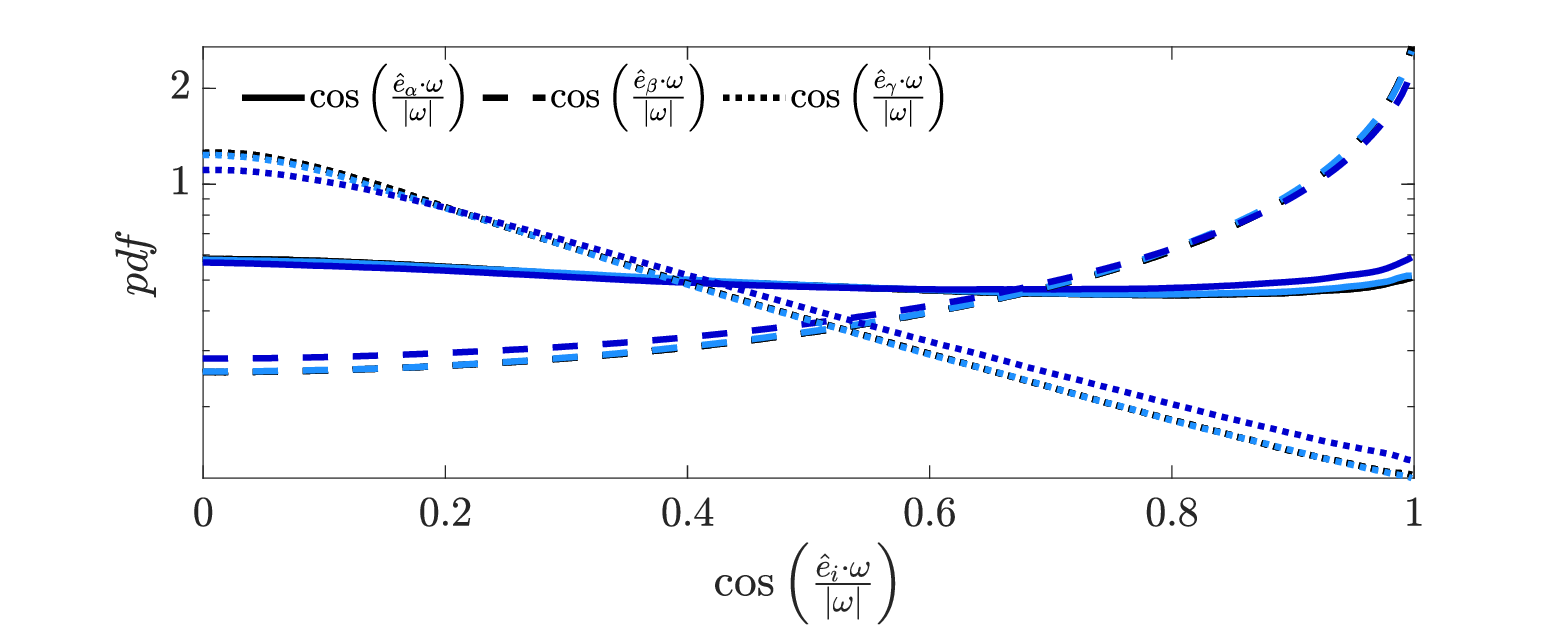}
\caption{(Top) Probability density function of the enstrophy. (bottom) Alignment of the vorticity vector $\bm{\omega}$ with the principal axes of strain. Here, $\hat{\bm{e}}_\alpha$ is aligned with the direction of maximum elongation of the flow, $\hat{\bm{e}}_\gamma$ is aligned with the direction of compression of the flow, and $\hat{\bm{e}}_\beta$ is orthogonal to the previous two directions. The data shown are for $\Phi_V = 10^{-3}$.}
\label{fig:omega}
\end{figure}
Figure \ref{fig:omega} shows the influence of the solid phase on (top) the square of the vorticity magnitude $\omega^2$, i.e. the enstrophy, and on (bottom) the alignment between the vorticity $\bm{\omega}$ and the principal axes of strain. The distribution of $\omega^2$ shows that the tail becomes longer and the probability of large events increases due to the presence of the particles. This agrees with the enhanced flow intermittency discussed in \S\ref{sec:strfun}. The tail of the distribution is longer for the $\rho_p/\rho_f = 100$ case, as the velocity gradients at the particles surface are more intense because of the larger relative velocity between the particles and the surrounding fluid phase. However, even the lighter particles with $\rho_p/\rho_f = 5$ are able to produce a non-negligible change of the distribution.
Instead, the bottom panel of figure \ref{fig:omega} shows that the alignment between $\bm{\omega}$ and the principal axes of strain are only slightly influenced by the presence of the particles. The results for the single-phase case perfectly overlaps with that of $\rho_p/\rho_f = 5$. 
The presence of the heavy particles, instead, slightly reduces the alignment between $\bm{\omega}$ and $\hat{\bm{e}}_\beta$, as well as the anti-alignment between $\bm{\omega}$ and $\hat{\bm{e}}_\gamma$. Our results suggest that for $D_p/\eta \approx 1$ the perturbation field induced by the heavy particles is characterised by events where vorticity is more aligned with the direction of extension ($\hat{\bm{e}}_\alpha$) and compression ($\hat{\bm{e}}_\gamma$), and more anti-aligned with the intermediate eigenvector ($\hat{\bm{e}}_\beta$). Interestingly, this differs from what observed by \cite{cannon-olivieri-rosti-2024}, who found the opposite trend for particles with size in the inertial range. We presume that the difference is due to the different particle Reynolds number, which results in a different kind of deformation of the local fluid flow.  

We now move and consider the entire velocity gradient tensor $A_{ij}$. Any second-order tensor possesses three invariants $P$, $Q$ and $R$, which are directly related to its eigenvalues $\lambda$ by the characteristic polynomial function
\begin{equation}
  \lambda^3 + P \lambda^2 + Q \lambda + R = 0.
  \label{eq:poly}
\end{equation}
Following \cite{davidson-2004,meneveau-2011}, it can be shown that
\begin{equation}
\begin{aligned}
P = & \alpha + \beta + \gamma, \\
Q = & - \frac{1}{2} \left( \alpha^2 + \beta^2 + \gamma^2 \right) + \frac{\omega^2}{4}, \\
R = & - \frac{1}{3} \left( \alpha^3 + \beta^3 + \gamma^3 \right) - \frac{1}{4} \omega_i \omega_j S_{ij}.
\end{aligned}
\end{equation}
Note that, $P=0$ due to the incompressibility constraint. The $Q$ and $R$ invariants are commonly used to distinguish between regions of intense vorticity and regions of strong strain. In particular, the discriminant of equation \ref{eq:poly} $\Delta = 27/4R^2 + Q^3$ is used to distinguish between regions where motions are mainly vortical (i.e. regions where $\Delta<0$, meaning that $A_{ij}$ has one real and two complex conjugate eigenvalues) or characterised by a node-saddle streamline pattern (i.e. regions where $\Delta >0$, and all the eigenvalues are real). When $Q$ is large and negative the strain is intense, while the vorticity is weak; in this case, $R \sim - (\alpha^3 + \beta^3 + \gamma^3) = - \alpha \beta \gamma$ \citep{davidson-2004}, and a positive $R$ implies a region of biaxial strain ($\gamma<0$, $\alpha> \beta > 0$), while a negative $R$ implies a region of axial strain ($\gamma < \beta < 0$ and $\alpha>0$). When instead $Q$ is large and positive the strain is locally weak and $R \sim - \omega_i \omega_j S_{ij}$. In this case, a positive $R$ implies vortex compression, while a negative $R$ implies vortex stretching. 

\begin{figure}
\centering
\includegraphics[width=0.49\textwidth]{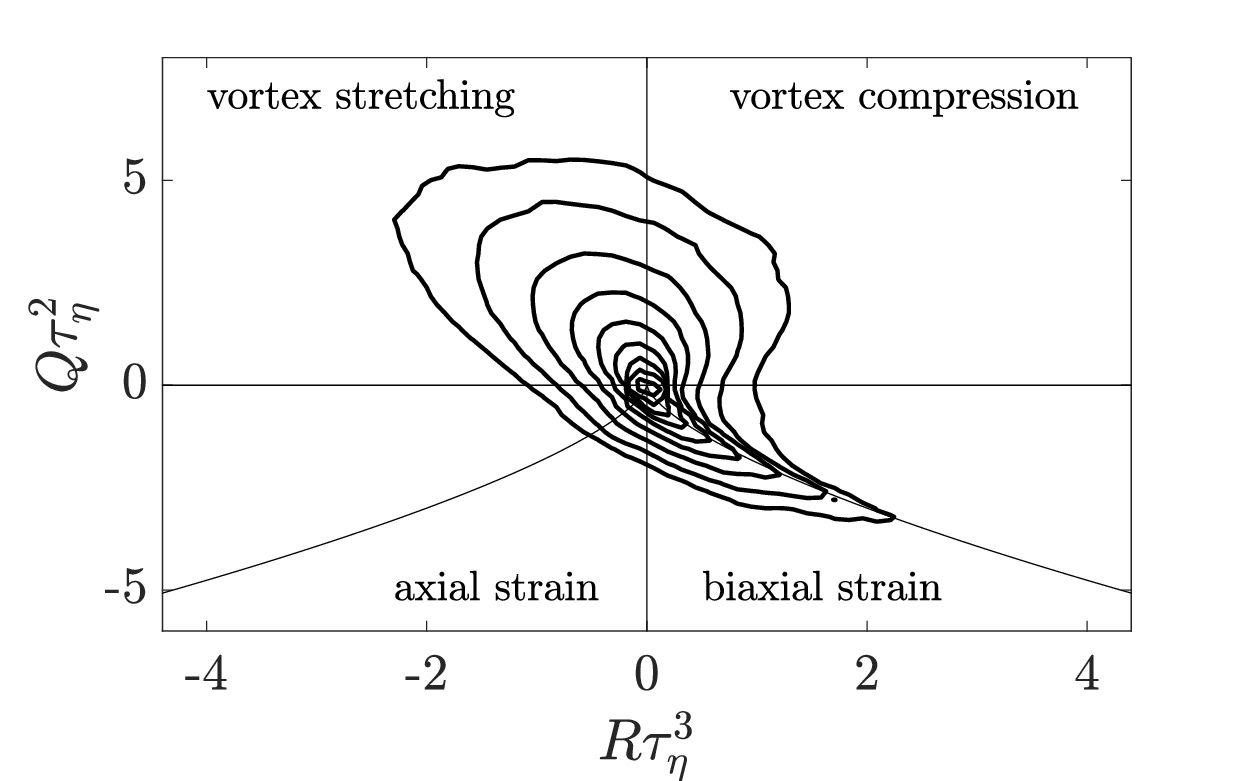}
\includegraphics[width=0.49\textwidth]{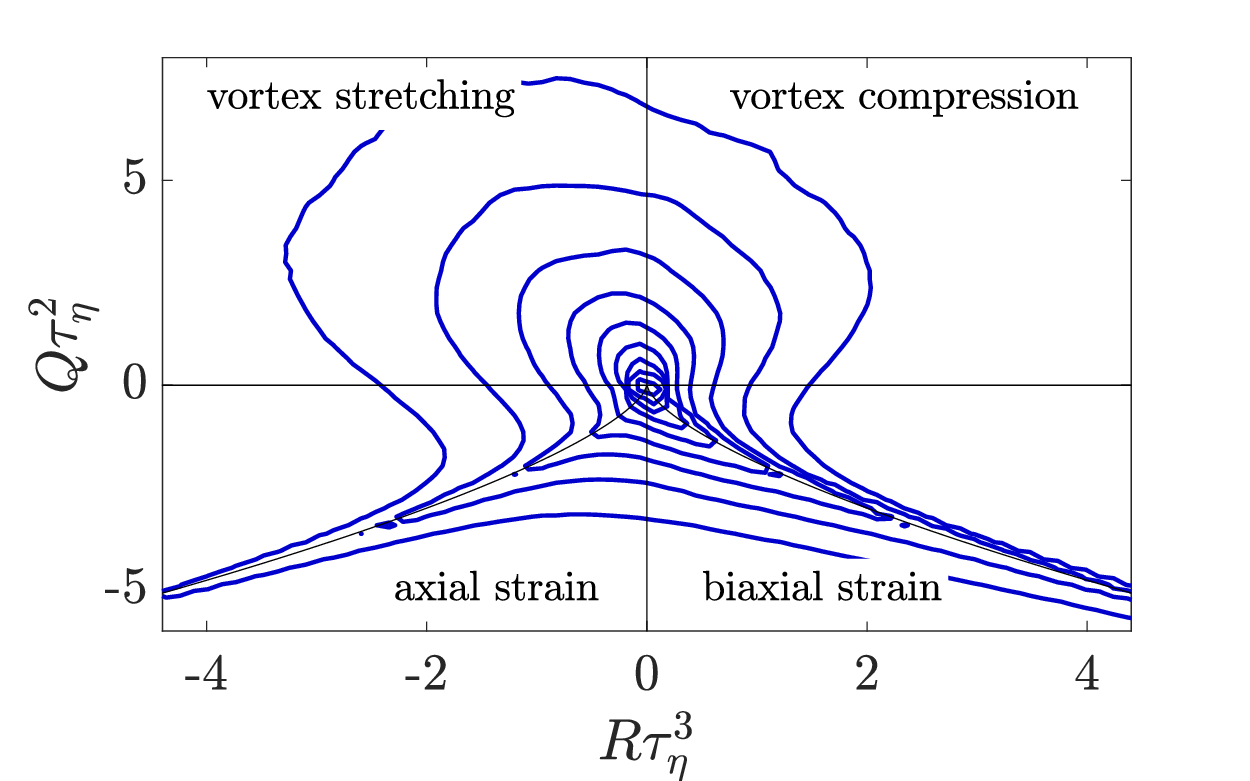}
\includegraphics[width=0.49\textwidth]{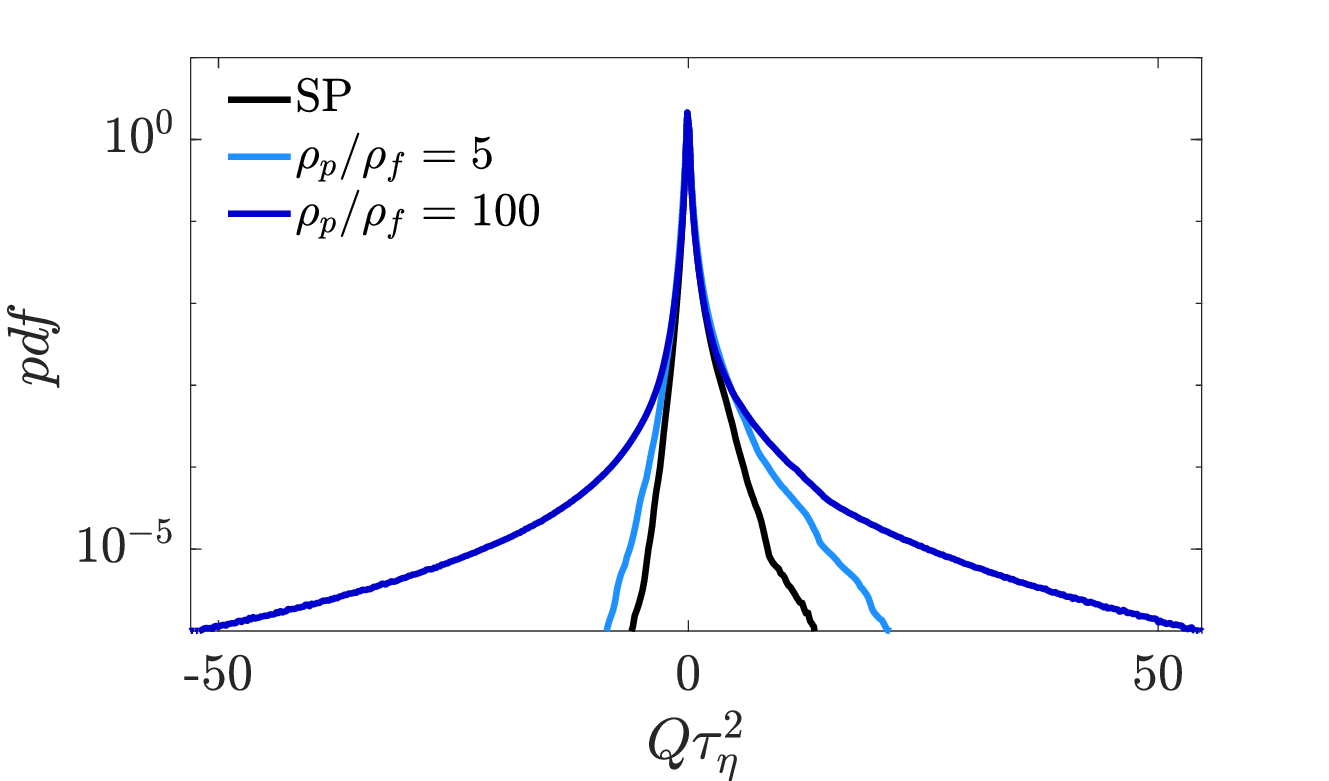}
\includegraphics[width=0.49\textwidth]{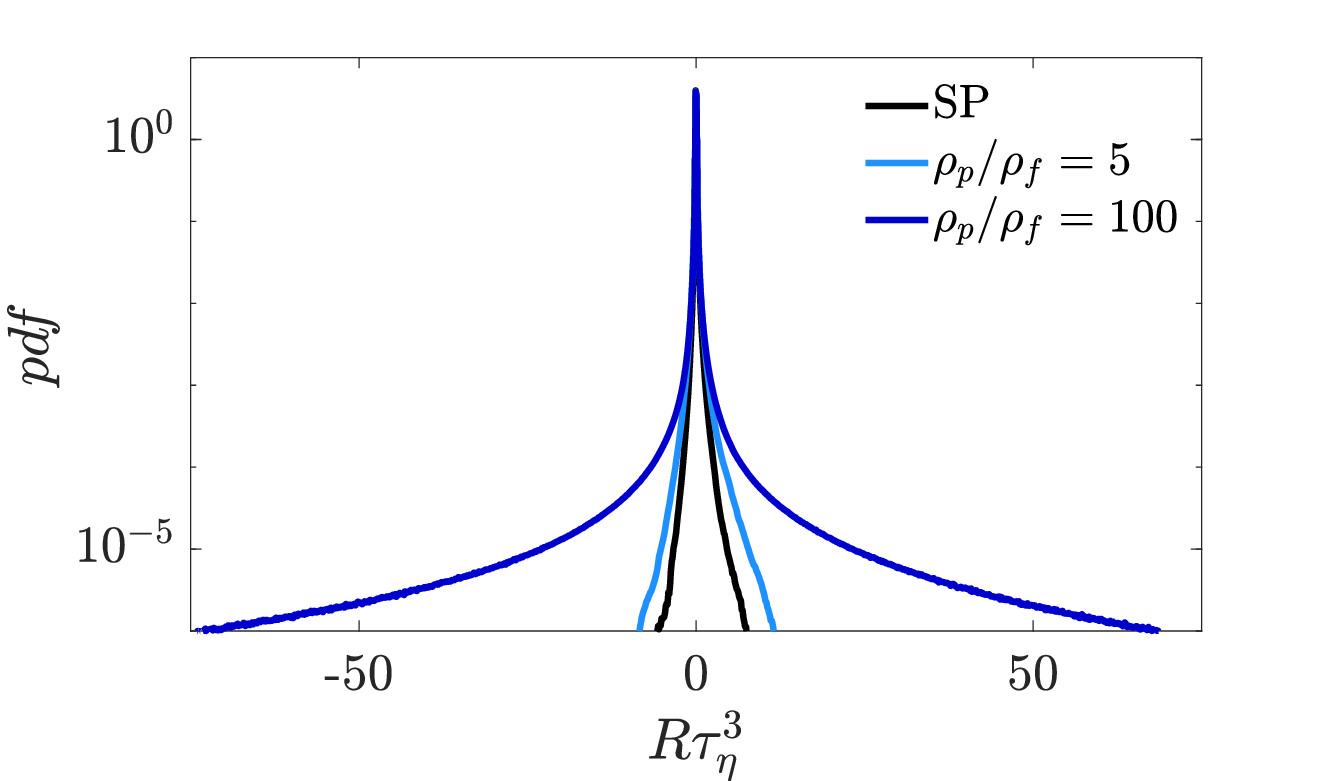}
\caption{Top: $Q$-$R$ map for the single-phase (left) and $\Phi_V=10^{-3}$ and $\rho_p/\rho_f = 100$ (right). For both panels the $11$ isolines have a logarithmic spacing between $5\times 10^{-5}$ and $10^2$. Bottom: distribution of $Q$ (left) and $R$ (right) for $\Phi_V=10^{-3}$.}
\label{fig:QRmap}
\end{figure}

\begin{figure}
\centering
\includegraphics[width=0.47\textwidth]{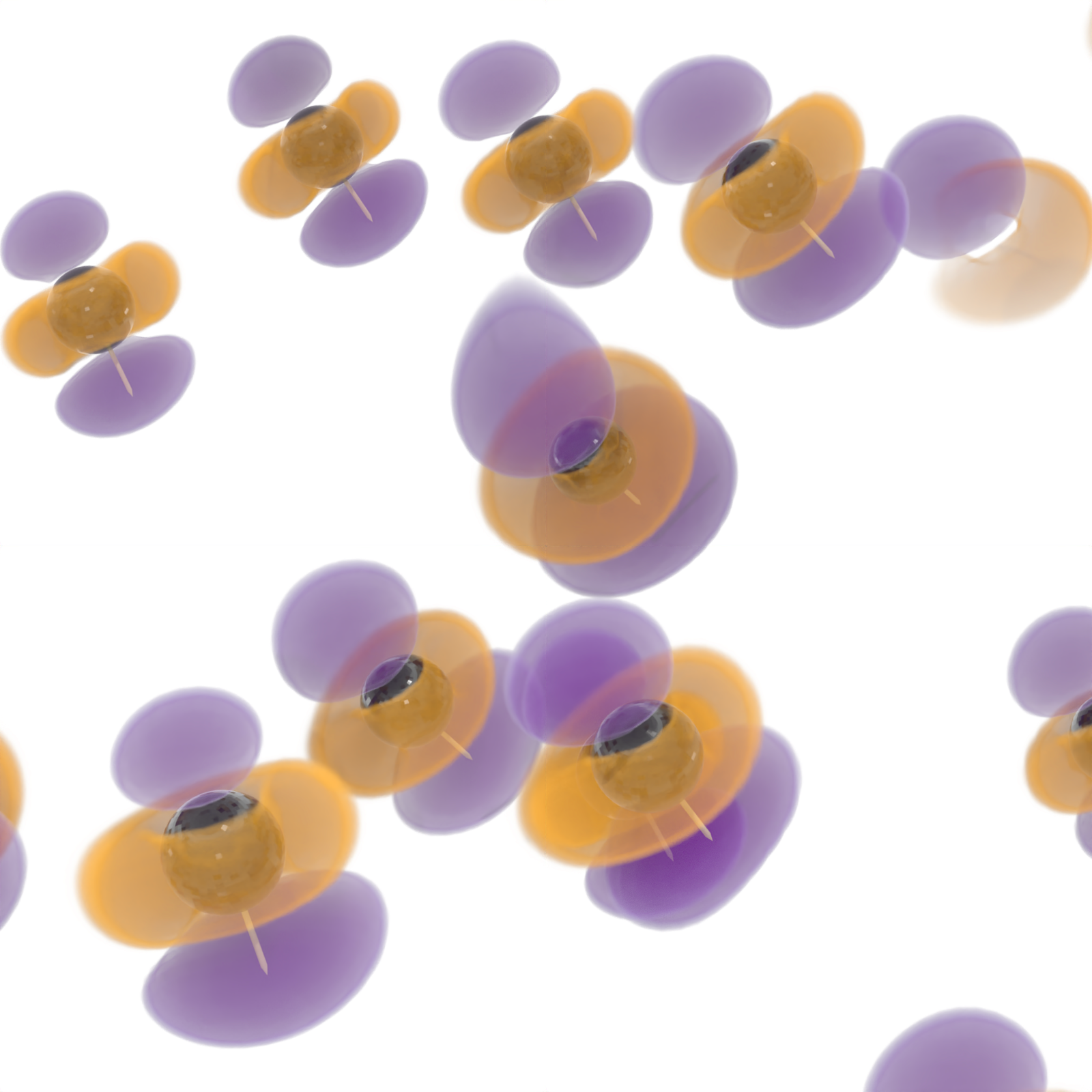}
\hfill
\includegraphics[width=0.47\textwidth]{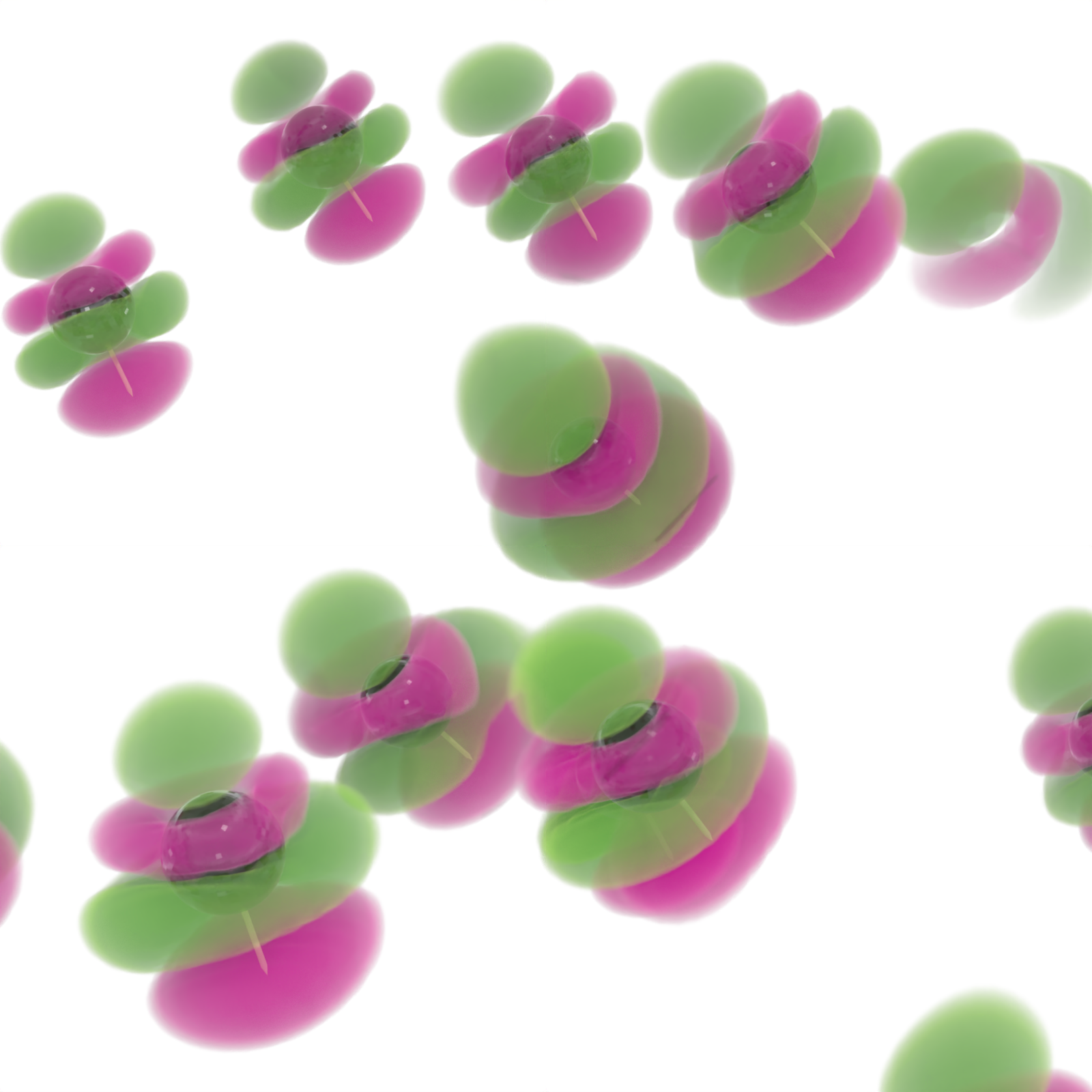}
\caption{Volumetric rendering of the (left) $Q$ and (right) $R$ fields for $\Phi_V = 10^{-3}$ and $\rho_p/\rho_f=100$. Orange and magenta regions are associated with large, positive values of $Q$ and $R$, i.e. $0.2072 \lessapprox Q \tau_\eta^2  \lessapprox 2.072$ and $0.2109 \lessapprox  R \tau_\eta^3  \lessapprox 0.4218$. Indigo and green regions indicate large, negative values of $Q$ and $R$, i.e. $-2.072 \lessapprox Q \tau_\eta^2 \lessapprox -0.2072$ and $-0.4218 \lessapprox R \tau_\eta^3 \lessapprox -0.2109$. The particle travelling direction relative to the local fluid velocity in a shell of radius $R_{sh}=5$ is indicated for each particle by a pointer. See figure \ref{fig:sketch_QRparticles} for a schematic representation of the flow. 
}
\label{fig:QRparticles}
\end{figure}

\begin{figure}
\centering
\includegraphics[width=0.6\textwidth]{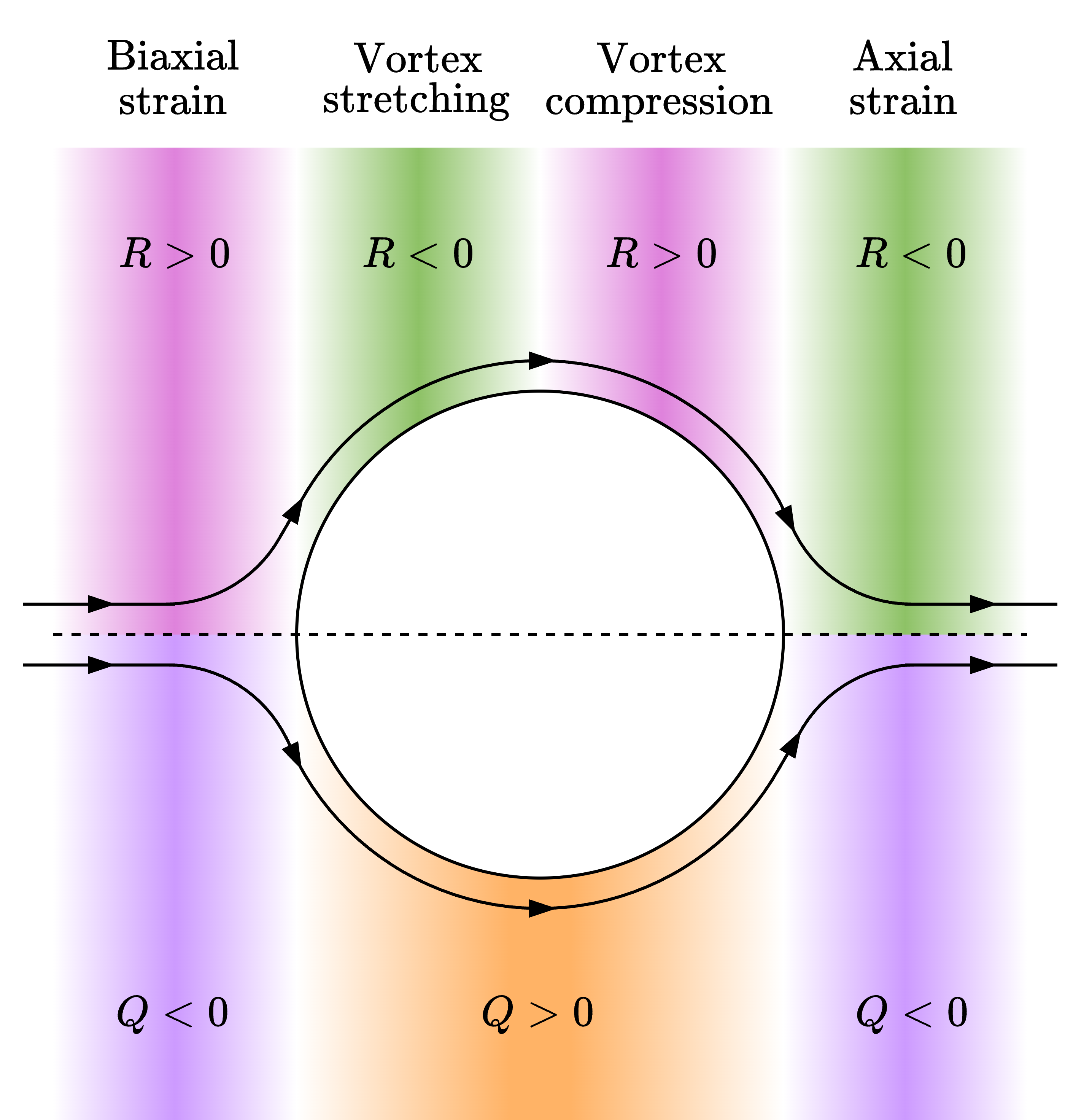}
\caption{Schematic representation of the distributions of $Q$ and $R$ around the particles based on figure \ref{fig:QRparticles}. The flow, represented by continuous stream lines, is incoming from the left in the particle's reference frame.}
\label{fig:sketch_QRparticles}
\end{figure}

Figure \ref{fig:QRmap}, plots the joint distribution of $Q$ and $R$ for the $\Phi_V = 10^{-3}$ and $\rho_p/\rho_f = 100$ case (top right) and for the single phase (top left); the black, thin, solid (curved) lines denote the left- and right-Vieillefosse tails with $\Delta=0$. For completeness the bottom panels report the distributions of $Q$ (left) and $R$ (right). In absence of the particles, the $Q$-$R$ joint distribution takes a tear-drop pattern, with a clear point at the right-Vieillefosse tail with $R>0$ and $Q<0$ \citep{meneveau-2011}. The distribution is skewed towards positive $Q$, but rather evenly distributed among positive and negative values of $R$ (see the bottom panels). The largest probability is observed in the second and fourth quadrants, i.e. $Q<0$ and $R>0$, and $Q>0$ and $R<0$. Thus, in a purely Newtonian turbulent flow there is a strong negative correlation between $Q$ and $R$, and the two most common states are vortex stretching $\omega_i \omega_j S_{ij}>0$, and biaxial strain $\alpha \beta \gamma <0$ \citep{betchov-1956,davidson-2004}. Also, the points in the $Q$-$R$ map are distributed around the origin, since the mean values of $Q$ and $R$ are zero in homogeneous flow \citep{nomura-post-1998}. The presence of the particles enlarges the range of possible $Q$ and $R$ values, in agreement with the increase of the probability of events with intense velocity gradients associated with the boundary conditions at the particles' surface. In particular, Kolmogorov-size particles mainly favour events that lay in the first and third quadrant, resulting in a joint distribution that is more symmetric with respect to an inversion of the $R$ axis (see the bottom right panel in figure \ref{fig:QRmap}). Compared to the unladen case, particles mainly promote events with axial strain $\alpha \beta \gamma > 0$ ($R<0$ and large negative $Q<0$) and with vortex compression $\omega_i \omega_j S_{ij}<0$ ($R>0$ and $Q>0$). The probability of events with $R<0$ and $Q<0$ is particularly enhanced, as visualised in the top right panel of figure \ref{fig:QRmap} by the occurrence of a point at the left-Vieillefosse tail. This is consistent with the increase of the probability of events with $s^*=-1$ shown in figure \ref{fig:sstar}. 
A visual investigation of the $Q$ and $R$ fields around the particles (figure \ref{fig:QRparticles}) helps to explain this effect by highlighting the local contribution of the particles. By comparing the two fields in the surroundings of the particles, we observe that they are both almost axisymmetric along the axis aligned with their travelling direction. However, across the velocity-normal median plane, $Q$ is symmetric, while $R$ is antisymmetric. This relation between the two fields implies contributions across all four quadrants of the $Q$-$R$ distribution (figure \ref{fig:QRmap}). The effect, however, is particularly apparent in the third quadrant, as the region is otherwise not explored by the single-phase flow. In particular, regions of $Q<0$ and $R<0$ are associated with axial strain and appear to be found at the downstream end of the particles' along their travelling direction. Consistently with the above-mentioned symmetries, a region of biaxial strain ($Q<0$ and $R>0$) is found at the upstream end of the particles. This is conveniently visualised in the schematic of figure \ref{fig:sketch_QRparticles}.

This scenario only partially agrees with the results of \cite{schneiders-etal-2017} for decaying homogeneous isotropic turbulence laden with Kolmogorov-size particles. They also found that particles favour events with axial strain ($Q<0$ and $R<0$) and biaxial strain ($Q<0$ and $R>0$), as shown by the occurrence of a point at the left-Vieillefosse tail and by the more pronounced right-Vieillefosse tail in figure 20 of their paper. However, they did not report an increase of the probability of events with $Q>0$ like in the present case. It is worth mentioning, however, that they do report a global increase of the vortex stretching. A possible explanation of this difference is the lower Reynolds number they considered ($Re_\lambda \approx 79$ at the initial time) that does not ensure a proper separation of scales, in particular at larger times when turbulence decays. Compared to larger particles with size in the inertial range of scales, the scenario is completely different: in fact, \cite{cannon-olivieri-rosti-2024} found that when large particles are added both $Q$ and $R$ are reduced. Besides energising the small scales, large particles indeed behave also as obstacles for the large flow structures, largely weakening thus the energy content at the large scales. \cite{cannon-olivieri-rosti-2024} only observed an increase of the probability in the strain-dominated region (that we also observe), which is an indication of the intense dissipation regions that arise around the particles.

\begin{figure}
\centering
\includegraphics[width=0.49\textwidth]{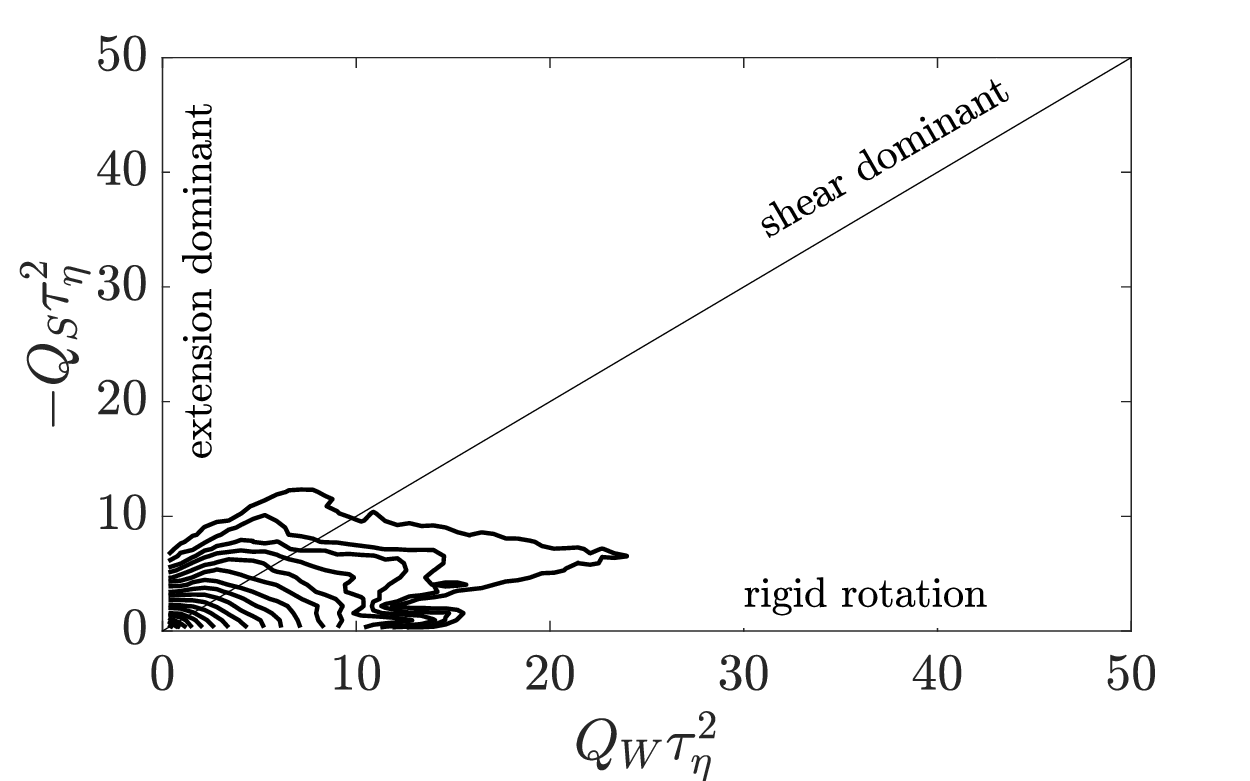}
\includegraphics[width=0.49\textwidth]{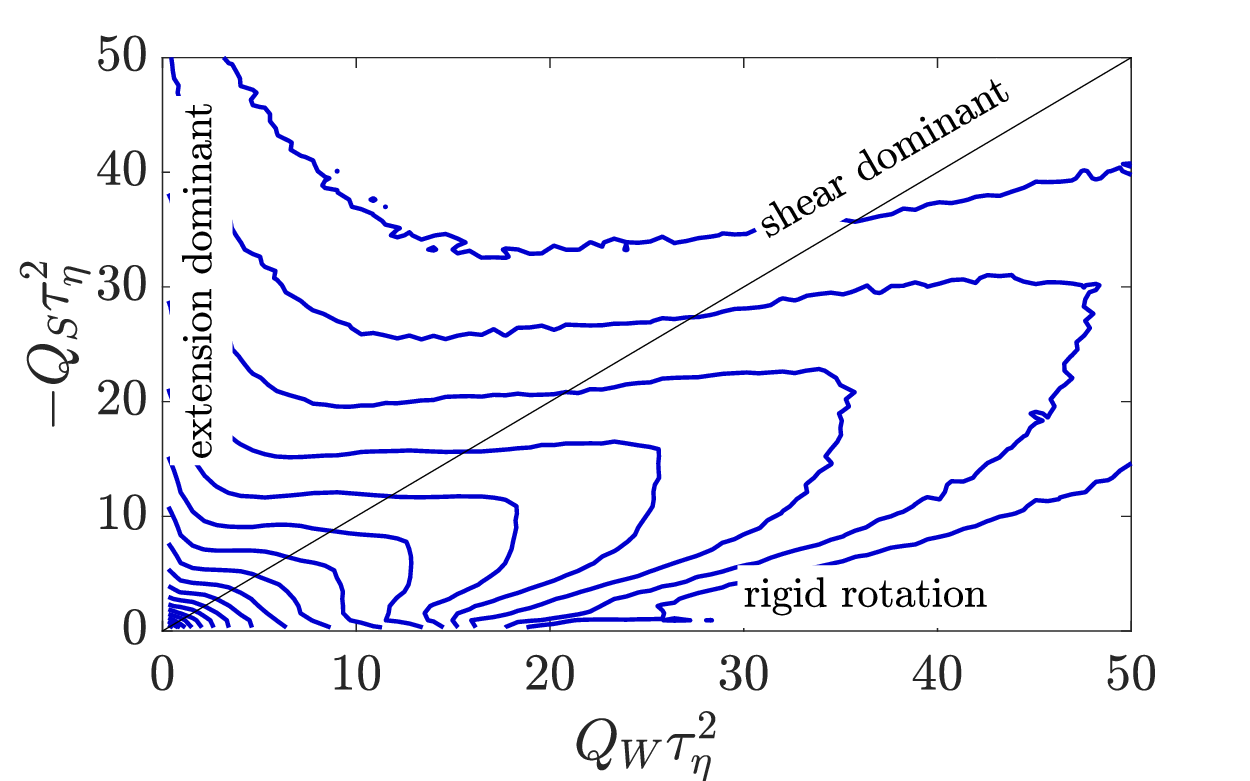}
\caption{$-Q_S$-$Q_W$ map for the single-phase case (left) and $\Phi_V=10^{-3}$ and $\rho_p/\rho_f=100$ (right). For both panels the $20$ isolines have a logarithmic spacing between $10^{-7}$ and $10^1$.}
\label{fig:QSQWmap}
\end{figure}

Additional insights are provided by looking separately at the invariants of $S_{ij}$ and $W_{ij}$; see figure \ref{fig:QSQWmap}. In particular, we consider their second invariants, i.e.
\begin{equation}
Q_S = - \frac{1}{2} \left( \alpha^2 + \beta^2 + \gamma^2 \right) \ \text{and} \
Q_W = \frac{1}{4} \omega^2.
\end{equation}
These invariants are related with the fluid dissipation $\epsilon = -4 \nu Q_S$ and with the fluid enstrophy $\omega^2 = 4Q_W$. Therefore, the $Q_S$-$Q_W$ joint distribution determines whether the flow is dominated by dissipation (extensional dominated regions with $Q_S>Q_W$) or by enstrophy (rigid rotation regions with $Q_W>Q_S$). In shear dominated regions, dissipation and enstrophy balance and $-Q_S = Q_W$ \citep{soria-etal-1994}. For simplicity, we follow \cite{truesdell-1954} and introduce $\mathcal{K} = (-Q_W/Q_S)^{1/2}$; when $\mathcal{K} = 0$ the flow is extension dominated, when $\mathcal{K} = \infty$ the flow is dominated by rigid rotation events, and when $\mathcal{K} = 1$ the rotation and the stretching are equal, as typical of vortex sheets and shear layers. In the unladen case, events with $Q_W>-Q_S$ ($\mathcal{K} \rightarrow \infty$) are more frequent, meaning that for purely Newtonian turbulence the flow is mainly dominated by rigid rotations. 
In the presence of the particles the scenario slightly changes. A first observation is that the probability of events with large $Q_S$ (or $\mathcal{K} = 0$) increases, indicating that the perturbation field induced by these small particles is extensional dominated. A second observation is that the solid phase favours also events with $-Q_S \approx Q_W$ ($\mathcal{K} \approx 1$), that agrees with the presence of the shear layers induced by the presence of the particles. 

%
%
%
%

\section{Particle dynamics}
\label{sec:particles}

This section is devoted to the dynamics of the particles, and we compare the results of the PR-DNS with those of the PP-DNS. Besides characterising the motion of the particles, indeed, the objective is to address the reliability of the one-way-coupled PP-DNS at the present parameters. 

\subsection{Lagrangian velocity increments}

\begin{figure}
  \centering
  \includegraphics[width=0.49\textwidth]{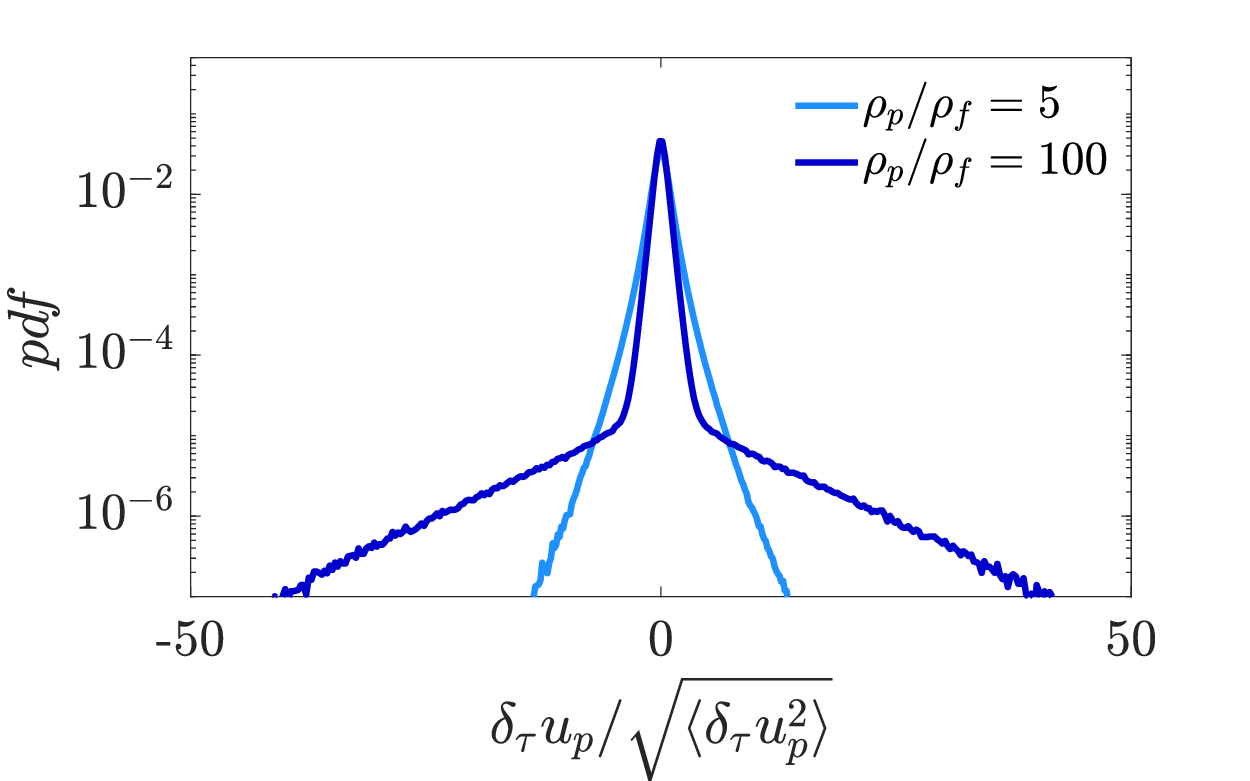}  
  \includegraphics[width=0.49\textwidth]{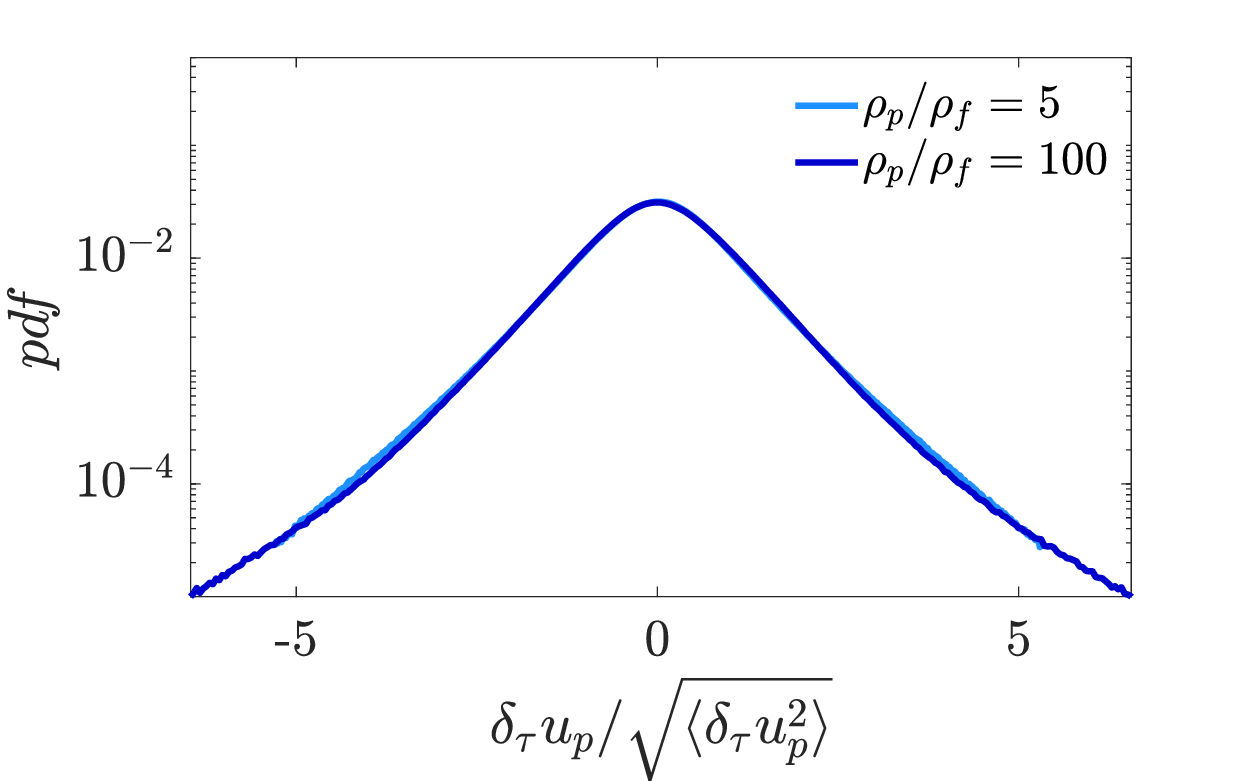} 
  \caption{Probability density function of the Lagrangian velocity increments of the particles $\delta_\tau u_p$ 
   for (left) $\tau = 0.2 \tau_\eta$, and (right) $\tau = 2 \tau_\eta$. The data are for the case with $\Phi_V=10^{-3}$ and $\rho_p/\rho_f=5$ and $\rho_p/\rho_f=100$}
  \label{fig:lagvelrhop}    
\end{figure}

\begin{figure}
\centerline{
\begin{tikzpicture}
  \node at (0,8.4) {\includegraphics[width=0.49\textwidth]{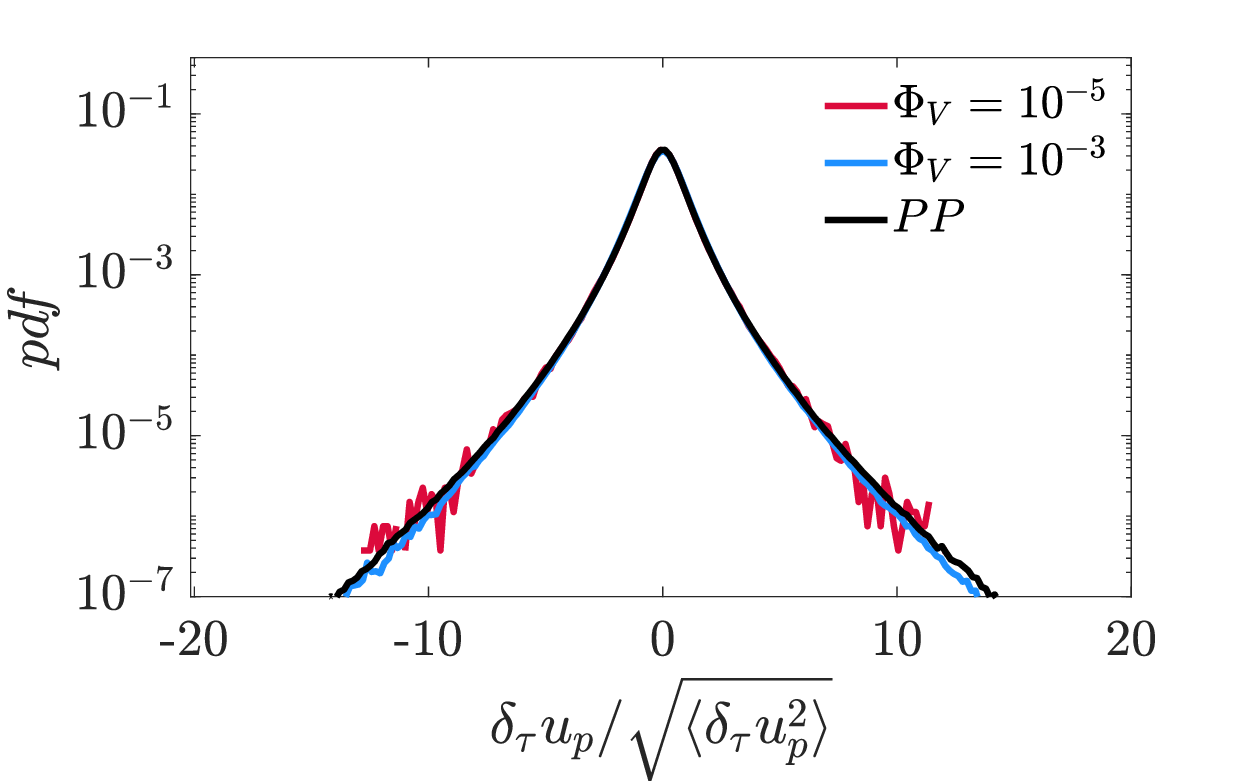}};
  \node at (6.5,8.4) {\includegraphics[width=0.49\textwidth]{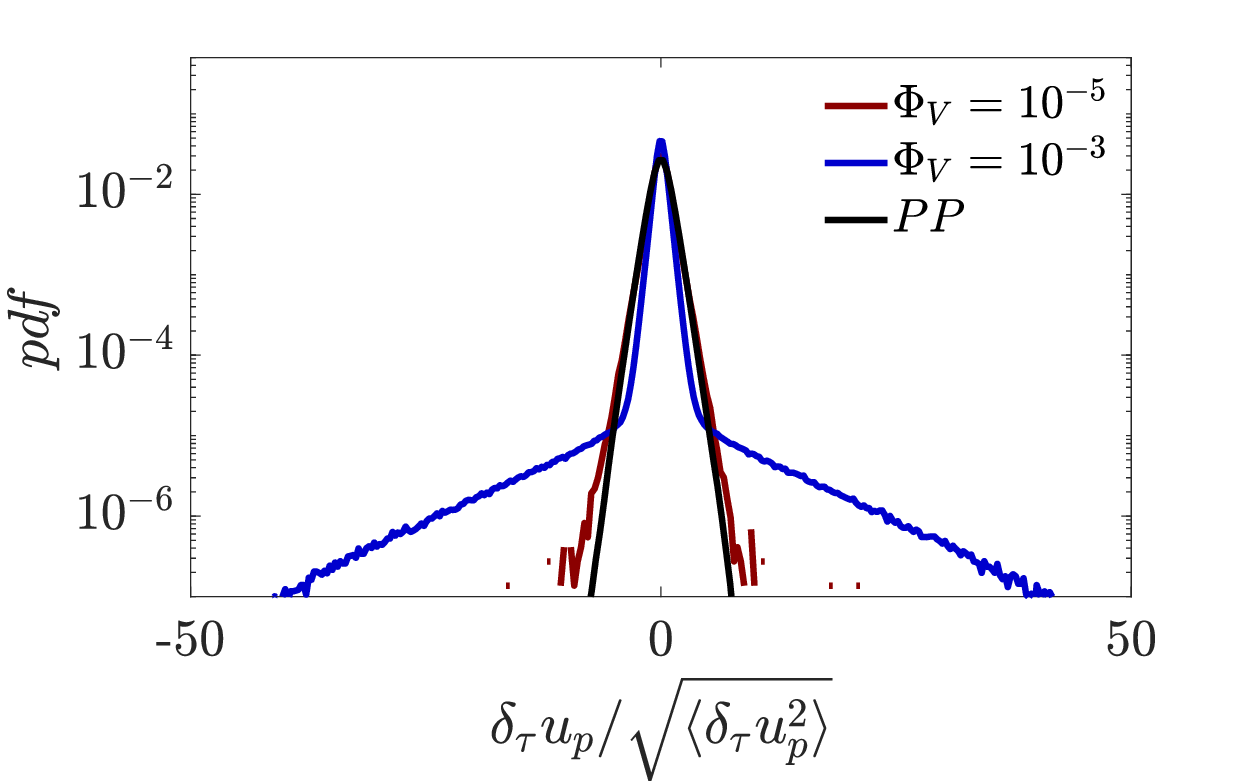}};
  \node at (0,4.2) {\includegraphics[width=0.49\textwidth]{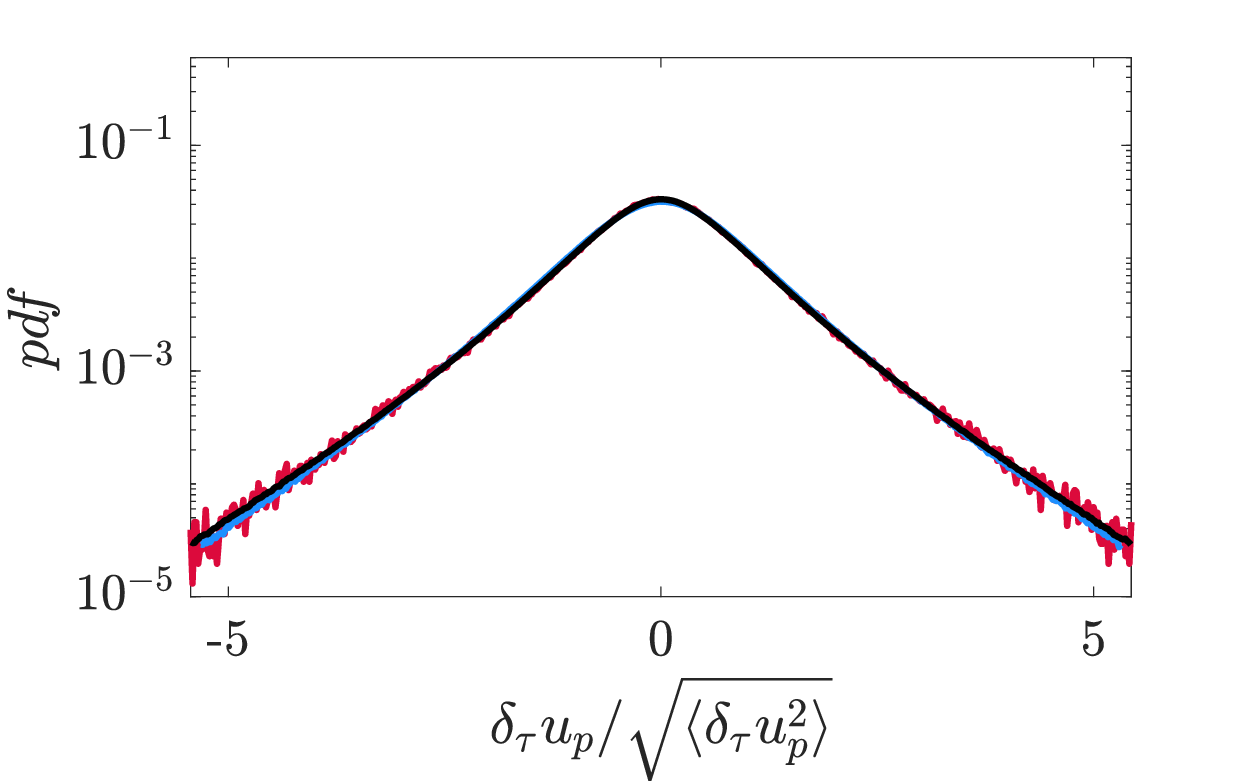}};
  \node at (6.5,4.2) {\includegraphics[width=0.49\textwidth]{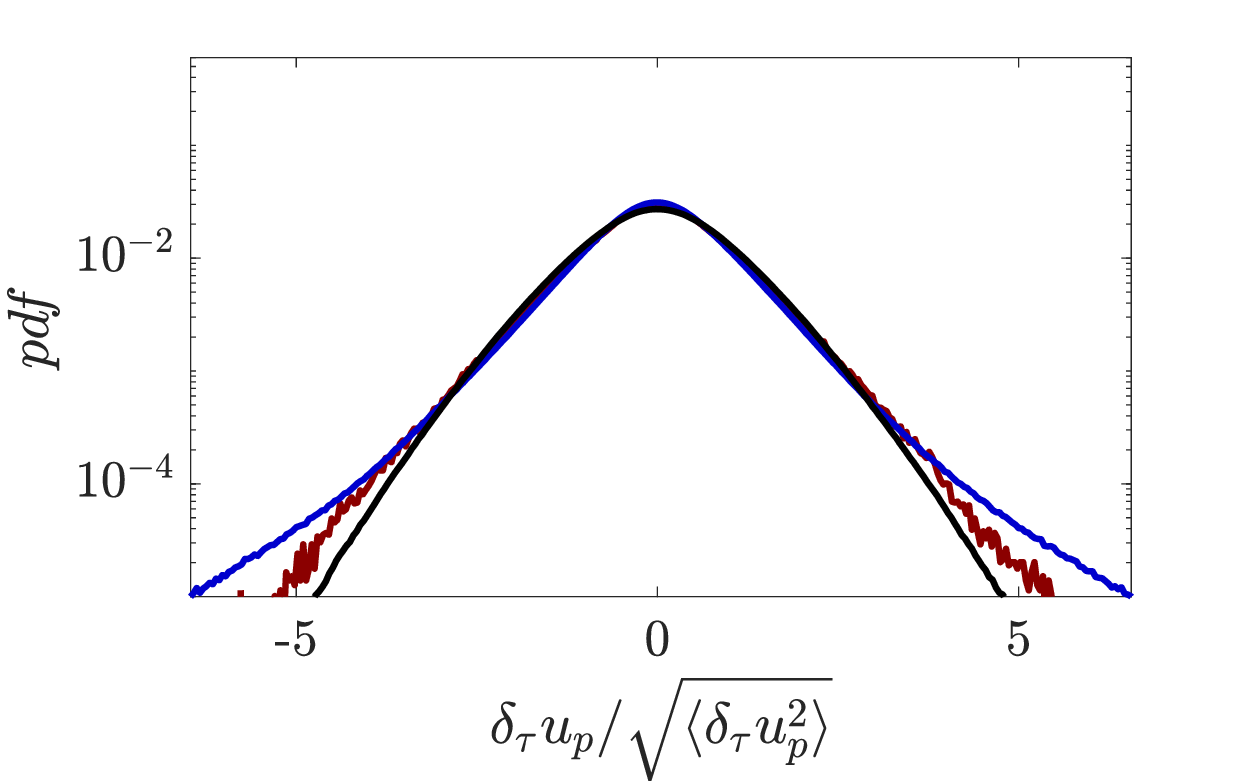}};
  \node at (0,0) {\includegraphics[width=0.49\textwidth]{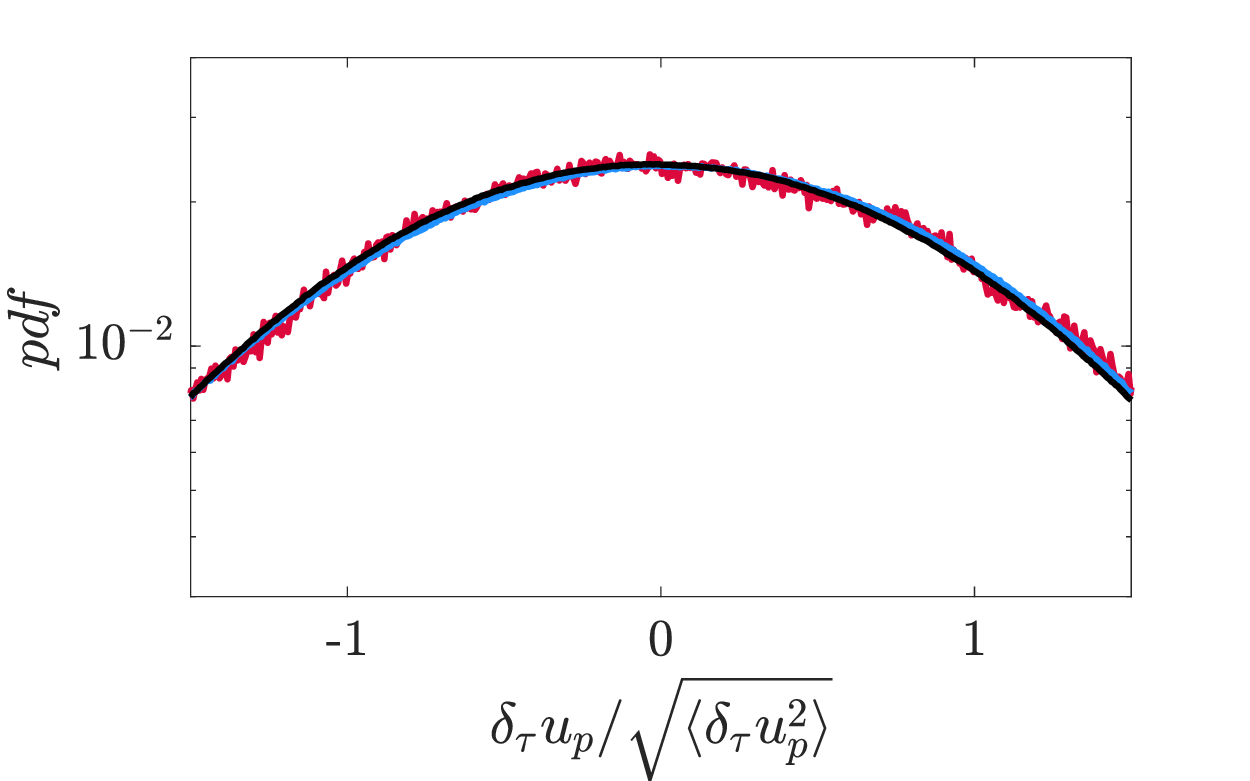}};
  \node at (6.5,0) {\includegraphics[width=0.49\textwidth]{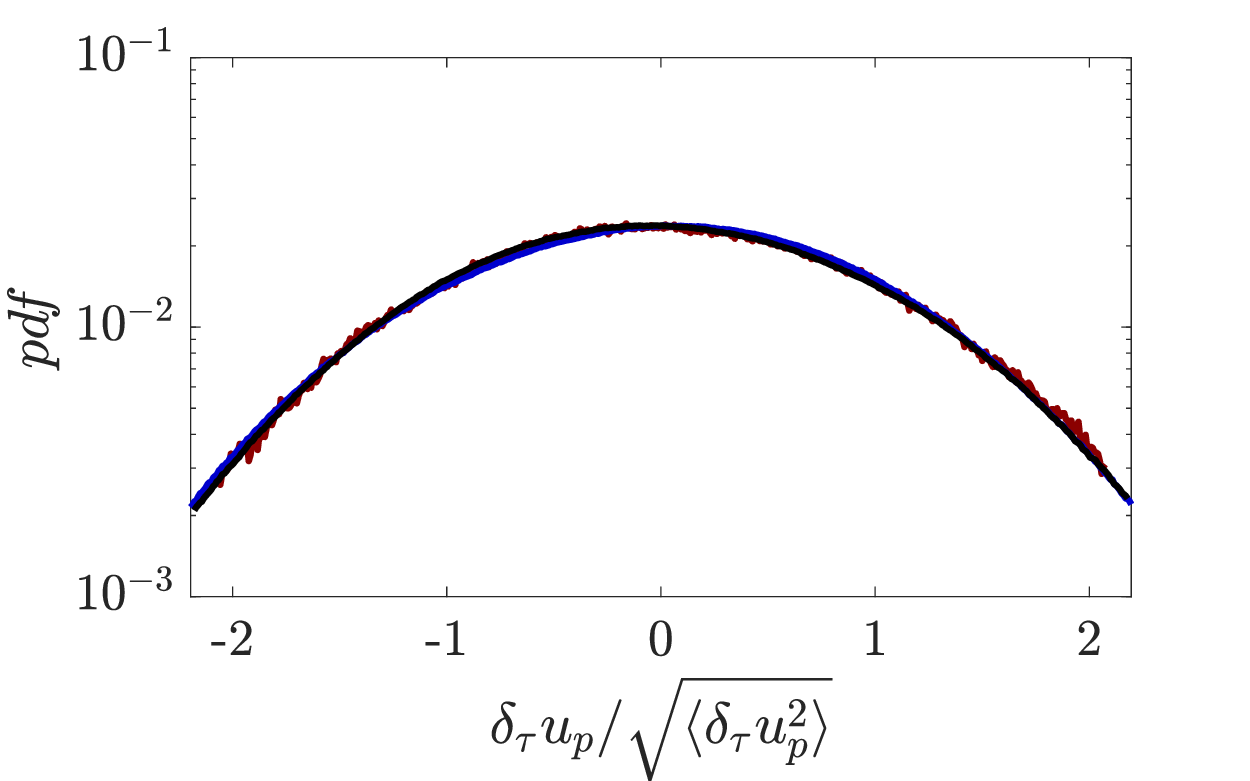}};
  \node at (10.3,8.75) {$\tau=0.2 \tau_\eta$};
  \node at (10.3,4.45) {$\tau = 2 \tau_\eta$};
  \node at (10.3,0.25) {$\tau =30 \tau_\eta$};
  \node at (0.5,10.7) {$\rho_p/\rho_f = 5$};
  \node at (6.5,10.7) {$\rho_p/\rho_f = 100$};
\end{tikzpicture}
}
\caption{Probability density function of the Lagrangian velocity increments of the particles $\delta_\tau u_p $
for (top) $\tau = 0.2 \tau_\eta$, (centre) $\tau = 2 \tau_\eta$, and (bottom) $\tau = 30 \tau_\eta$. Left panels are for $\rho_p/\rho_f = 5$, while the right ones for $\rho_p/\rho_f = 100$ . The black solid lines are the results of the PP-DNS simulations.}
\label{fig:LagVelInv}
\end{figure}

The Lagrangian statistics of the particles motion are of fundamental importance in the understanding of transport and mixing.
In order to investigate this, we study the Lagrangian velocity increments, defined here as $\delta_\tau u_{p,i} = u_{p,i} (t + \tau) - u_{p,i}(t)$, with $u_{p,i}(t)$ being the instantaneous velocity of a particle along direction $i$ at time $t$. The symmetries of the present problem make the statistics of $\delta_\tau u_{p,i}$ independent of both $t$ and $i$; for simplicity hereafter we drop the $i$ index. Figures \ref{fig:lagvelrhop} and \ref{fig:LagVelInv} describe the particle dynamics at different time scales, by plotting $\delta_\tau u_p$ for different values of the time scale $\tau$ in the $0.2 \tau_\eta \le \tau \le 30 \tau_\eta$ range, where $\tau_\eta = (\nu/\epsilon)^{1/2}$ is the Kolmogorov time scale. For small time scales, the velocity increment provides information about the particle acceleration, i.e. $\delta_\tau u_p \sim a_p\tau$. A first observation is that the distributions are symmetric, in agreement with the symmetries of the flow. The probability density function of $\delta_\tau u_p$ continuously deforms from the Gaussian at large time scales (see $\tau \approx 30 \tau_\eta$) to the development of stretched exponential tails at dissipative time scales (see $\tau \approx 0.2 \tau_\eta$), which are the statistical signature of an intermittent Lagrangian dynamics; see \cite{mordant-etal-2001}, \cite{laporta-etal-2001} and \cite{chevillard-etal-2003} for small tracers and \cite{qureshi-etal-2007} for finite-size neutrally buoyant particles. The wide stretched exponential tails for the smallest $\tau$ show that the finite-size particles with $D_p \approx \eta$ experience very high acceleration events, with a probability which is higher than Gaussian, similarly to what was found for small tracers by \cite{laporta-etal-2001} and for finite-size particles by \cite{qureshi-etal-2007}. 

We start looking at the influence of $\rho_p/\rho_f$ and $\Phi_V$ on the Lagrangian intermittency of Kolmogorov-size particles. 
The left panels of figure \ref{fig:lagvelrhop} show that heavier particles ($\rho_p/\rho_f=100$) are less likely to experience intermediate values of the acceleration compared to lighter particles ($\rho_p/\rho_f=5$). In contrast, they are more likely to exhibit very low or very intense accelerations. On one side, the larger inertia of these particles opposes to large accelerations and favours small values of $a_p$. On the other side, heavy particles enhance the flow intermittency (see \S\ref{sec:flow}), promoting extreme events in the flow that are in turn responsible for rare but large particle accelerations. It is worth noticing that for $\Phi_V=10^{-5}$ the latter effect is barely visible (see  figure \ref{fig:LagVelInv}), in agreement with the weak flow modulation shown in \S\ref{sec:flow}. For light particles $\rho_p/\rho_f=5$ figure \ref{fig:LagVelInv} shows that the distribution of $\delta_\tau u_p$ obtained for $\Phi_V=10^{-5}$ and $10^{-3}$ overlap almost perfectly, in agreement with the low level of backreaction. 

\begin{figure}
  \centering
  \includegraphics[width=0.49\textwidth]{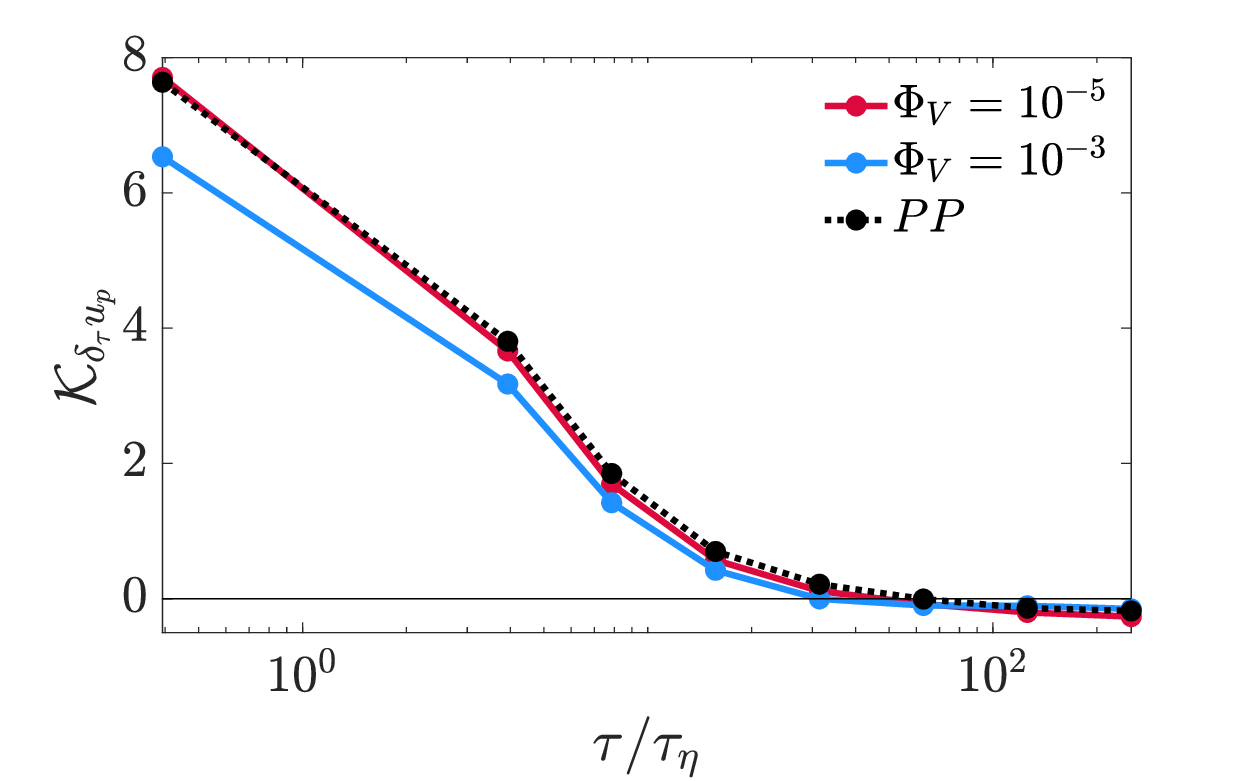}
  \includegraphics[width=0.49\textwidth]{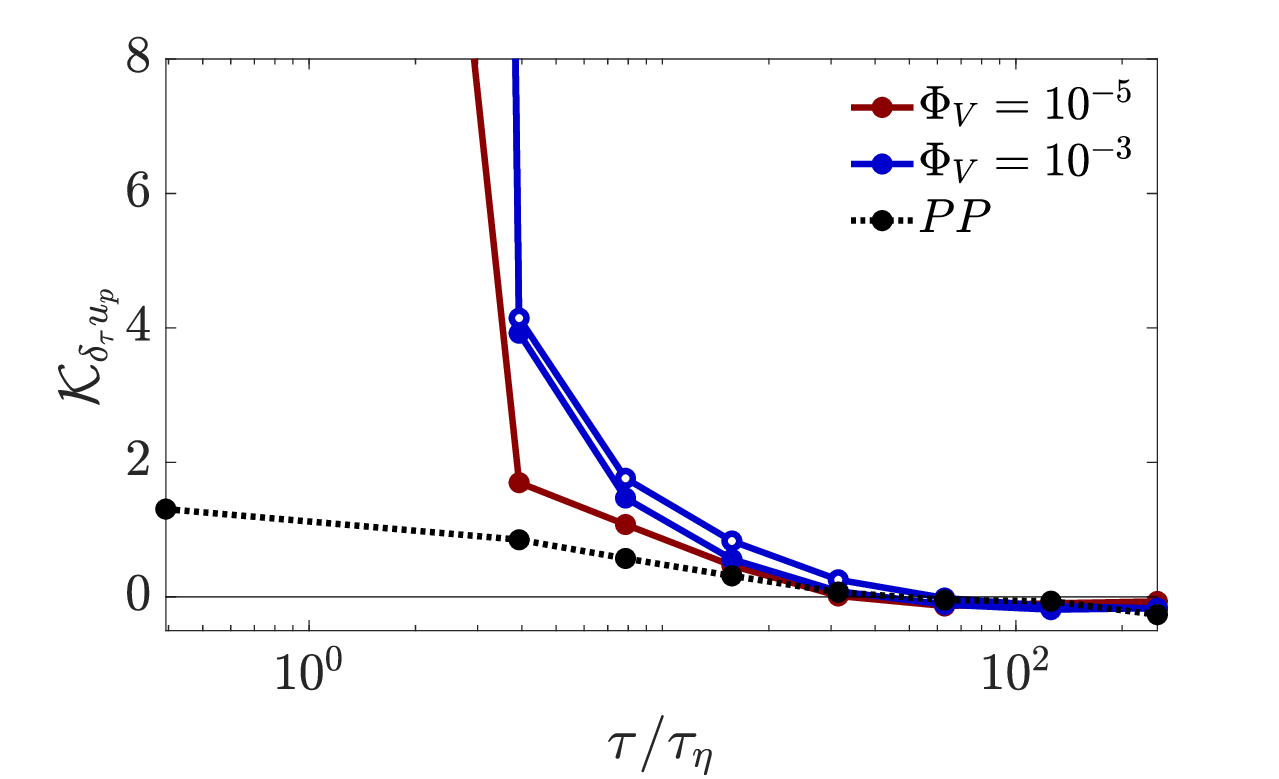}
  \caption{Evolution of the excess kurtosis factor $\mathcal{K}_{\delta_\tau u_p}(\tau) $ 
   for the distributions of the time increments of the particle velocity. Left: $\rho_p/\rho_f=5$. Right: $\rho_p/\rho_f=100$. The empty circles in the right panel are from the simulation with the coarser grid at $\Phi_V = 10^{-3}$ and $\rho_p/\rho_f = 100$, and are shown for validation purpose.}
  \label{fig:kurt}
\end{figure}

The left panels of figure \ref{fig:LagVelInv} show that for $\rho_p/\rho_f=5$ the distributions obtained by means of PP-DNS and PR-DNS almost perfectly overlap for all time scales $\tau$: for light particles the complete MRG equation properly predicts the Lagrangian intermittency of the particles dynamics. For heavier particles the match between the PP-DNS and the PR-DNS is rather good at large time scales, but differences are observed at small $\tau$, where the PP-DNS does not predict the large tails for $\tau \lessapprox 2 \tau_\eta$. As discussed above, these extreme events are associated with the flow modulation which is not modelled in our PP-DNS.
The comparison between the PP-DNS and PR-DNS results is further detailed in figure \ref{fig:kurt} where the evolution of the $\delta_\tau u_p$ distribution with $\tau$ is quantified by means of the excess kurtosis $\mathcal{K}_{\delta_\tau u_p}(\tau) = \aver{ \delta_\tau u_p^4}/\aver{ \delta_\tau u_p^2 }^2 -3$. At large scales $\mathcal{K}_{\delta_\tau u_p} \approx 0$ in agreement with the Gaussian-like shape of the distribution, while it steeply increases at small scales. For $\rho_p/\rho_f=5$ the good agreement between the PP-DNS and the PR-DNS is again clear, with a small deviation for the $\Phi_V=10^{-3}$ case, which is due to the non-zero flow modulation. For $\rho_p/\rho_f=100$, instead, the agreement is good at large scales, while the three curves substantially deviate for small $\tau$, accordingly with the larger tails of $\delta_\tau u_p$ found in the PR-DNS. 

\subsection{The particle-velocity Structure function}

We now consider the statistics of the particle-particle relative velocity $\delta \bm{u}_p = \bm{u}_p(\bm{x}_{p,j}(t),t) - \bm{u}_p(\bm{x}_{p,i}(t),t)$, where $\bm{x}_{p,i}(t)$ and $\bm{x}_{p,j}(t)$ denote the position of any two particles $i$ and $j$ at time $t$. The distribution of $\delta \bm{u}_p$ across all particle couples plays a key role in several theories regarding the tendency of particles to form clusters; see for example \cite{gustavsson-mehlig-2011}, \cite{bragg-collins-2014} and \cite{bragg-ireland-collins-2015}.

\begin{figure}
  \centering
  \includegraphics[width=0.85\textwidth]{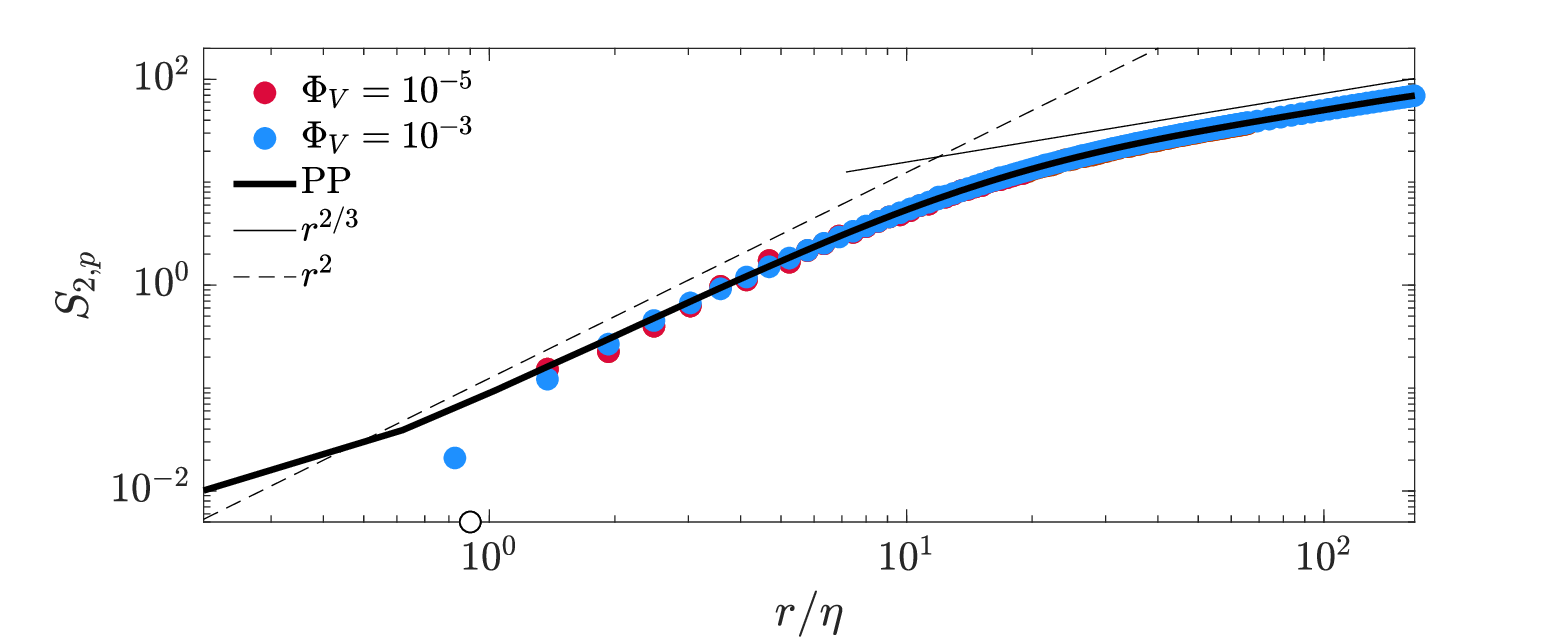}
  \includegraphics[width=0.85\textwidth]{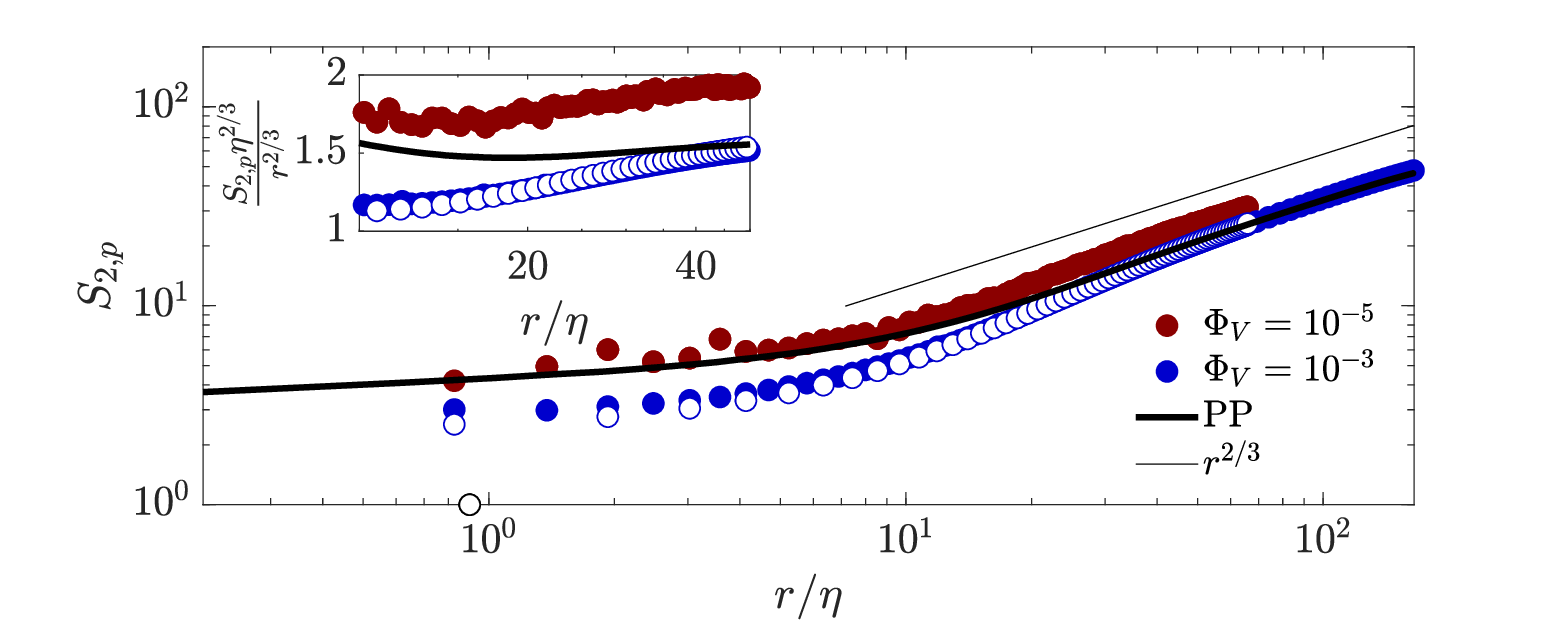}
  \caption{Second order structure function based on the particle velocity field $S_{2,p}$. Top: $\rho_p/\rho_f=5$. Bottom: $\rho_p/\rho_f = 100$. The empty circles in the bottom panel are from the simulation with the coarser grid at $\Phi_V = 10^{-3}$ and $\rho_p/\rho_f = 100$, and are shown for validation purpose. The inset in the bottom panel shows the compensated $S_{2,p}r^{-2/3}$ in the $10 \le r/\eta \le 50$ range.}
  \label{fig:Pstrfun}
\end{figure}
Figure \ref{fig:Pstrfun} plots the second-order structure function of the particle velocity, i.e.
\begin{equation}
  S_{2,p}(r) = \aver{ \left( \delta \bm{u}_p(\bm{r}) \cdot \frac{\bm{r}}{r} \right)^2 },
\end{equation}
where $\bm{r}$ is the separation vector between particle $i$ and $j$, and $r = |\bm{r}|$. The top panel is for $\rho_p/\rho_f=5$, while the bottom panel is for $\rho_p/\rho_f=100$. 
%
For $\rho_p/\rho_f=5$, $S_{2,p}$ resembles the fluid second-order structure function $S_2$ (see the top panel in figure \ref{fig:strfun}): light particles have small inertia and closely follow the fluid motion. $S_{2,p}$ exhibits the $r^2$ scaling at small scales and the $r^{2/3}$ scaling predicted by the Kolmogorov theory in the inertial range of scales. In agreement with the negligible flow modulation, the results of the PP-DNS match almost perfectly those of the PR-DNS at these parameters.

The bottom panel of figure \ref{fig:Pstrfun} deals with the $\rho_p/\rho_f=100$ cases. A first observation is that the results from the PR-DNS with $\Phi_V=10^{-5}$ and $\Phi_V=10^{-3}$ do not collapse; this is consistent with the larger flow modulation observed for the larger volume fraction, and agrees with the above discussed results for the single-particle statistics. Notably, for heavy particles $S_{2,p}$ differs from the fluid structure function $S_2$ at the small scales. According to both the PP-DNS and the PR-DNS, $S_{2,p}$ does not exhibit a $r^2$ scaling at the smallest scales, being substantially flat at small $r$. The relative motion between couples of heavy particles placed at a small distances $r$ is substantially uncorrelated as well as decoupled from the small scale fluid motion due to their large inertia. Note that, the absence of the $S_{2,p} \sim r^2$ scaling indicates that at small scales the Eulerian particle velocity field cannot be described with a Taylor expansion. Notably, figure \ref{fig:Pstrfun} shows that $S_{2,p}$ recovers the $S_2-$slope at larger $r$, exhibiting the classical Kolmogorov $r^{2/3}$ scaling in the inertial range of scales. This suggests that, despite the large inertia, the relative particle-particle velocity $\delta \bm{u}_p$ between two particles is driven by turbulent eddies having size comparable to $r$, provided that $r$ is large enough.
When comparing the results of the PP-DNS with those of the PR-DNS, we note that the slope of $S_{2,p}$ matches for small ($ r/\eta \lessapprox 5$) and large ($r \gtrapprox 30$) scales. For intermediate scales $10 \lessapprox r/\eta \lessapprox 30$, instead, the PR-DNS predict a steeper slope for both $\Phi_V=10^{-5}$ and $\Phi_V=10^{-3}$ (see the inset in the bottom panel of figure \ref{fig:Pstrfun}). The finite size of the particles does not influence $S_{2,p}$ for large and small scales where $r/D_p = \mathcal{O}(100)$ and $r/D_p =\mathcal{O}(1)$, but it does for intermediate scales $r/D_p = \mathcal{O}(10)$. 

\begin{figure}
  \centering
  \includegraphics[width=0.49\textwidth]{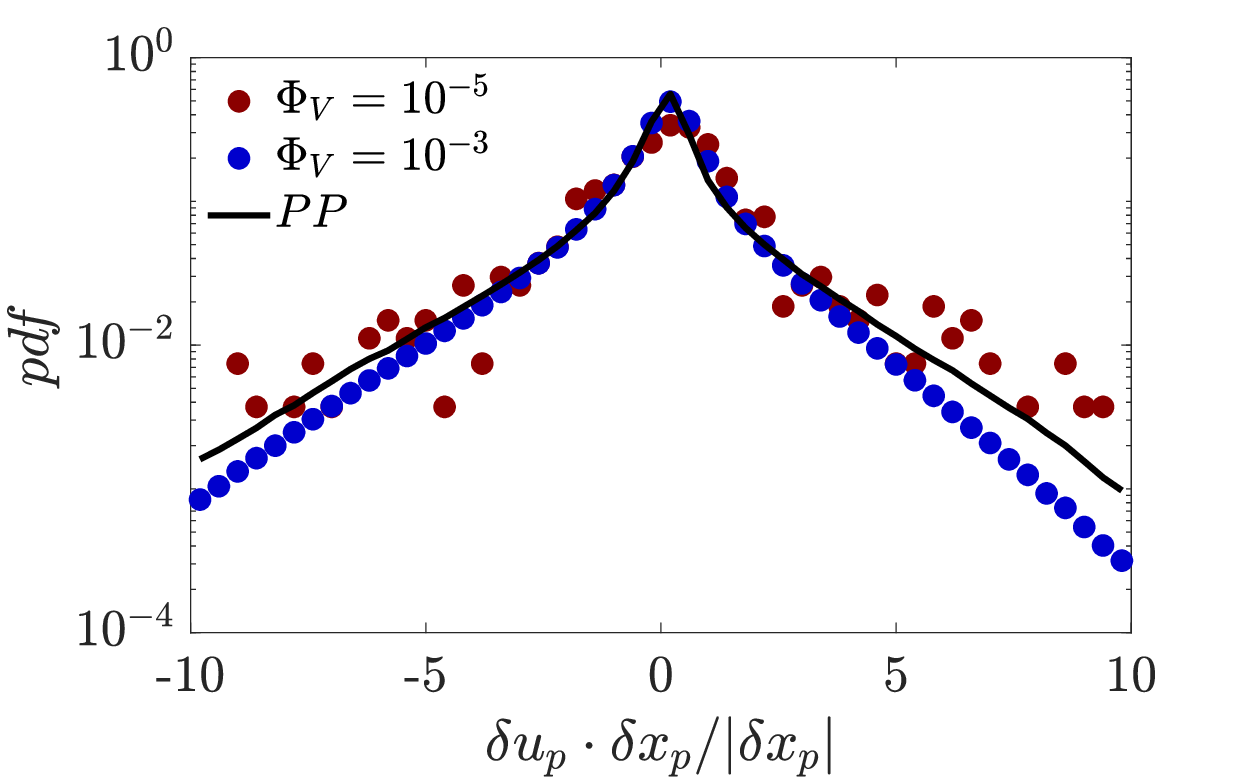}
  \includegraphics[width=0.49\textwidth]{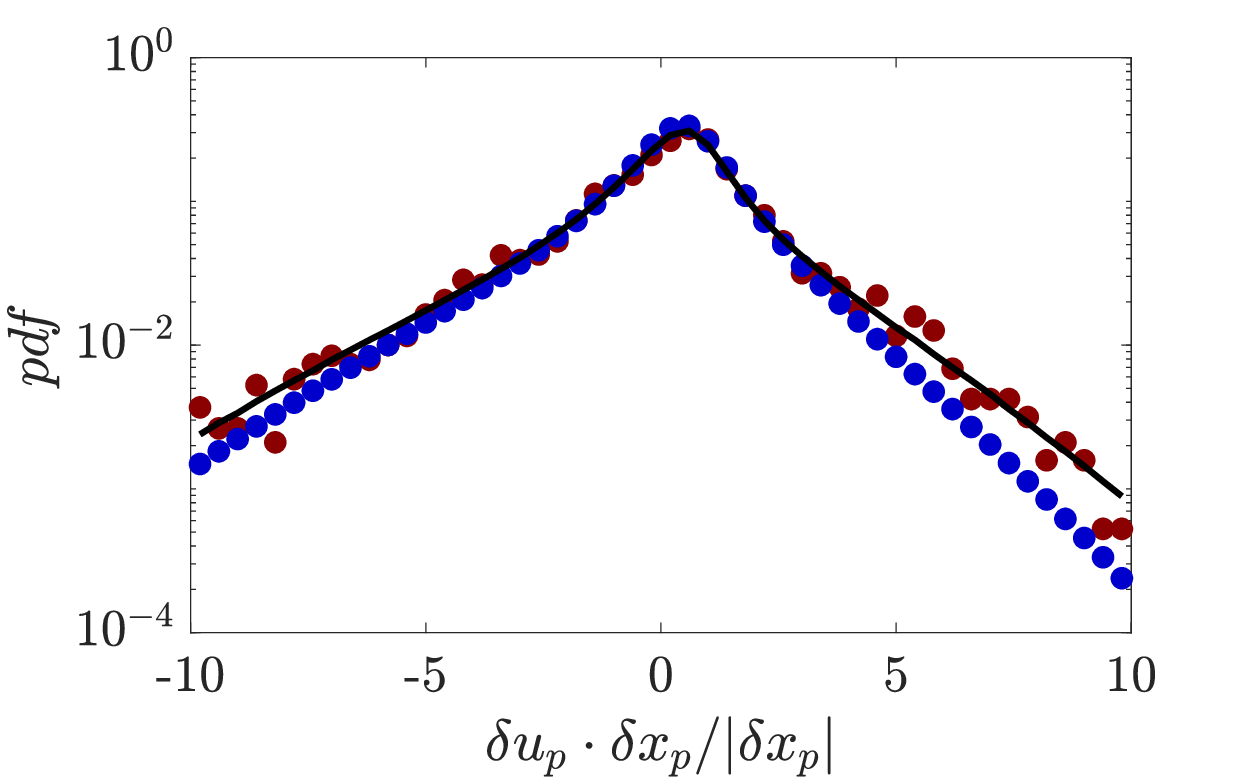}  
  \includegraphics[width=0.49\textwidth]{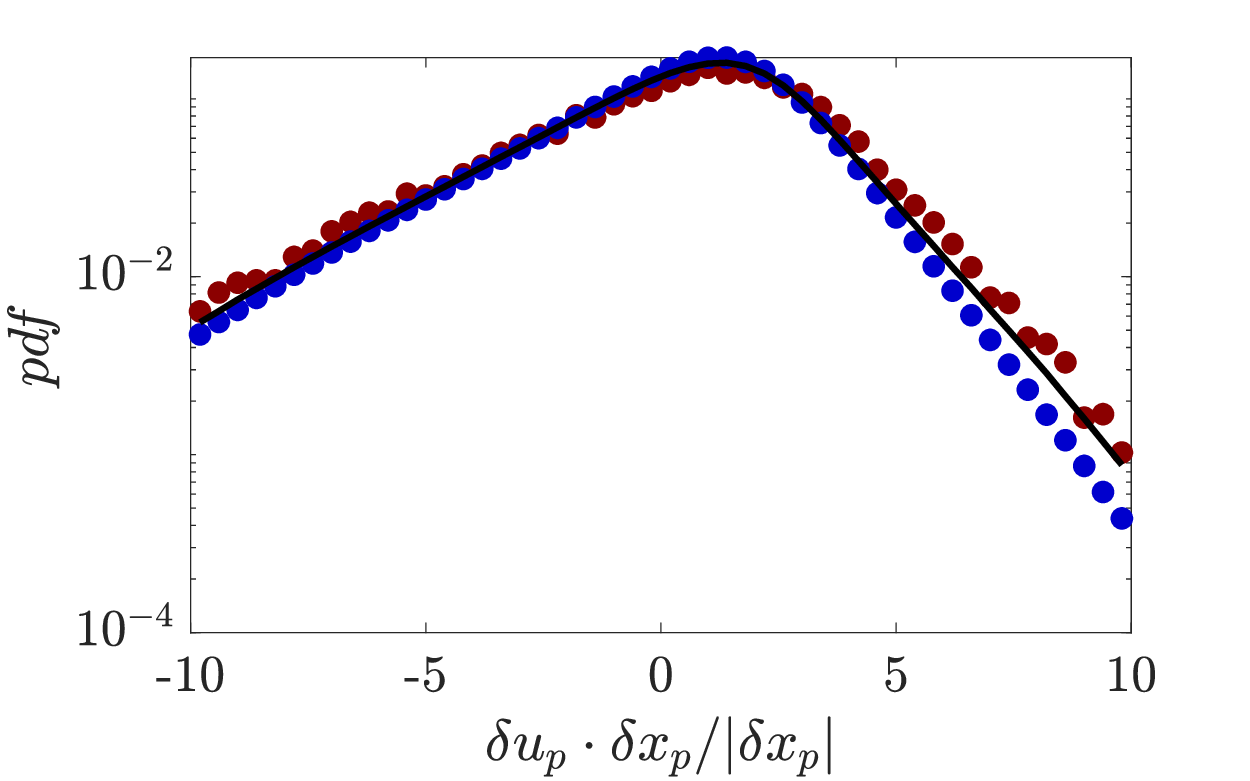}
  \includegraphics[width=0.49\textwidth]{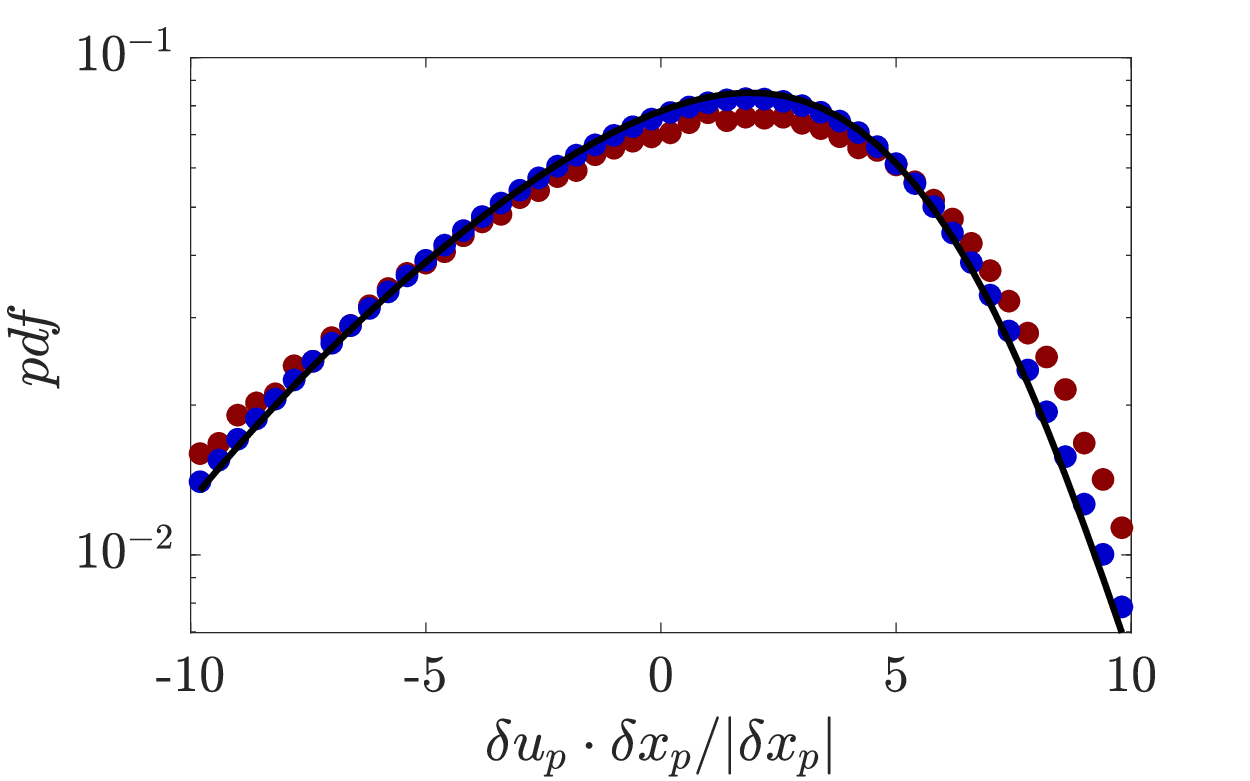}  
  \caption{Probability density function of the radial particle-particle relative velocity $\delta \bm{u}_p \cdot \bm{r}/r$ for $\rho_p/\rho_f=100$. The panels are for (top left) $r \approx 3.5D_p = 0.0627$, (top right) $r \approx 7D_p =0.1254$, (bottom left) $r \approx 21D_p = 0.3761$, and (bottom right) $r \approx 67D_p = 1.2$.}
  \label{fig:dupPDF}
\end{figure}
%
%
Figure \ref{fig:dupPDF} sheds further light on the relative particle-particle velocity by plotting the distribution of $\delta \bm{u}_p \cdot \bm{r}/r$, i.e. the component of $\delta \bm{u}_p$ projected along the vector separating the two particles, for different values of $r$. When $\delta \bm{u}_p \cdot \bm{r}/r>0$, the two particles depart, whereas they get closer when $\delta \bm{u}_p \cdot \bm{r}/r<0$. We consider the case with $\rho_p/\rho_f = 100$ to provide further insights of the distribution of $S_{2,p}$, shown in the bottom panel of figure \ref{fig:Pstrfun}. A first observation is that, similarly to what found for larger particles by \cite{chiarini-rosti-2024}, the distribution of $\delta \bm{u}_p \cdot \bm{r}/r$ is left skewed, with a slightly positive mode and a long negative tail. The distribution becomes progressively more flat when increasing $r$, in agreement with a lower level of the correlation between the velocity of the two particles. When comparing the distributions for $\Phi_V = 10^{-5}$ and $\Phi_V = 10^{-3}$, figure \ref{fig:dupPDF} shows that for all $r$ the tails are shorter for the larger $\Phi_V$, with the difference decreasing as $r$ increases. This is consistent with the stronger flow modulation that globally leads to a weaker level of the flow fluctuations; see table \ref{tab:simulations}. Also, in agreement with the evolution of $S_{2,p}$ with $r$ (see figure \ref{fig:Pstrfun}), for $\Phi_V=10^{-5}$ the distribution obtained with the PP-DNS collapses nicely with that obtained with the PR-DNS at small scales (see the top panels), with some substantial differences arising for intermediate scales when the finite-size effects are relevant.

\section{The collective motion of the particles}
\label{sec:collective}

In this section we focus on the collective motion of the particles. First, we investigate whether Kolmogorov-size particles agglomerate and form clusters. Then, we relate the presence of the clusters with the tendency of particles to preferentially sample particular regions of the flow.

\subsection{Clustering}

Over the years, several tools have been used to characterise the spatial arrangement of the particles in the flow \citep[see][for an overview]{brandt-coletti-2022}. We use the Vorono\"{i} tessellation, which has been extensively used by several authors \citep[see for example][]{monchaux-etal-2010,monchaux-etal-2012}. The position of each particle is identified with its centre and the computational domain is divided in subdomains, such that each grid cell is associated with the closest particle. The Vorono\"{i} volume $V_V$ of each particle is thus defined as the collective volume of grid cells that are closer to it than to other particles. The inverse of the Vorono\"{i} volumes provides a measure of the local concentration: particles placed in void regions possess a large Vorono\"{i} volume, while particles that are part of a cluster have a small Vorono\"{i} volume. Based on these observations, the intensity of the clustering of a suspension can be measured by comparing the distribution of its Vorono\"{i} volumes to that of a control consisting of an equivalent, uniformly random suspension of particles. More intense clustering leads to a variance of the distribution of $V_V$ that is larger than that of the control. In the context of PR-DNS, the overlap between particles is not allowed also in the reference random arrangement.

\begin{figure}
\centering
\includegraphics[width=0.49\textwidth]{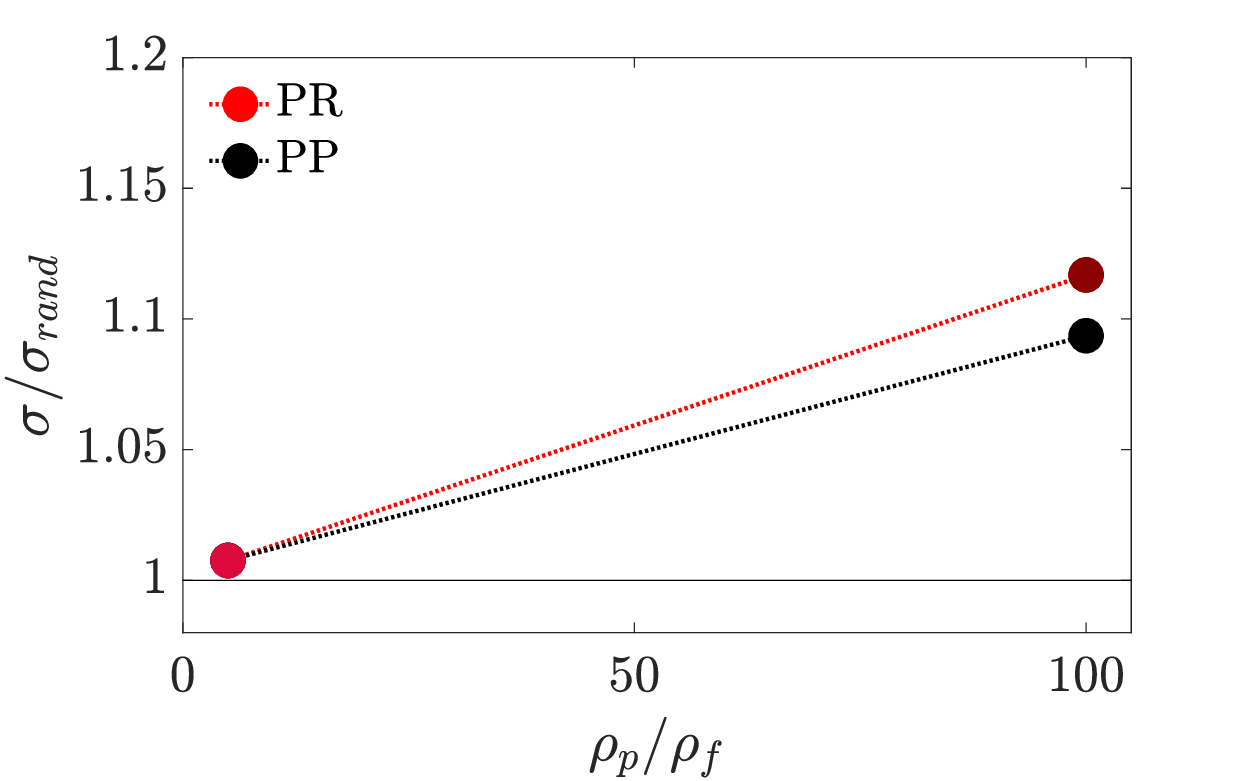}
\includegraphics[width=0.49\textwidth]{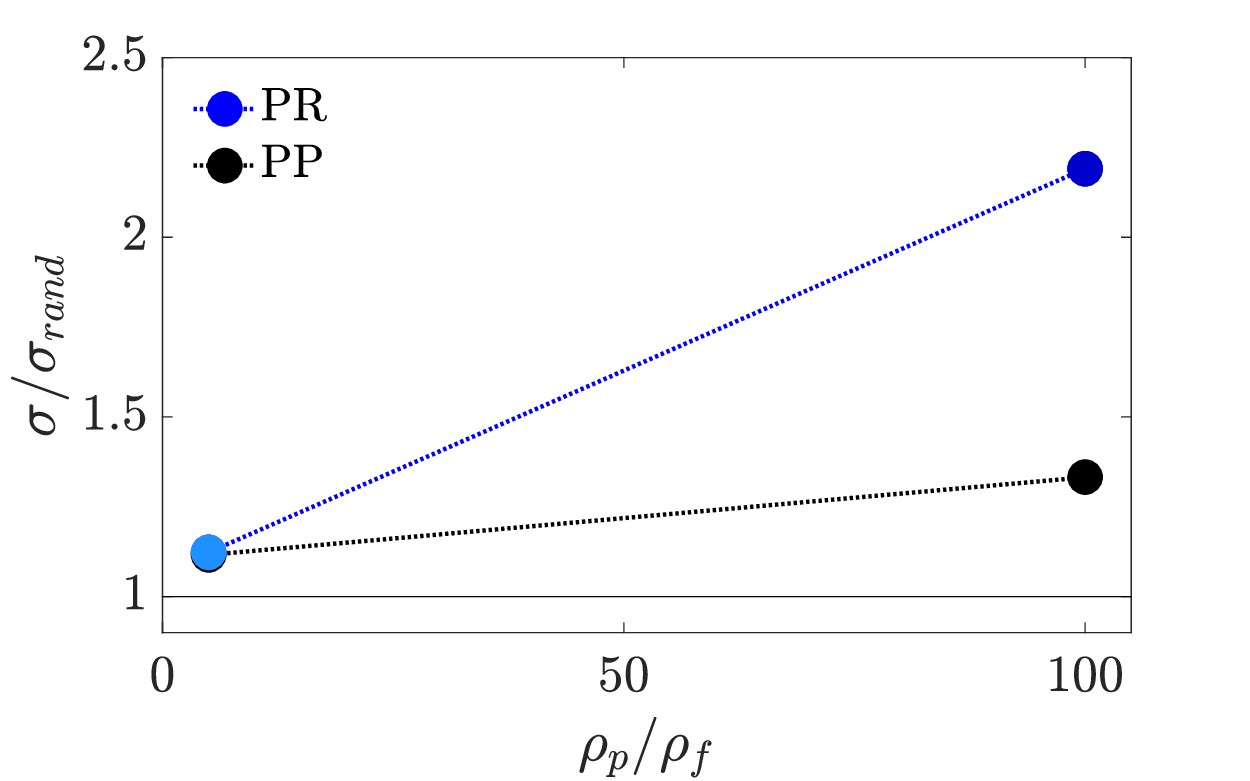}
\caption{Comparison of the variance of the Voronoi Volumes obtained with PR-DNS and PP-DNS. Left: $\Phi_V=10^{-5}$. Right: $\Phi_V = 10^{-3}$.}
\label{fig:Voronoi-sigma}
\end{figure}
Figure \ref{fig:Voronoi-sigma} presents the clustering intensity for the different values of $\Phi_V$ and $\rho_p/\rho_f$ considered. A first observation is that the PP-DNS underestimate the level of clustering for all cases. Our computations show that the discrepancy between the PP-DNS and the PR-DNS increases with $\rho_p/\rho_f$ and/or $\Phi_V$ (see \S\ref{sec:sampling} for further details). 

We now move to the effect of the volume fraction $\Phi_V$ and of the particle density $\rho_p/\rho_f$. As expected, the level of clustering increases with $\Phi_V$. When fixing $\Phi_V$, instead, figure \ref{fig:Voronoi-sigma} shows that heavier particles with $\rho_p/\rho_f = 100$ cluster more than lighter particles with $\rho_p/\rho_f =5$. For light particles, the low level of clustering $\sigma/\sigma_{rand} \approx 1$ agrees with the previous results of \cite{fiabane-etal-2012}, \cite{uhlmann-chouippe-2017} and \cite{chiarini-rosti-2024}, who considered larger particles $ 5 \le D_p/\eta \le 123$ over a wide range of Reynolds numbers $105 \le Re_\lambda \le 430$. Light particles have small inertia and are less likely to drift from the trajectories of the fluid elements. In contrast, the larger level of clustering observed when increasing the particle density from $\rho_p/\rho_f = 5$ to $\rho_p/\rho_f=100$ is not consistent with what found for larger particles. For suspensions of particles with size in the inertial range of scales, indeed, \cite{chiarini-rosti-2024} found that the level of clustering exhibits a non-monotonous dependence on $\rho_p/\rho_f$, with the maximum occurring at intermediate densities (see figure 24 of their paper), and the minimum being at the largest density ratio they considered, i.e. $\rho_p/\rho_f = 100$. However, they found that in the $D_p-\rho_p$ space of parameters the maximum level of clustering moves towards larger $\rho_p$ as $D_p$ decreases, suggesting that the tendency of particles to cluster is driven by the Stokes number of the particles, rather than by their density. Accordingly, their data show that the level of clustering is maximum when $St = \mathcal{O}(1-10)$. This agrees with the early works of \cite{wang-maxey-1993}, \cite{fessler-etal-1994} and \cite{aliseda-etal-2002}, and it is consistent with our results (see table \ref{tab:simulations} for the particles' Stokes number). 

\begin{figure}
\centering
\includegraphics[width=0.49\textwidth]{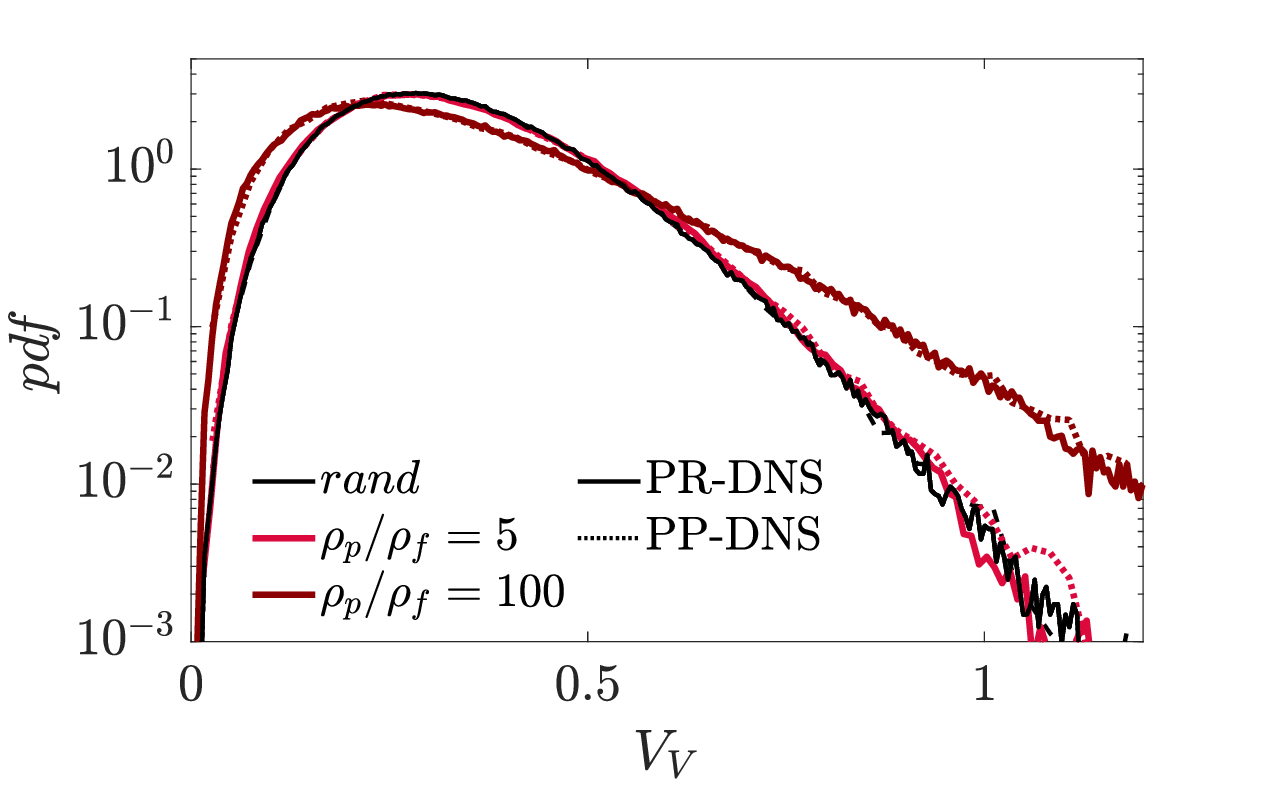}
\includegraphics[width=0.49\textwidth]{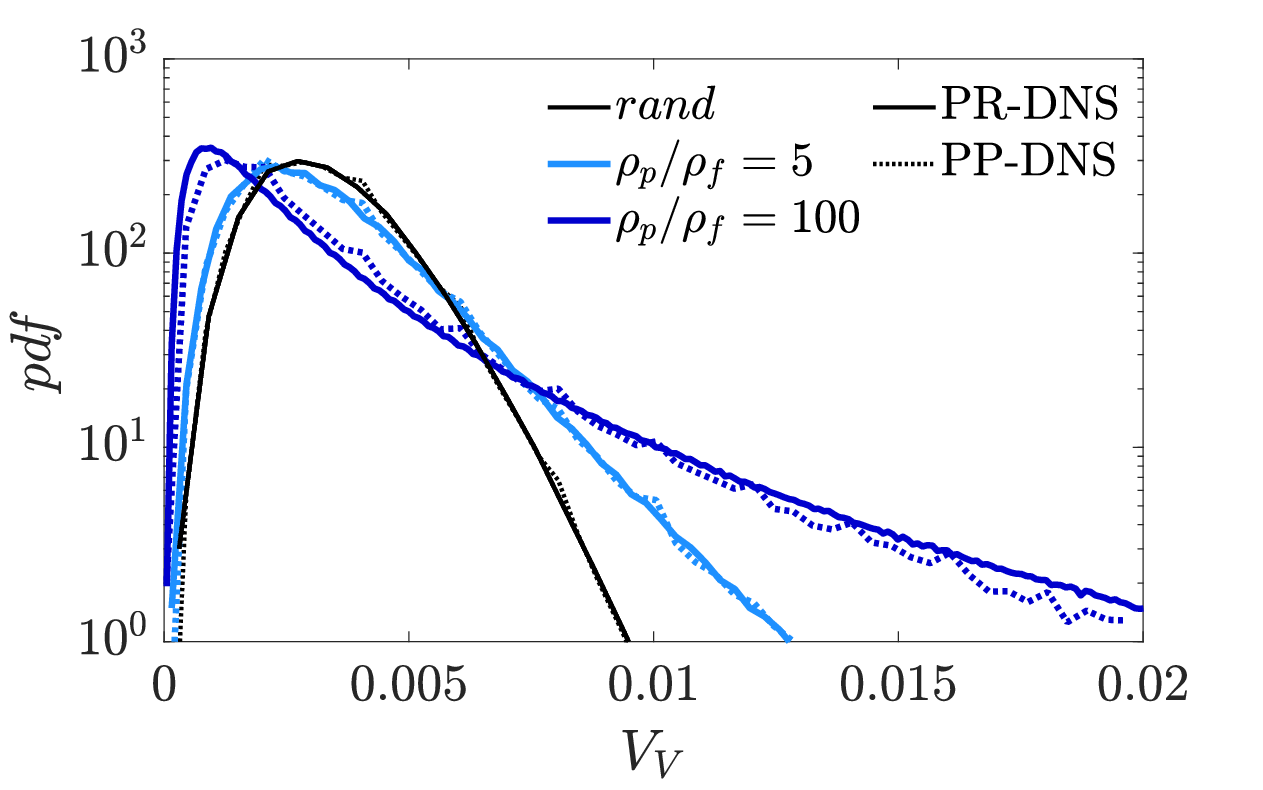}
\caption{Probability density functions of the Voronoi volumes for (left) $\Phi_V = 10^{-5}$ and (right) $\Phi_V = 10^{-3}$ for all $\rho_p/\rho_f$. In the right panel, the dashed lines are used for the PP-DNS. The black lines refer to the Voronoi volumes for a random distribution of particles. Note that, particles are not overlapping for the PR-DNS also in the random distribution.}
\label{fig:Voronoi-pdf}
\end{figure}
The complete distributions of the Vorono\"{i} volumes are provided in figure \ref{fig:Voronoi-pdf}. Compared to the corresponding random arrangement of particles, the tails of the $V_V$ distribution are longer, and grow ever more so with increasing $\Phi_V$ and/or $\rho_p/\rho_f$. This is in agreement with the above discussion, since stronger clustering corresponds to a larger number of small and large Vorono\"{i} volumes. Note that, the PP-DNS underestimation of the level of clustering is visualised in figure \ref{fig:Voronoi-pdf} with the shorter tails. Figure \ref{fig:Voronoi-pdf} can be used to determine which particles are part of clusters and which are part of void regions \citep{monchaux-etal-2010}. This information is used in \S\ref{sec:sampling}, when discussing the particle preferential sampling. In presence of clusters, two cross-over points arise between the $V_V$ distributions of the actual suspension and that of the corresponding random arrangement of particles. Particles with a Vorono\"{i} volume smaller than the left cross-over point $V_{th,l}$ are part of a cluster, while those with a Vorono\"{i} volume larger than the right cross-over point $V_{th,r}$ are part of void regions. Particles that are part of a cluster and have Vorono\"{i} volumes that share at least one vertex are part of the same cluster. Note that, as the level of clustering increases, the threshold of the Vonoro\"{i} volume that delimits the particles entrapped in clusters decreases. 

\begin{figure}
\centering
\includegraphics[width=0.49\textwidth]{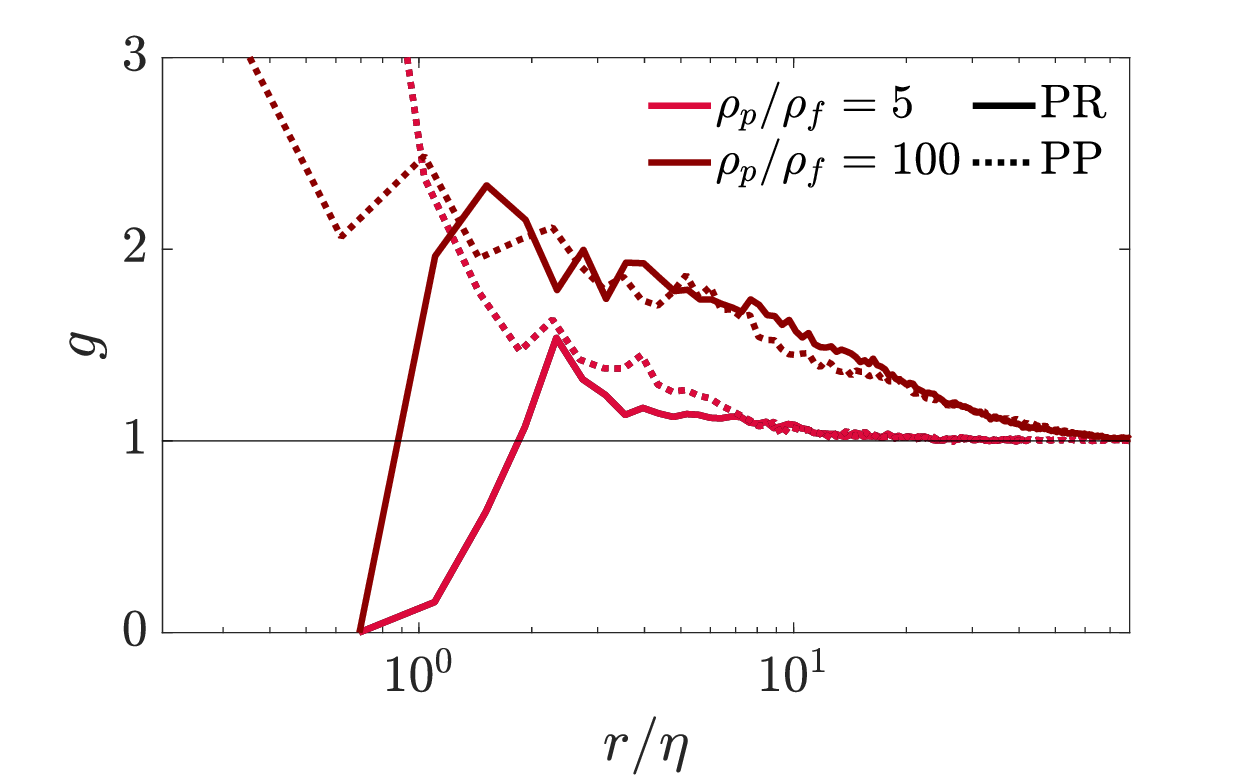}
\includegraphics[width=0.49\textwidth]{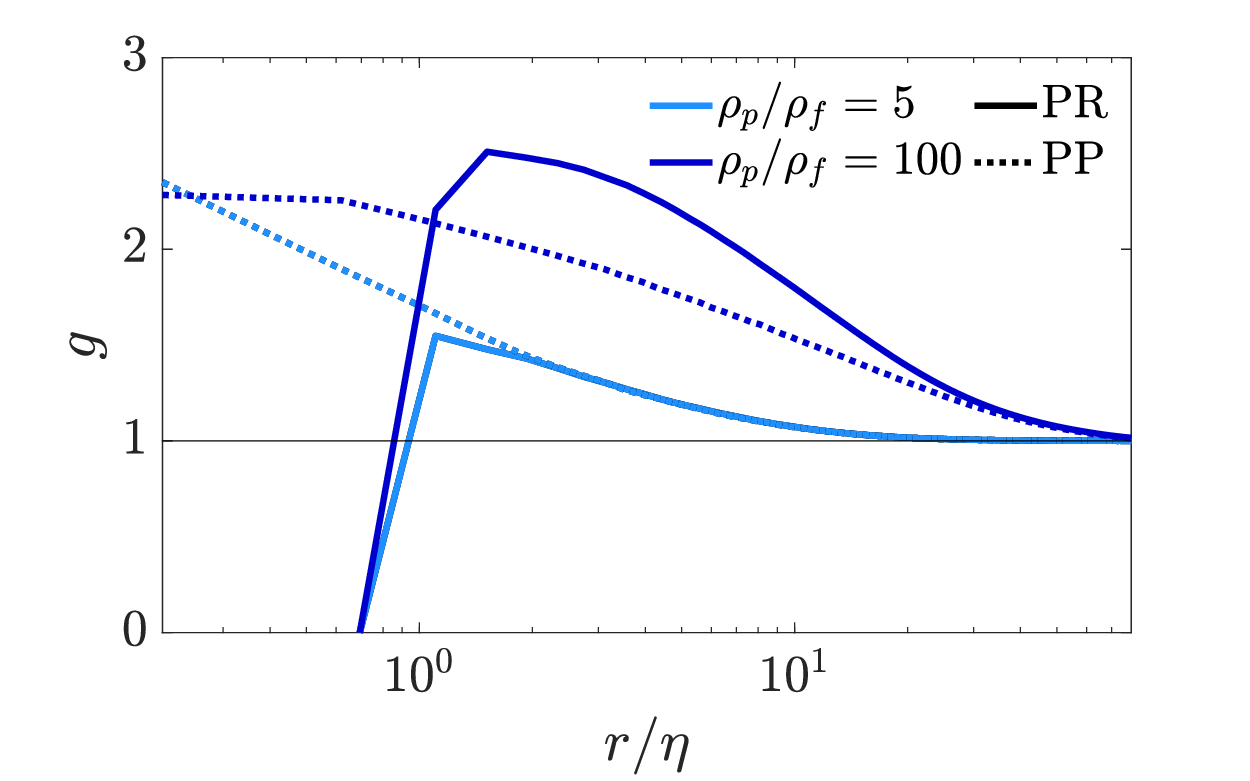}
\caption{Radial distribution function for (left) $\Phi_V = 10^{-5}$ and (right) $\Phi_V = 10^{-3}$. Solid lines are for PR-DNS, while dashed lines are for PP-DNS.}
\label{fig:rdf}
\end{figure}

A different type of information regarding the spatial arrangement of the particles can be provided by means of the radial distribution function $g(r)$, also referred to as pair correlation function; see figure \ref{fig:rdf}. It describes how the particles' density varies as a function of the distance away from a reference particle. In other words, it is a measure of the probability of finding particles at a distance $r$ relative to that of a homogeneous distribution. Following \cite{salazar-etal-2008} and \cite{saw-etal-2008}, the radial distribution function is defined as
\begin{equation}
g(r) = \frac{ N_s(r)/\Delta V(r) }{ N_{pa}/V },
\end{equation}
where $N_s(r)$ is the number of particle pairs separated by a distance between $r-\Delta r$ and $r+\Delta r$, $\Delta V(r)$ is the volume of the spherical shell of inner and outer radius $r-\Delta r$ and $r+\Delta r$ respectively, $N_{pa}$ is the total number of particle pairs present in the system $N_{pa} = N(N-1)/2$, and $V$ is the volume of the computational domain. In a uniform distribution where overlapping between particles is allowed $g(r)=1$ for all $r$.

The radial distribution function (see figure \ref{fig:rdf}) shows that for all cases the accumulation is maximum at the smallest distances. Note that, the maximum of $g(r)$ occurs at $r \approx D_p$ for PR-DNS, as the overlap between particle is not allowed. In agreement with the above discussion, the heavy particles with $\rho_p/\rho_f = 100$ show a larger level of clustering compared to the lighter ones with $\rho_p/\rho_f = 5$. The level of accumulation is also slightly larger for $\Phi_V = 10^{-3}$ at all $r$. Similarly to what observed with the Vorono\"{i} tessellation, figure \ref{fig:rdf} shows that the PP-DNS underpredict the level of particle accumulation. As clearly visible for $\Phi_V = 10^{-3}$ and $\rho_p/\rho_f = 100$, the discrepancy between the PP-DNS and PR-DNS is maximum at the smallest separations. 

\subsection{Preferential sampling}
\label{sec:sampling}

In the previous section we have shown that the solid phase is not homogeneously distributed in space, and that the particles exhibit a mild level of clustering. In this section we relate the presence of the clusters with the tendency of the particles to preferentially sample certain regions of the flow. In doing this, we also provide a possible explanation of the different level of clustering predicted by the PR-DNS and PP-DNS for the $\Phi_V = 10^{-3}$ and $\rho_p/\rho_f = 100$ case. Over the years several mechanisms have been proposed as governing the particles' preferential sampling of the flow, most of them justified using the MRG equation and thus, strictly speaking, valid only in the context of sub-Kolmogorov particles. 
In the following we use the centrifuge mechanism \citep{maxey-1987} to explain the tendency of Kolmogorov-size particles to form clusters. 

In the limit where the point-particle approximation holds, \cite{maxey-1987} has shown that particles with large $\rho_p/\rho_f$ tend to collect in regions of high strain rate and low vorticity. In the presence of a vortex, indeed, heavy particles cannot  follow the flow streamlines because of their large inertia, and tend to drift from the vortex core. Similarly, in the case of a pure straining flow, heavy particles drift towards the stagnation point at the centre. 
We quantify the tendency of particles to sample regions of high strain by using the second invariant of the deformation tensor (see \S\ref{sec:qr}), i.e. $Q = - S_{ij}S_{ij}/2 + \omega^2/4$; recall that regions where $Q$ is large and positive are regions of high vorticity ($Q \sim \omega^2/4$), while regions where $Q$ is large and negative are regions of high strain ($ Q \sim - S_{ij}S_{ij} $). We investigate the particle preferential sampling by computing the probability density function of $Q$ at the particles position \citep{squires-eaton-1990}. For the sake of brevity, in this section we limit the investigation to the largest volume fraction $\Phi_V = 10^{-3}$. For PP-DNS the value of $Q$ at each particle position is obtained after linear interpolation. For PR-DNS, instead, the value of $Q$ seen by each particle is estimated as the average value within a spherical shell centred with the particle and having a radius $R_{sh}>R_p$, where $R_p=D_p/2$ is the radius of the particles. It is important to note, however, that due to the particles' backreaction, in the case of PR-DNS, the value of $Q$ seen by each particle is actually the result of three different effects: \textit{(i)} the larger scale flow properties of the region that the particle is sampling, \textit{(ii)} the smaller scale influence of the particle on the surrounding flow, and \textit{(iii)} the effect of nearby particles on the flow. Also, a suitable choice of $R_{sh}$ should be made: when $R_{sh}$ is too small, only the influence of the particle on the surrounding flow is considered \citep[see for example][]{kidanemariam-etal-2013,oka-goto-2022}, while when $R_{sh}$ is too large, spurious contributions that do not affect the particle location are instead captured. In order to obtain a complete picture, we have tested different values of $R_{sh}$.

\begin{figure}
  \centering
  \includegraphics[width=0.85\textwidth]{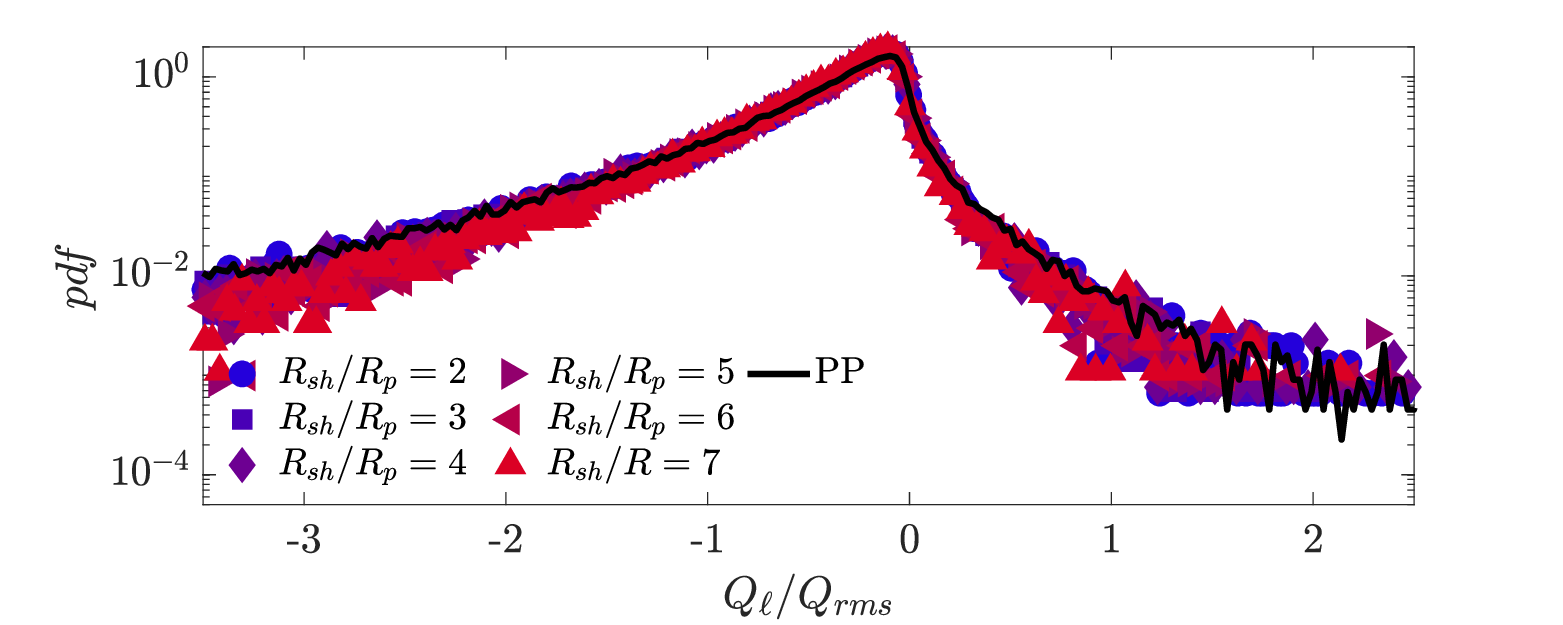}
  \includegraphics[width=0.85\textwidth]{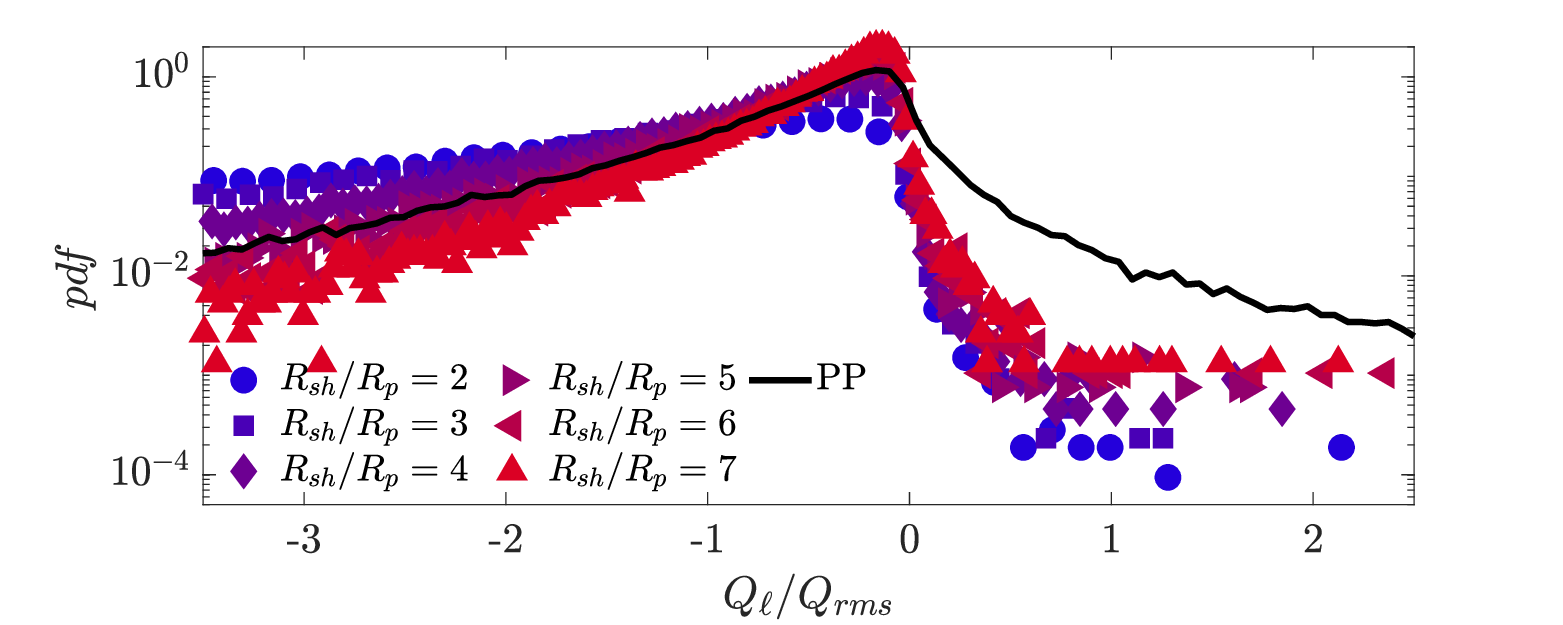}
  \caption{Probability density function of the second invariant $Q$ of the velocity gradient tensor evaluated at the particle position. The top panel is for $\Phi_V=10^{-3}$ and $\rho_p/\rho_f=5$. The bottom panel is for $\Phi_V=10^{-3}$ and $\rho_p/\rho_f = 100$. $R_{sh}$ indicates the radius of the spherical shell used to estimate the value of $Q$ seen by the particles in the PR-DNS. The black line indicates the distribution according to the PP-DNS.}
  \label{fig:Qloc}
\end{figure}
Figure \ref{fig:Qloc} shows the probability density function of $Q_\ell$, i.e. $Q$ computed at the particle position. We start by looking at the dependence of the PR-DNS results on the radius $R_{sh}$ of the shell. For $\rho_p/\rho_f = 5$, the curves computed for values of $R_{sh}$ between $2 \le R_{sh}/R_p \le 7$ show an almost perfect overlap. This is consistent with the low level of modulation discussed in \S\ref{sec:flow}, and indicates that for these light particles the main contribution to $Q_\ell$ comes from the larger scale properties of the flow region sampled by the particles (note that due to the low volume fraction the influence of the particle-particle interaction is negligible). For heavy particles with $\rho_p/\rho_f = 100$, instead, the distribution of $Q_\ell$ largely varies with $R_{sh}$: as $R_{sh}$ decreases, the left tail of the distribution becomes longer, meaning that particles are more likely to see large negative values of $Q$. These large negative values of $Q$ are, at least partially, the result of the influence of the particles on the neighbouring flow; see the $Q-R$ map in figures \ref{fig:QRmap}. Note that, for $R_{sh}/R_p \gtrapprox 5$ the distribution of $Q_\ell$ shows only marginal variations, as for these $R_{sh}$ the large scale flow contribution dominates. This suggests that for $\rho_p/\rho_f = 100$ the influence of particles on the surrounding flow extends for less than $5R_p$. Overall, for both light and heavy particles the distribution of $Q_\ell$ is left skewed and shows an almost null probability of positive $Q_\ell$. At the present parameters, both PP-DNS and PR-DNS give evidence that Kolmogorov-size particles preferentially sample regions of high strain rate. This is also visible in the instantaneous snapshot shown in figure \ref{fig:setup}, with particles sampling regions with low $\omega^2$. In other words, in the context of Kolmogorov-size particles the formation of clusters is, at least partially, governed by the centrifuge mechanism.

Let us now focus on the differences between the PR-DNS and PP-DNS results. For particles with $\rho_p/\rho_f=5$, the $Q_\ell$ distribution obtained with PP-DNS overlaps almost perfectly with that obtained with PR-DNS; the point-particle approximation predicts fairly well the tendency of light particles to sample the $Q<0$ regions of the flow. Note that this is consistent with the good agreement found in figure \ref{fig:Voronoi-pdf} when discussing the distribution of the Vorono\"{i} volumes. For $\rho_p/\rho_f = 100$, instead, the $Q_\ell$ distribution obtained with the PP-DNS significantly deviates from that obtained with the PR-DNS. For all $R_{sh}$, the $Q_\ell$ distribution obtained with PR-DNS shows a shorter right tail and predicts a higher probability of negative $Q$: finite-size heavy particles are less/more likely to see positive/negative values of $Q$. Based on this, one may conclude that, at the present parameters, the PP-DNS underestimates the tendency of the particles to preferentially sample regions of high-strain, and this may explain the larger level of clustering predicted by the PR-DNS for the $\Phi_V = 10^{-3}$ and $\rho_p/\rho_f = 100$ case (see figure \ref{fig:Voronoi-sigma}). 
We speculate that the discrepancy between the PR-DNS and the PP-DNS observed for the $\Phi_V=10^{-3}$ and $\rho_p/\rho_f=100$ case is due to the flow modulation, rather than to the particles' finite-size effects. Unlike for the light particles ($\rho_p/\rho_f=5$), indeed, at the present parameters the influence of the heavy particles ($\rho_p/\rho_f = 100$) on the fluid phase is not negligible (see \S\ref{sec:flow}).

A last comment regards the influence of $\rho_p/\rho_f$ on the distribution of $Q_\ell$. According to both PP-DNS and PR-DNS, heavier particles exhibit a larger tendency to sample regions with more negative $Q$, as shown by the left tail being longer for the $\rho_p/\rho_f=100$ case. Due to their larger inertia, indeed, heavier particles enhance the centrifuge mechanism, being more likely to drift from the high-vorticity regions of the flow.

\begin{figure}
\centerline{
\begin{tikzpicture}
  \node at (0,4.2) {\includegraphics[width=0.49\textwidth]{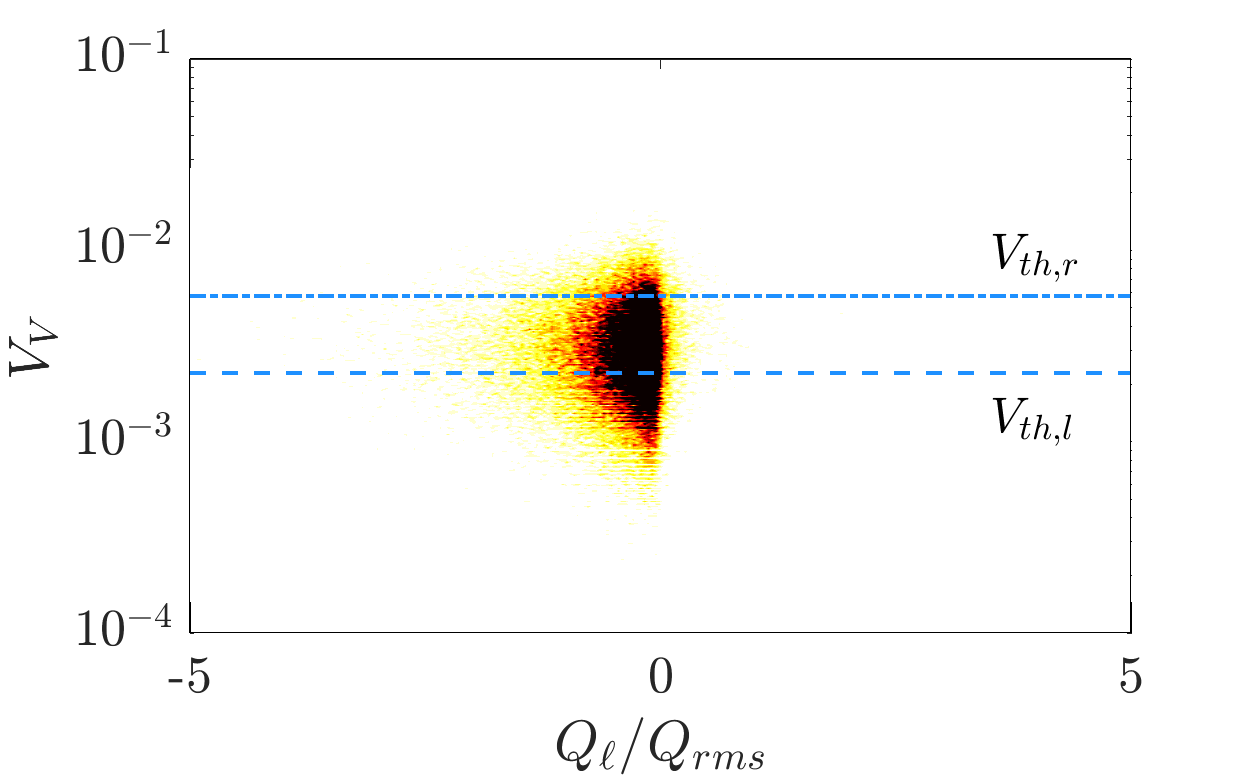}};
  \node at (6.5,4.2) {\includegraphics[width=0.49\textwidth]{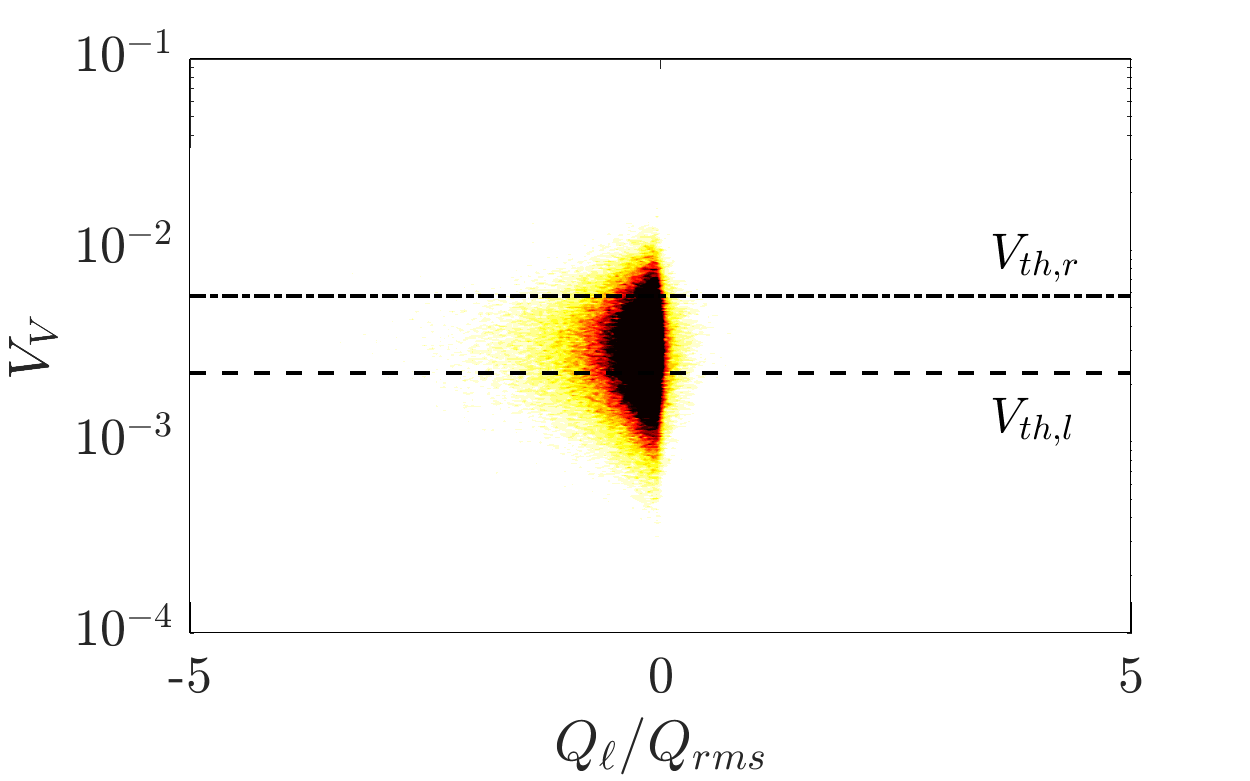}};
  \node at (0,0) {\includegraphics[width=0.49\textwidth]{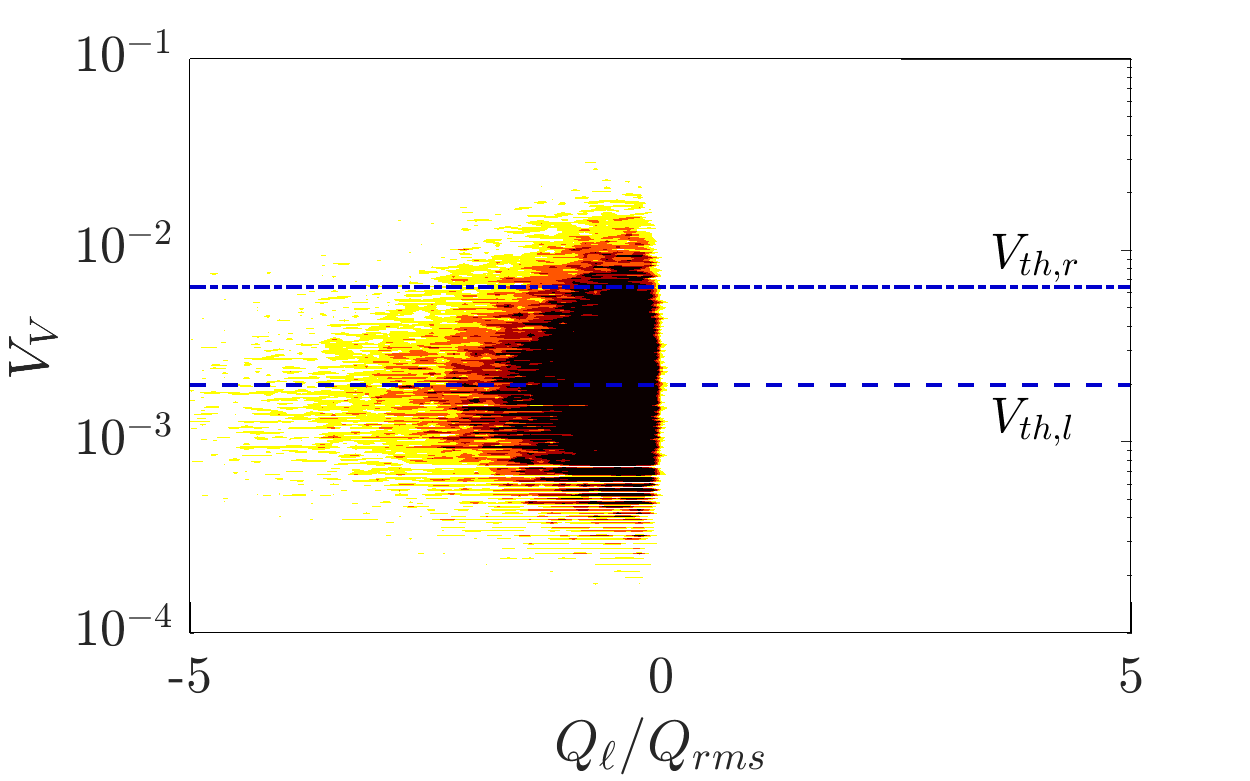}};
  \node at (6.5,0) {\includegraphics[width=0.49\textwidth]{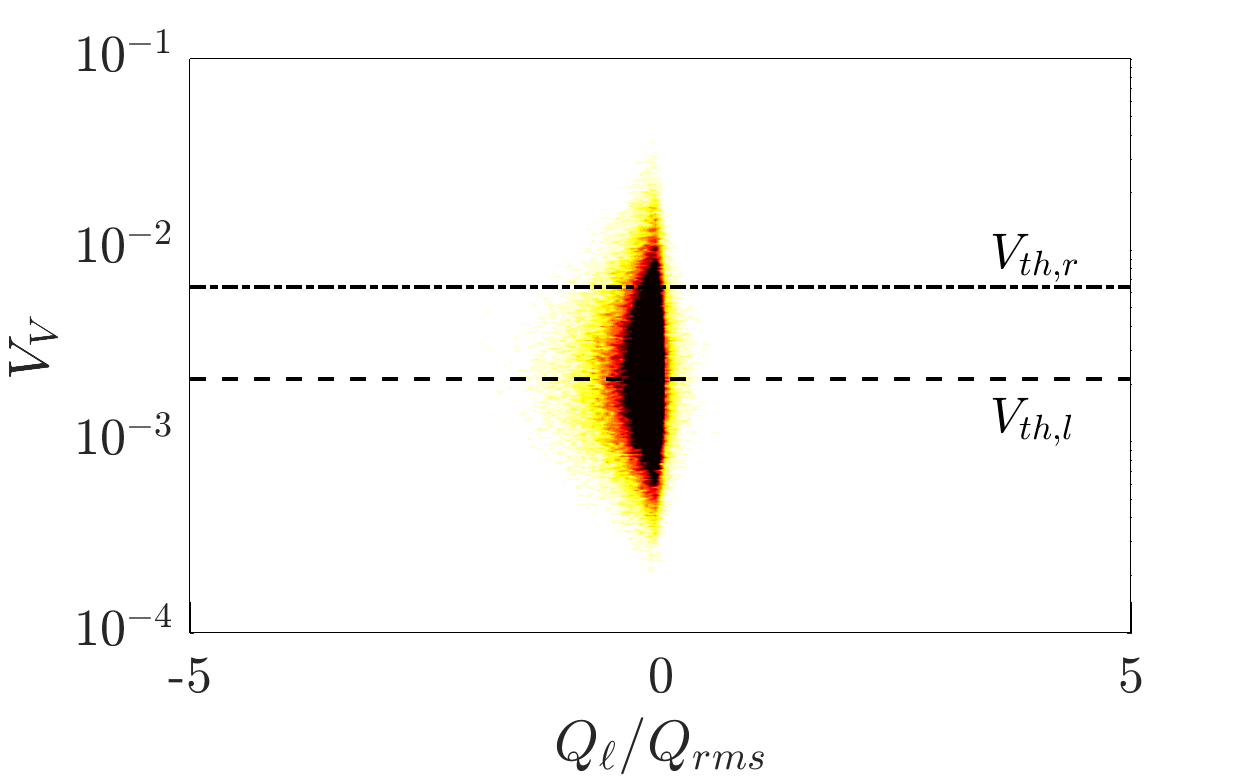}};
  \node at (10.3,4.45) {$\Phi_V=10^{-5}$};
  \node at (10.3,0.25) {$\Phi_V=10^{-3}$};
  \node at (0.3,6.5) {PR-DNS};
  \node at (6.7,6.5) {PP-DNS};
\end{tikzpicture}
}
  \caption{Joint probability density function of $Q$ and $V_V$ for (top) $\Phi_V = 10^{-3}$ and for (bottom) $\rho_p/\rho_f=5$ and $\rho_p/\rho_f=100$. The left panels are for the PR-DNS, while the right panels are for the PP-DNS. White/black colour denotes minimum/maximum probability. The dashed lines represent $V_{th,l}$, while the dash-dotted lines $V_{th,r}$.}
  \label{fig:Q_V_jpdf}
\end{figure}

To provide additional insights regarding the relation between the presence of clusters and the particle preferential sampling, figure \ref{fig:Q_V_jpdf} shows the joint probability density function of $Q_\ell$ and $V_V$ for $\Phi_V = 10^{-3}$. Based on the above discussion, here we set $R_{sh}/R_p = 3$ for the computation of $Q_\ell$, since it is large enough to account for the particle preferential location and small enough to avoid spurious contributions. We recall that according to \cite{monchaux-etal-2010}, particles with $V_V \le V_{th,l}$ are part of a cluster, while particles with $V_V \ge V_{th,r}$ are in void regions of the flow. For all cases, the most negative values of $Q_\ell$ well correlate with small and intermediate Vorono\"{i} volumes, with $V_V \lessapprox V_{th,r}$. Particles that are in void regions and are not part of a cluster are less likely to see large negative values of $Q$. This agrees with the above observation that the tendency of Kolmogorov-size particles to form clusters is governed by the centrifuge mechanism. We now focus on $\rho_p/\rho_f=100$ (see the bottom panels). The joint distribution shows that the larger probability of negative $Q_\ell$ predicted by the PR-DNS is concentrated at the smallest Vorono\"{i} volumes with $V_V \lessapprox V_{th,l}$. Again, this shows that the higher level of clustering detected in this case with the PR-DNS well correlates with the larger tendency of finite-size particles to sample regions of the flow with intense strain.

\section{Conclusion}
\label{sec:conclusions}

We have investigated by direct numerical simulations the fluid-solid interaction of suspensions of Kolmogorov-size spherical particles moving in homogeneous isotropic turbulence. The work is based on both interface-resolved (PR-DNS) and one-way-coupled point-particle (PP-DNS) direct numerical simulations. In PR-DNS the presence of the particles is dealt with the immersed boundary method introduced by \cite{hori-rosti-takagi-2022}. In PP-DNS the motion of the particles is described by solving the complete Maxey-Riley-Gatignol equation \citep{maxey-riley-1983,gatignol-1983}, including the time-history Basset term and the Fax\'en correction. The objective of the study is twofold. On one side, we aim to shed light on how Kolmogorov-size particles influence the organisation of the velocity fluctuations. Few work have indeed considered particles with $D_p/\eta \approx 1$ at a Reynolds number that is large enough to ensure a separation of scales due to the intrinsic complexity of the problem: in experiments it requires the access to sub-Kolmogorov measurements, and in simulations it requires an extremely fine grid with a resulting prohibitive computational cost. On the other side, we aim to assess the limits of the one-way-coupled PP-DNS that, despite the large number of works present in literature, have not been completely addressed yet. For this reason, we consider a portion of the parameter space that is on the edge of the range of validity of the one-way-coupled point-particle models \citep[see][]{brandt-coletti-2022}. The micro-scale Reynolds number is $Re_\lambda \approx 140$, being large enough to ensure a proper separation of scales. The volume fraction of the suspension has been set to the small values of $\Phi_V=10^{-5}$ and $\Phi_V = 10^{-3}$ to guarantee that the backreaction of the solid phase on the carrier flow is low. Two solid-to-fluid density ratios are considered, i.e. $\rho_p/\rho_f = 5$ and $\rho_p/\rho_f = 100$, to investigate the role of inertia. 

The PR-DNS shows that at the present parameters the modulation of the flow is rather low and mainly involves the smallest scales. The modulated energy spectrum $\mathcal{E}(\kappa)$ shows a multi-scaling behaviour: the classical $\kappa^{-5/3}$ scaling in the inertial range of scales is indeed followed by a steeper $\kappa^{-4}$ scaling, that resembles what has been observed by several authors in the context of bubbly flows \citep[see][]{pandey-mitra-perlekar-2023}. Accordingly, the scale-by-scale energy budget shows two different regimes: in the inertial range of scales the fluid-structure interaction term is negligible and the nonlinear term equals the dissipation rate, i.e. $\Pi(\kappa) \sim \epsilon$; at these scales energy is transferred from larger to smaller scales by means of the classical energy cascade described by Richardson and Kolmogorov. At small scales, where $\mathcal{E}(\kappa) \sim \kappa^{-4}$, the nonlinear term is negligible and the fluid-structure interaction term is balanced by the viscous dissipation, i.e. $\Pi_{fs}(\kappa) \sim D_v(\kappa)$; at these scales the energy injected into the flow by the particles is directly dissipated by viscosity. 
The small-scale topology of the flow has been investigated by inspecting the influence of the particles on the invariants of the velocity gradient tensor $A_{ij} = \partial u_i/\partial x_j$ \citep{meneveau-2011}. The effect of the solid phase on the eigenvalues of the strain-rate tensor shows that the presence of the particles favours axisymmetric compression rather than axisymmetric extension. The joint probability density function of the second and third invariants of $A_{ij}$ reveals that particles mainly enhance events with axial strain and vortex compression. Accordingly, the inspection of the joint probability density function of the second invariants of the symmetric and antisymmetric parts of $A_{ij}$ indicates that the presence of the particles favours dissipation events dominated by extensional events rather than rotational ones. Our findings show that the modulation of homogeneous isotropic turbulence by Kolmogorov-size particles largely differs from what has been observed in previous studies with larger particles.

The limits of the one-way-coupled point-particle models have been addressed by looking at the dynamics of the particles and at their collective motion. We find that the PP-DNS predicts fairly well the Lagrangian and Eulerian statistics of the particles velocity field for the low-density case. For heavy particles, however, some discrepancies are observed, particularly for the larger volume fraction. These differences are due to a combination of the finite-size of the particles and of the flow modulation, that are not accounted for in the PP-DNS. By using the Vorono\"{i} tessellation method and the radial distribution function, we find that the PP-DNS underpredicts the level of clustering; the discrepancy with the PR-DNS results increases with the volume fraction and the particle density. In the attempt to have a clearer picture, we have investigated the tendency of the particles to preferentially sample particular regions of the flow. By plotting the distribution of the second invariant $Q$ of the fluid velocity gradient tensor at the particle position, we find that, according to both PP-DNS and PR-DNS, the particles preferentially sample regions of high strain rate. This suggests that the presence of the clusters is driven by the centrifuge mechanism introduced by \cite{maxey-1987}. Accordingly with the larger level of clustering, we find that PR-DNS shows a larger tendency of the particles to sample these regions of the flow compared to PP-DNS. Note, however, that some care is needed when dealing with these results. In PR-DNS, indeed, the value of $Q$ seen by each particle is the result of three different contributions that cannot be easily isolated, i.e. \textit{(i)} the larger scale flow properties of the region that the particle is sampling, \textit{(ii)} the smaller scale influence of the particle on the surrounding flow, and \textit{(iii)} the effect of nearby particles on the flow.
%

By characterising the fluid-solid interaction of Kolmogorov-size particles in homogeneous isotropic turbulence, the present study aims to serve as a stepping stone for further investigations. A natural extension of this work is to use the present PR-DNS database to assess the limits of two-way-coupled and four-way-coupled PP-DNS that account for the backreaction of the solid phase to the carrier flow \citep[see for instance the models introduced by][]{gualtieri-etal-2015,vreman-2016}, and possibly the particle-particle hydrodynamic interaction. In addition, the present results may be used as a ground truth for studies in the spirit of \cite{olivieri-etal-2014}, that investigate the relevance of each term at the right-hand-sides of the MRG equation in predicting the different statistics of the particles. This knowledge will help guide the choice of suitable models for engineering applications. Eventually, it would be of interest to investigate whether Kolmogorov-size particles modulate the energy spectrum also in the inertial range of scales at larger volume fractions, influencing thus the $\kappa^{-5/3}$ scaling range and the classical energy cascade as observed for larger particles \citep{chiarini-rosti-2024}. Despite the computational challenges such a study would present, the field would benefit from the investigation. Overall, the present results can be exploited for the development of improved point-particle models for the one- and two-way-coupling regimes.



\section*{Acknowledgments} 
The authors acknowledge the computer time provided by the Scientific Computing and Data Analysis section of the Core Facilities at OIST and the computational resources on Fugaku provided by HPCI (project ID: hp230536). A.C. acknowledges Marildo Kola for discussions and suggestions.
  
\section*{Funding} 
The research was supported by the Okinawa Institute of Science and Technology Graduate University (OIST) with subsidy funding to M.E.R. from the Cabinet Office, Government of Japan.
  
\section*{Declaration of Interests} 
The authors report no conflict of interest.

\appendix
\section{The Basset time history force}
\label{sec:basset}

In the PP-DNS we resort on the second-order and memory-efficient algorithm developed by \citet{hinsberg-boonkkamp-clercx-2011} to deal with the Basset time history force. 

The Basset force is split into two parts, denoted as window and tail. In particular, at time $\tilde{t}$ the first part consists of a numerical integration over the $\tilde{t}-t_{w} \le t \le \tilde{t}$ interval, considering thus $N_{w}=t_w/\Delta t$ previous steps. The second integral, instead, considers the $-\infty \le t \le \tilde{t} - t_w$ interval and is approximated using recursive exponential functions.
The kernel $K_B(t)$ in equation \ref{eq: PP-momentum-setup} is thus replaced with an approximated kernel $K(t)$ such that 
\begin{equation}
    K(t) = \begin{cases} 
                  K_B(t) & \text{if } t < t_{win} \\
                  K_{tail}(t) & \text{if } t \geq t_{twin},\end{cases}
\end{equation}
with
\begin{equation*}
    \lim_{t \rightarrow +\infty} K_{tail}(t) = 0.
\end{equation*}
The Basset force, therefore, reads 
\begin{equation*}
\mathbf{F}_B(t) = \underbrace{c_B \int_{t-t_{win}}^{t} K_B(t-\tau) g(\tau) d\tau}_{\mathbf{F}_{B-win}(t)} +
\underbrace{c_B \int_{-\infty}^{t-t_{win}} K_{tail}(t-\tau) g(\tau) d\tau}_{\mathbf{F}_{B-tail}(t)}, 
\end{equation*}
where $c_B = 3/2 D_p^2 \rho_f \sqrt{\pi \nu} $ and $ g(t) = \text{d}(\mathbf{u}-\mathbf{u}_p + (1/6) (D_p/2)^2 \mathbf{\nabla}^2 \mathbf{u}) / \text{d}t $. The window term is integrated in time using the diffusive Basset kernel. The integration exploits a modified trapezoidal method, which allows the kernel's singularity to be taken into account. Thus, following the work of \citet{olivieri-etal-2014}, the window contribution reads:
\begin{equation}
\begin{aligned}
    \mathbf{F}_{B-win} &= \frac{4}{3} c_B \sqrt{\Delta t} \mathbf{g}_0 + \frac{4}{3} c_B \sqrt{\Delta t}\sum_{n = 1}^{N_w-1} \left[  (n-1)\sqrt{n-1} - 2n\sqrt{n} + (n+1)\sqrt{n+1}\right] \mathbf{g}_n \\
    &+ c_B \sqrt{\Delta t} \left[ \frac{4}{3}(N_w - 1)\sqrt{N_w-1} + (2-\frac{4}{3}N_w)\sqrt{N_w} \right]\mathbf{g}_{N_w},
\end{aligned}
\end{equation}    
where $g_n = g(t - n \Delta t )$ with  $n = 0,1,..., N_w$. Here $\Delta t = t_{win}/N_w$ and $N_w$ is the number interval in which the window is discretised.

As stated above, the tail term is integrated in a recursive manner and, exploiting exponential kernels, it reads:
\begin{equation}
    \mathbf{F}_{B-tail}(t) = \sum_{i=1}^m a_i \mathbf{F}_i(t) = \sum_{i=1}^m  a_i \left( \mathbf{F}_{i-di}(t) + \mathbf{F}_{i-re}(t) \right),
\end{equation}
where $\mathbf{F}_{i-di}$ is computed directly as
\begin{equation}
    \mathbf{F}_{i-di}(t) =  2 c_B \sqrt{e t_i} \exp\left(-\frac{t_{win}}{2t_i}\right) \left\{ \mathbf{g}_N \left[ 1 - \phi\left( \frac{\Delta t}{2t_i}\right) \right] +  \mathbf{g}_{N+1} \left[\phi\left(- \frac{\Delta t}{2t_i}\right) - 1 \right]  \right\},
\end{equation}
and $\mathbf{F}_{i-re}$ is computed recursively as
\begin{equation}
\mathbf{F}_{i-re}(t) = \exp\left( -\frac{\Delta t}{2 t_i} \right) \mathbf{F}_i(t-\Delta t ).
\end{equation}
Here, $\phi(z) = (\exp(z)-1)/z$, and for a given value of $m$ the coefficients $\left\{a_i,t_i \right\}_{i=1}^{m}$ are chosen to minimise the error. For a detailed explanation, the reader is referred to \cite{hinsberg-boonkkamp-clercx-2011} and \cite{casas-ferrer-onate-2017}. We choose $m=10$ and set the values of the $a_i$ and $t_i$ parameters to the ones proposed in the work of \citet{hinsberg-boonkkamp-clercx-2011}. As suggested by \citet{hinsberg-boonkkamp-clercx-2011} the point-particle equation is written in a semi-implicit manner to guarantee numerical stability when integrating in time.

\bibliographystyle{jfm}

\begin{thebibliography}{113}
\expandafter\ifx\csname natexlab\endcsname\relax\def\natexlab#1{#1}\fi
\def\au#1{#1} \def\ed#1{#1} \def\yr#1{#1}\def\at#1{#1}\def\jt#1{\textit{#1}}
  \def\bt#1{#1}\def\bvol#1{\textbf{#1}} \def\vol#1{#1} \def\pg#1{#1}
  \def\publ#1{#1}\def\arxiv#1{#1}\def\org#1{#1}\def\st#1{\textit{#1}}

\bibitem[Aliseda {\em et~al.\/}(2002)Aliseda, Cartellier, Hainaux \&
  Lasheras]{aliseda-etal-2002}
{\sc \au{Aliseda, A.}, \au{Cartellier, A.}, \au{Hainaux, F.} \& \au{Lasheras,
  J.C.}} \yr{2002}  \at{Effect of preferential concentration on the settling
  velocity of heavy particles in homogeneous isotropic turbulence}.  \jt{J.
  Fluid Mech.}  \bvol{468},  \pg{77–105}.

\bibitem[Alméras {\em et~al.\/}(2017)Alméras, Mathai, Lohse \&
  Sun]{almeras-etal-2017}
{\sc \au{Alméras, E.}, \au{Mathai, V.}, \au{Lohse, D.} \& \au{Sun, C.}}
  \yr{2017}  \at{Experimental investigation of the turbulence induced by a
  bubble swarm rising within incident turbulence}.  \jt{J. Fluid Mech.}
  \bvol{825},  \pg{1091–1112}.

\bibitem[Auton {\em et~al.\/}(1988)Auton, Hunt \&
  Prud’Homme]{auton-hunt-1988}
{\sc \au{Auton, T.~R.}, \au{Hunt, J. C.~R.} \& \au{Prud’Homme, M.}} \yr{1988}
   \at{The force exerted on a body in inviscid unsteady non-uniform rotational
  flow}.  \jt{J. Fluid Mech.}  \bvol{197},  \pg{241–257}.

\bibitem[Balachandar(2009)]{balachandar-2009}
{\sc \au{Balachandar, S.}} \yr{2009}  \at{A scaling analysis for
  point–particle approaches to turbulent multiphase flows}.  \jt{Int. J.
  Multiph. Flow}  \bvol{35},  \pg{801--810}.

\bibitem[Balachandar \& Eaton(2010)]{balachandar-eaton-2010}
{\sc \au{Balachandar, S.} \& \au{Eaton, John~K.}} \yr{2010}  \at{Turbulent
  {Dispersed} {Multiphase} {Flow}}.  \jt{Annu. Rev. Fluid Mech.}
  \bvol{42}~(1),  \pg{111--133}.

\bibitem[Benzi {\em et~al.\/}(1993)Benzi, Ciliberto, Tripiccione, Baudet,
  Massaioli \& Succi]{benzi-1993}
{\sc \au{Benzi, R.}, \au{Ciliberto, S.}, \au{Tripiccione, R.}, \au{Baudet, C.},
  \au{Massaioli, F.} \& \au{Succi, S.}} \yr{1993}  \at{Extended self-similarity
  in turbulent flows}.  \jt{Phys. Rev. E}  \bvol{48}~(1),  \pg{R29--R32}.

\bibitem[Betchov(1956)]{betchov-1956}
{\sc \au{Betchov, R.}} \yr{1956}  \at{An inequality concerning the production
  of vorticity in isotropic turbulence}.  \jt{J. Fluid Mech.}  \bvol{1}~(5),
  \pg{497--504}.

\bibitem[Boivin {\em et~al.\/}(1998)Boivin, Simonin \&
  Squires]{boivin-simonin-squires-1998}
{\sc \au{Boivin, M.}, \au{Simonin, O.} \& \au{Squires, K.D.}} \yr{1998}
  \at{Direct numerical simulation of turbulence modulation by particles in
  isotropic turbulence}.  \jt{J. Fluid Mech.}  \bvol{375},  \pg{235–263}.

\bibitem[Bragg \& Collins(2014)]{bragg-collins-2014}
{\sc \au{Bragg, A.~D.} \& \au{Collins, L.~R.}} \yr{2014}  \at{New insights from
  comparing statistical theories for inertial particles in turbulence: {I}.
  {Spatial} distribution of particles}.  \jt{New J. Phys.}  \bvol{16}~(5),
  \pg{055013}, publisher: IOP Publishing.

\bibitem[Bragg {\em et~al.\/}(2015)Bragg, Ireland \&
  Collins]{bragg-ireland-collins-2015}
{\sc \au{Bragg, A.~D.}, \au{Ireland, P.~J.} \& \au{Collins, L.~R.}} \yr{2015}
  \at{Mechanisms for the clustering of inertial particles in the inertial range
  of isotropic turbulence}.  \jt{Phys. Rev. E.}  \bvol{92}~(2),  \pg{023029},
  publisher: American Physical Society.

\bibitem[Brandt \& Coletti(2022)]{brandt-coletti-2022}
{\sc \au{Brandt, L.} \& \au{Coletti, F.}} \yr{2022}  \at{Particle-{{Laden
  Turbulence}}: {{Progress}} and {{Perspectives}}}.  \jt{Annu. Rev. Fluid
  Mech.}  \bvol{54}~(1),  \pg{159--189}.

\bibitem[Breugem(2012)]{breugem-2012}
{\sc \au{Breugem, W.P.}} \yr{2012}  \at{A second-order accurate immersed
  boundary method for fully resolved simulations of particle-laden flows}.
  \jt{J. Comput. Phys.}  \bvol{231},  \pg{4469--4498}.

\bibitem[Burton \& Eaton(2005)]{burton-eaton-2005}
{\sc \au{Burton, T.M.} \& \au{Eaton, J.K.}} \yr{2005}  \at{Fully resolved
  simulations of particle-turbulence interaction}.  \jt{J. Fluid Mech.}
  \bvol{545},  \pg{67--111}.

\bibitem[Cannon {\em et~al.\/}(2024)Cannon, Olivieri \&
  Rosti]{cannon-olivieri-rosti-2024}
{\sc \au{Cannon, I.}, \au{Olivieri, S.} \& \au{Rosti, M.E.}} \yr{2024}
  \at{Spheres and fibers in turbulent flows at various reynolds numbers}.
  \jt{Phys. Rev. Fluids}  \bvol{9},  \pg{064301}.

\bibitem[Casas {\em et~al.\/}(2018)Casas, Ferrer \&
  Oñate]{casas-ferrer-onate-2017}
{\sc \au{Casas, G.}, \au{Ferrer, A.} \& \au{Oñate, E.}} \yr{2018}
  \at{Approximating the basset force by optimizing the method of van hinsberg
  et al.}  \jt{J. Comput. Phys.}  \bvol{352},  \pg{142--171}.

\bibitem[Chevillard {\em et~al.\/}(2003)Chevillard, Roux, Lev{\^e}que, Mordant,
  Pinton \& Arneodo]{chevillard-etal-2003}
{\sc \au{Chevillard, L.}, \au{Roux, S.~G.}, \au{Lev{\^e}que, E.}, \au{Mordant,
  N.}, \au{Pinton, J.-F.} \& \au{Arneodo, A.}} \yr{2003}  \at{Lagrangian
  {{Velocity Statistics}} in {{Turbulent Flows}}: {{Effects}} of
  {{Dissipation}}}.  \jt{Phys. Rev. Lett.}  \bvol{91}~(21),  \pg{214502}.

\bibitem[Chiarini {\em et~al.\/}(2024)Chiarini, Cannon \&
  Rosti]{chiarini-etal-2023}
{\sc \au{Chiarini, A.}, \au{Cannon, I.} \& \au{Rosti, M.~E.}} \yr{2024}
  \at{Anisotropic mean flow enhancement and anomalous transport of finite-size
  spherical particles in turbulent flows}.  \jt{Phys. Rev. Lett.}  \bvol{132},
  \pg{054005}.

\bibitem[Chiarini \& Rosti(2024)]{chiarini-rosti-2024}
{\sc \au{Chiarini, A.} \& \au{Rosti, M.E.}} \yr{2024}  \at{Finite-size inertial
  spherical particles in turbulence}.  \jt{J. Fluid Mech.}  \bvol{988},
  \pg{A17}.

\bibitem[Coleman \& Vassilicos(2009)]{coleman-vassilicos-2009}
{\sc \au{Coleman, S.~W.} \& \au{Vassilicos, J.~C.}} \yr{2009}  \at{A unified
  sweep-stick mechanism to explain particle clustering in two- and
  three-dimensional homogeneous, isotropic turbulence}.  \jt{Phys. Fluids}
  \bvol{21}~(11),  \pg{113301}.

\bibitem[Costa {\em et~al.\/}(2020)Costa, Brandt \&
  Picano]{costa-brandt-picano-2020}
{\sc \au{Costa, P.}, \au{Brandt, L.} \& \au{Picano, F.}} \yr{2020}
  \at{Interface-resolved simulations of small inertial particles in turbulent
  channel flow}.  \jt{J. Fluid Mech.}  \bvol{883},  \pg{A54}.

\bibitem[Cundall \& Strack(1979)]{cundall-strack-1979}
{\sc \au{Cundall, P.~A.} \& \au{Strack, O.~D.L.}} \yr{1979}  \at{A discrete
  numerical model for granular assemblies}.  \jt{Geotechnique}  \bvol{29}~(1),
  \pg{47--65}.

\bibitem[Davidson(2004)]{davidson-2004}
{\sc \au{Davidson, P.A.}} \yr{2004} {\em Turbulence: {{An Introduction}} for
  {{Scientists}} and {{Engineers}}\/}.  \publ{{Oxford University Press}}.

\bibitem[Davidson \& Pearson(2005)]{davidson-pearson-2005}
{\sc \au{Davidson, P.~A.} \& \au{Pearson, B.~R.}} \yr{2005}  \at{Identifying
  {Turbulent} {Energy} {Distributions} in {Real}, {Rather} than {Fourier},
  {Space}}.  \jt{Phys. Rev. Lett.}  \bvol{95}~(21),  \pg{214501}, publisher:
  American Physical Society.

\bibitem[De~Lillo {\em et~al.\/}(2014)De~Lillo, Cencini, Durham, Barry,
  Stocker, Climent \& Boffetta]{delillo-etal-2014}
{\sc \au{De~Lillo, F.}, \au{Cencini, M.}, \au{Durham, W.~M.}, \au{Barry, M.},
  \au{Stocker, R.}, \au{Climent, E.} \& \au{Boffetta, G.}} \yr{2014}
  \at{Turbulent {{Fluid Acceleration Generates Clusters}} of {{Gyrotactic
  Microorganisms}}}.  \jt{Phys. Rev. Lett.}  \bvol{112}~(4),  \pg{044502}.

\bibitem[Dorgan \& Loth(2007)]{dorgan-loth-2007}
{\sc \au{Dorgan, A.J.} \& \au{Loth, E.}} \yr{2007}  \at{Efficient calculation
  of the history force at finite reynolds numbers}.  \jt{Int. J. Multiphase
  Flow}  \bvol{33},  \pg{833--848}.

\bibitem[Druzhinin(2001)]{druzhinin-2001}
{\sc \au{Druzhinin, O.~A.}} \yr{2001}  \at{The influence of particle inertia on
  the two-way coupling and modification of isotropic turbulence by
  microparticles}.  \jt{Phys. Fluids}  \bvol{13}~(12),  \pg{3738--3755}.

\bibitem[Dung {\em et~al.\/}(2023)Dung, Waasdorp, Sun, Lohse \&
  Huisman]{dung-etal-2022}
{\sc \au{Dung, O.-Y.}, \au{Waasdorp, P.}, \au{Sun, C.}, \au{Lohse, D.} \&
  \au{Huisman, S.G.}} \yr{2023}  \at{The emergence of bubble-induced scaling in
  thermal spectra in turbulence}.  \jt{J. Fluid Mech.}  \bvol{958},  \pg{A5}.

\bibitem[Elghobashi \& Truesdell(1993)]{elghobashi-truesdell-1993}
{\sc \au{Elghobashi, S.} \& \au{Truesdell, G.~C.}} \yr{1993}  \at{On the
  two-way interaction between homogeneous turbulence and dispersed solid
  particles. {{I}}: {{Turbulence}} modification}.  \jt{Phys. Fluids A: Fluid
  Dyn.}  \bvol{5}~(7),  \pg{1790--1801}.

\bibitem[Faxén(1922)]{faxen-1922}
{\sc \au{Faxén, H.}} \yr{1922}  \at{Der widerstand gegen die bewegung einer
  starren kugel in einer zähen flüssigkeit, die zwischen zwei parallelen
  ebenen wänden eingeschlossen ist}.  \jt{Annalen der Physik}
  \bvol{373}~(10),  \pg{89--119}.

\bibitem[Ferrante \& Elghobashi(2003)]{ferrante-elghobashi-2003}
{\sc \au{Ferrante, A.} \& \au{Elghobashi, S.}} \yr{2003}  \at{On the physical
  mechanisms of two-way coupling in particle-laden isotropic turbulence}.
  \jt{Phys. Fluids}  \bvol{15}~(2),  \pg{315--329}.

\bibitem[Fessler {\em et~al.\/}(1994)Fessler, Kulick \&
  Eaton]{fessler-etal-1994}
{\sc \au{Fessler, J.~R.}, \au{Kulick, J.D.} \& \au{Eaton, J.K.}} \yr{1994}
  \at{{Preferential concentration of heavy particles in a turbulent channel
  flow}}.  \jt{Phys. Fluids}  \bvol{6}~(11),  \pg{3742--3749}.

\bibitem[Fiabane {\em et~al.\/}(2012)Fiabane, Zimmermann, Volk, Pinton \&
  Bourgoin]{fiabane-etal-2012}
{\sc \au{Fiabane, L.}, \au{Zimmermann, R.}, \au{Volk, R.}, \au{Pinton, J.-F.}
  \& \au{Bourgoin, M.}} \yr{2012}  \at{Clustering of finite-size particles in
  turbulence}.  \jt{Phys. Rev. E}  \bvol{86}~(3),  \pg{035301}.

\bibitem[Frisch(1995)]{frisch-1995}
{\sc \au{Frisch, U.}} \yr{1995} {\em Turbulence: {{The Legacy}} of {{A}}.
  {{N}}. {{Kolmogorov}}\/}.  \publ{{Cambridge University Press}}.

\bibitem[Gatignol(1983)]{gatignol-1983}
{\sc \au{Gatignol, R.}} \yr{1983}  \at{The faxen formulae for a rigid particle
  in an unsteady non-uniform stokes flow}.  \jt{J. méc. théor. appl.}
  \bvol{2}~(2),  \pg{143--160}.

\bibitem[Gore \& Crowe(1989)]{gore-crowe-1989}
{\sc \au{Gore, R.~A.} \& \au{Crowe, C.~T.}} \yr{1989}  \at{Effect of particle
  size on modulating turbulent intensity}.  \jt{Int. J. Multiph. Flow.}
  \bvol{15}~(2),  \pg{279--285}.

\bibitem[Goto \& Vassilicos(2008)]{goto-vassilicos-2008}
{\sc \au{Goto, S.} \& \au{Vassilicos, J.~C.}} \yr{2008}  \at{Sweep-{{Stick
  Mechanism}} of {{Heavy Particle Clustering}} in {{Fluid Turbulence}}}.
  \jt{Phys. Rev. Lett.}  \bvol{100}~(5),  \pg{054503}.

\bibitem[Gualtieri {\em et~al.\/}(2015)Gualtieri, Picano, Sardina \&
  Casciola]{gualtieri-etal-2015}
{\sc \au{Gualtieri, P.}, \au{Picano, F.}, \au{Sardina, G.} \& \au{Casciola,
  C. M.}} \yr{2015}  \at{Exact regularized point particle method for
  multiphase flows in the two-way coupling regime}.  \jt{J. Fluid Mech.}
  \bvol{773},  \pg{520–561}.

\bibitem[Gustavsson \& Mehlig(2011)]{gustavsson-mehlig-2011}
{\sc \au{Gustavsson, K.} \& \au{Mehlig, B.}} \yr{2011}  \at{Distribution of
  relative velocities in turbulent aerosols}.  \jt{Phys. Rev. E.}
  \bvol{84}~(4),  \pg{045304}, publisher: American Physical Society.

\bibitem[Hassaini \& Coletti(2022)]{hassaini-coletti-2022}
{\sc \au{Hassaini, R.} \& \au{Coletti, F.}} \yr{2022}  \at{Scale-to-scale
  turbulence modification by small settling particles}.  \jt{J. Fluid Mech.}
  \bvol{949},  \pg{A30}.

\bibitem[Hassaini {\em et~al.\/}(2023)Hassaini, Petersen \&
  Coletti]{hassaini-etal-2023}
{\sc \au{Hassaini, R.}, \au{Petersen, A.J.} \& \au{Coletti, F.}} \yr{2023}
  \at{Effect of two-way coupling on clustering and settling of heavy particles
  in homogeneous turbulence}.  \jt{J. Fluid Mech.}  \bvol{976},  \pg{A12}.

\bibitem[van Hinsberg {\em et~al.\/}(2011)van Hinsberg, ten Thije~Boonkkamp \&
  Clercx]{hinsberg-boonkkamp-clercx-2011}
{\sc \au{van Hinsberg, M.A.T.}, \au{ten Thije~Boonkkamp, J.H.M.} \& \au{Clercx,
  H.J.H.}} \yr{2011}  \at{An efficient, second order method for the
  approximation of the basset history force}.  \jt{J. Comput. Phys.}
  \bvol{230},  \pg{1465--1478}.

\bibitem[Homann \& Bec(2010)]{homann-bec-2010}
{\sc \au{Homann, H.} \& \au{Bec, J.}} \yr{2010}  \at{Finite-size effects in the
  dynamics of neutrally buoyant particles in turbulent flow}.  \jt{J. Fluid
  Mech.}  \bvol{651},  \pg{81--91}.

\bibitem[Hori {\em et~al.\/}(2022)Hori, Rosti \&
  Takagi]{hori-rosti-takagi-2022}
{\sc \au{Hori, N.}, \au{Rosti, M.~E.} \& \au{Takagi, S.}} \yr{2022}  \at{An
  {{Eulerian-based}} immersed boundary method for particle suspensions with
  implicit lubrication model}.  \jt{Comput. Fluids}  \bvol{236},  \pg{105278}.

\bibitem[Horne \& Mahesh(2019)]{horne-mahesh-2019}
{\sc \au{Horne, W.J.} \& \au{Mahesh, K.}} \yr{2019}  \at{A massively-parallel,
  unstructured overset method to simulate moving bodies in turbulent flows}.
  \jt{J. Comput. Phys.}  \bvol{397},  \pg{108790}.

\bibitem[Hu {\em et~al.\/}(2001)Hu, Patankar \& Zhu]{hu-patankar-zhu-2001}
{\sc \au{Hu, H.H.}, \au{Patankar, N.A.} \& \au{Zhu, M.Y.}} \yr{2001}
  \at{Direct numerical simulation of ﬂuid-solid systems using the arbitrary
  lagrangian-eulerian technique}.  \jt{J. Comput. Phys.}  \bvol{169},
  \pg{427--462}.

\bibitem[Huang {\em et~al.\/}(2007)Huang, Shin \& Sung]{huang-etal-2007}
{\sc \au{Huang, W.-X.}, \au{Shin, S.~J.} \& \au{Sung, H.~J.}} \yr{2007}
  \at{Simulation of flexible filaments in a uniform flow by the immersed
  boundary method}.  \jt{J. Comput. Phys.}  \bvol{226}~(2),  \pg{2206--2228}.

\bibitem[Hwang \& Eaton(2006)]{hwang-eaton-2006}
{\sc \au{Hwang, W.} \& \au{Eaton, J.~K.}} \yr{2006}  \at{Homogeneous and
  isotropic turbulence modulation by small heavy particles}.  \jt{J. Fluid
  Mech.}  \bvol{564},  \pg{361--393}.

\bibitem[Kajishima {\em et~al.\/}(2001)Kajishima, Takiguchi, Hamasaki \&
  Miyake]{kajishima-etal-2001}
{\sc \au{Kajishima, T.}, \au{Takiguchi, S.}, \au{Hamasaki, H.} \& \au{Miyake,
  Y.}} \yr{2001}  \at{Turbulence {{Structure}} of {{Particle-Laden Flow}} in a
  {{Vertical Plane Channel Due}} to {{Vortex Shedding}}}.  \jt{JSME Int. J.
  B-Fluid T.}  \bvol{44}~(4),  \pg{526--535}.

\bibitem[Kempe \& Fröhlich(2012)]{kempe-frolich-2012}
{\sc \au{Kempe, T.} \& \au{Fröhlich, J.}} \yr{2012}  \at{An improved immersed
  boundary method with direct forcing for the simulation of particle laden
  flows}.  \jt{J. Comput. Phys.}  \bvol{231},  \pg{3663--3684}.

\bibitem[Kidanemariam {\em et~al.\/}(2013)Kidanemariam, {Chan-Braun}, Doychev
  \& Uhlmann]{kidanemariam-etal-2013}
{\sc \au{Kidanemariam, A.~G.}, \au{{Chan-Braun}, C.}, \au{Doychev, T.} \&
  \au{Uhlmann, M.}} \yr{2013}  \at{Direct numerical simulation of horizontal
  open channel flow with finite-size, heavy particles at low solid volume
  fraction}.  \jt{New J. Phys.}  \bvol{15}~(2),  \pg{025031}.

\bibitem[Koblitz {\em et~al.\/}(2017)Koblitz, Lovett, Nikiforakis \&
  Henshaw]{koblitz-etal-2017}
{\sc \au{Koblitz, A.}, \au{Lovett, S.}, \au{Nikiforakis, N.} \& \au{Henshaw,
  W.D.}} \yr{2017}  \at{Direct numerical simulation of particulate flows with
  an overset grid method}.  \jt{J. Comput. Phys.}  \bvol{343},  \pg{414--431}.

\bibitem[Kolmogorov(1941)]{kolmogorov-1941}
{\sc \au{Kolmogorov, A.N.}} \yr{1941}  \at{The {{Local Structure}} of
  {{Turbulence}} in an {{Incompressible Viscous Fluid}} for {{Very Large
  Reynolds Numbers}}}.  \jt{Dokl. Akad. Nauk. SSSR}  \bvol{30},  \pg{301--305}.

\bibitem[La~Porta {\em et~al.\/}(2001)La~Porta, Voth, Crawford, Alexander \&
  Bodenschatz]{laporta-etal-2001}
{\sc \au{La~Porta, A.}, \au{Voth, Greg~A.}, \au{Crawford, Alice~M.},
  \au{Alexander, Jim} \& \au{Bodenschatz, Eberhard}} \yr{2001}  \at{Fluid
  particle accelerations in fully developed turbulence}.  \jt{Nature}
  \bvol{409}~(6823),  \pg{1017--1019}.

\bibitem[Lance \& Bataille(1991)]{lance-betaille-1991}
{\sc \au{Lance, M.} \& \au{Bataille, J.}} \yr{1991}  \at{Turbulence in the
  liquid phase of a uniform bubbly air–water flow}.  \jt{J. Fluid Mech.}
  \bvol{222},  \pg{95–118}.

\bibitem[Lucci {\em et~al.\/}(2010)Lucci, Ferrante \&
  Elghobashi]{lucci-etal-2010}
{\sc \au{Lucci, F.}, \au{Ferrante, A.} \& \au{Elghobashi, S.}} \yr{2010}
  \at{Modulation of isotropic turbulence by particles of {{Taylor}}
  length-scale size}.  \jt{J. Fluid Mech.}  \bvol{650},  \pg{5--55}.

\bibitem[Lucci {\em et~al.\/}(2011)Lucci, Ferrante \&
  Elghobashi]{lucci-etal-2011}
{\sc \au{Lucci, F.}, \au{Ferrante, A.} \& \au{Elghobashi, S.}} \yr{2011}
  \at{Is {{Stokes}} number an appropriate indicator for turbulence modulation
  by particles of {{Taylor-length-scale}} size?}  \jt{Phys. Fluids}
  \bvol{23}~(2),  \pg{025101}.

\bibitem[Lund \& Rogers(1994)]{lund-rogers-1994}
{\sc \au{Lund, Thomas~S.} \& \au{Rogers, Michael~M.}} \yr{1994}  \at{An
  improved measure of strain state probability in turbulent flows}.  \jt{Phys.
  Fluids}  \bvol{6}~(5),  \pg{1838--1847}.

\bibitem[Martinez~Mercado {\em et~al.\/}(2010)Martinez~Mercado, Chehata~Gomez,
  Van~Gils, Sun \& Lohse]{mercado-etal-2010}
{\sc \au{Martinez~Mercado, J.}, \au{Chehata~Gomez, D.}, \au{Van~Gils, D.},
  \au{Sun, C.} \& \au{Lohse, D.}} \yr{2010}  \at{On bubble clustering and
  energy spectra in pseudo-turbulence}.  \jt{J. Fluid Mech.}  \bvol{650},
  \pg{287–306}.

\bibitem[Matsuda {\em et~al.\/}(2024)Matsuda, Yoshimatsu \&
  Schneider]{matsuda-etal-2024}
{\sc \au{Matsuda, K.}, \au{Yoshimatsu, K.} \& \au{Schneider, K.}} \yr{2024}
  \at{Heavy particle clustering in inertial subrange of high--reynolds number
  turbulence}.  \jt{Phys. Rev. Lett.}  \bvol{132},  \pg{234001}.

\bibitem[Maxey(1987)]{maxey-1987}
{\sc \au{Maxey, M.~R.}} \yr{1987}  \at{The gravitational settling of aerosol
  particles in homogeneous turbulence and random flow fields}.  \jt{J. Fluid
  Mech.}  \bvol{174},  \pg{441--465}, publisher: Cambridge University.

\bibitem[Maxey \& Riley(1983)]{maxey-riley-1983}
{\sc \au{Maxey, M.~R.} \& \au{Riley, J.~J.}} \yr{1983}  \at{Equation of motion
  for a small rigid sphere in a nonuniform flow}.  \jt{Phys. Fluids}
  \bvol{26}~(4),  \pg{883--889}.

\bibitem[McLaughlin(1991)]{mclaughlin-1991}
{\sc \au{McLaughlin, J.B.}} \yr{1991}  \at{Inertial migration of a small sphere
  in linear shear flows}.  \jt{J. Fluid Mech.}  \bvol{224},  \pg{261–274}.

\bibitem[Mehrabadi {\em et~al.\/}(2018)Mehrabadi, Horwitz, Subramaniam \&
  Mani]{mehrabadi-etal-2018}
{\sc \au{Mehrabadi, M.}, \au{Horwitz, J. A.~K.}, \au{Subramaniam, S.} \&
  \au{Mani, A.}} \yr{2018}  \at{A direct comparison of particle-resolved and
  point-particle methods in decaying turbulence}.  \jt{J. Fluid Mech.}
  \bvol{850},  \pg{336–369}.

\bibitem[Mei(1992)]{mei-1992}
{\sc \au{Mei, R.}} \yr{1992}  \at{An approximate expression for the shear lift
  force on a spherical particle at finite reynolds number}.  \jt{Int. J.
  Multiph. Flow}  \bvol{18},  \pg{145--147}.

\bibitem[Mei \& Adrian(1992)]{mei-adrian-1992}
{\sc \au{Mei, R.} \& \au{Adrian, R.~J.}} \yr{1992}  \at{Flow past a sphere with
  an oscillation in the free-stream velocity and unsteady drag at finite
  reynolds number}.  \jt{J. Fluid Mech.}  \bvol{237},  \pg{323–341}.

\bibitem[Meneveau(2011)]{meneveau-2011}
{\sc \au{Meneveau, C.}} \yr{2011}  \at{Lagrangian {{Dynamics}} and {{Models}}
  of the {{Velocity Gradient Tensor}} in {{Turbulent Flows}}}.  \jt{Annu. Rev.
  Fluid Mech.}  \bvol{43}~(1),  \pg{219--245}.

\bibitem[Michaelides(1992)]{michaelides-1992}
{\sc \au{Michaelides, E.E.}} \yr{1992}  \at{A novel way of computing the basset
  term in unsteady multiphase flow computations}.  \jt{Phys. Fluids A}
  \bvol{4}~(7),  \pg{1579--1582}.

\bibitem[Monchaux {\em et~al.\/}(2010)Monchaux, Bourgoin \&
  Cartellier]{monchaux-etal-2010}
{\sc \au{Monchaux, R.}, \au{Bourgoin, M.} \& \au{Cartellier, A.}} \yr{2010}
  \at{Preferential concentration of heavy particles: {{A Vorono\"i}} analysis}.
   \jt{Phys. Fluids}  \bvol{22}~(10),  \pg{103304}.

\bibitem[Monchaux {\em et~al.\/}(2012)Monchaux, Bourgoin \&
  Cartellier]{monchaux-etal-2012}
{\sc \au{Monchaux, R.}, \au{Bourgoin, M.} \& \au{Cartellier, A.}} \yr{2012}
  \at{Analyzing preferential concentration and clustering of inertial particles
  in turbulence}.  \jt{Int. J. Multiph. Flow}  \bvol{40},  \pg{1--18}.

\bibitem[Monti {\em et~al.\/}(2021)Monti, Rathee, Shen \&
  Rosti]{monti-etal-2021}
{\sc \au{Monti, A.}, \au{Rathee, V.}, \au{Shen, A.~Q.} \& \au{Rosti, M.~E.}}
  \yr{2021}  \at{A fast and efficient tool to study the rheology of dense
  suspensions}.  \jt{Phys. Fluids}  \bvol{33}~(10),  \pg{103314}.

\bibitem[Mordant {\em et~al.\/}(2001)Mordant, Metz, Michel \&
  Pinton]{mordant-etal-2001}
{\sc \au{Mordant, N.}, \au{Metz, P.}, \au{Michel, O.} \& \au{Pinton, J.-F.}}
  \yr{2001}  \at{Measurement of {{Lagrangian Velocity}} in {{Fully Developed
  Turbulence}}}.  \jt{Phys. Rev. Lett.}  \bvol{87}~(21),  \pg{214501}.

\bibitem[Brändle~de Motta {\em et~al.\/}(2016)Brändle~de Motta, Estivalezes,
  Climent \& Vincent]{brandle-etal-2016}
{\sc \au{Brändle~de Motta, J.~C.}, \au{Estivalezes, J.~L.}, \au{Climent, E.}
  \& \au{Vincent, S.}} \yr{2016}  \at{Local dissipation properties and
  collision dynamics in a sustained homogeneous turbulent suspension composed
  of finite size particles}.  \jt{Int. J. Multiph. Flow}  \bvol{85},
  \pg{369--379}.

\bibitem[Nomura \& Post(1998)]{nomura-post-1998}
{\sc \au{Nomura, K.K.} \& \au{Post, G.K.}} \yr{1998}  \at{The structure and
  dynamics of vorticity and rate of strain in incompressible homogeneous
  turbulence}.  \jt{J. Fluid Mech.}  \bvol{377},  \pg{65–97}.

\bibitem[Oka \& Goto(2022)]{oka-goto-2022}
{\sc \au{Oka, S.} \& \au{Goto, S.}} \yr{2022}  \at{Attenuation of turbulence in
  a periodic cube by finite-size spherical solid particles}.  \jt{J. Fluid
  Mech.}  \bvol{949},  \pg{A45}.

\bibitem[Olivieri {\em et~al.\/}(2022{\natexlab{{\em a\/}}})Olivieri, Cannon \&
  Rosti]{olivieri-cannon-rosti-2022}
{\sc \au{Olivieri, S.}, \au{Cannon, I.} \& \au{Rosti, M.~E.}}
  \yr{2022{\natexlab{{\em a\/}}}}  \at{The effect of particle anisotropy on the
  modulation of turbulent flows}.  \jt{J. Fluid Mech.}  \bvol{950},  \pg{R2}.

\bibitem[Olivieri {\em et~al.\/}(2022{\natexlab{{\em b\/}}})Olivieri, Mazzino
  \& Rosti]{olivieri-mazzino-rosti-2022}
{\sc \au{Olivieri, S.}, \au{Mazzino, A.} \& \au{Rosti, M.~E.}}
  \yr{2022{\natexlab{{\em b\/}}}}  \at{On the fully coupled dynamics of
  flexible fibres dispersed in modulated turbulence}.  \jt{J. Fluid Mech.}
  \bvol{946},  \pg{A34}.

\bibitem[Olivieri {\em et~al.\/}(2014)Olivieri, Picano, Sardina, Iudicone \&
  Brandt]{olivieri-etal-2014}
{\sc \au{Olivieri, S.}, \au{Picano, F.}, \au{Sardina, G.}, \au{Iudicone, D.} \&
  \au{Brandt, L.}} \yr{2014}  \at{The effect of the basset history force on
  particle clustering in homogeneous and isotropic turbulence}.  \jt{Phys.
  Fluids}  \bvol{26}~(4),  \pg{041704}.

\bibitem[Pandey {\em et~al.\/}(2022)Pandey, Mitra \&
  Perlekar]{pandey-mitra-perlekar-2022}
{\sc \au{Pandey, V.}, \au{Mitra, D.} \& \au{Perlekar, P.}} \yr{2022}
  \at{Turbulence modulation in buoyancy-driven bubbly flows}.  \jt{J. Fluid
  Mech.}  \bvol{932},  \pg{A19}.

\bibitem[Pandey {\em et~al.\/}(2023)Pandey, Mitra \&
  Perlekar]{pandey-mitra-perlekar-2023}
{\sc \au{Pandey, V.}, \au{Mitra, D.} \& \au{Perlekar, P.}} \yr{2023}
  \at{Kolmogorov turbulence coexists with pseudo-turbulence in buoyancy-driven
  bubbly flows}.  \jt{Phys. Rev. Lett.}  \bvol{131},  \pg{114002}.

\bibitem[Pandey {\em et~al.\/}(2020)Pandey, Ramadugu \&
  Perlekar]{pandey-ramadugu-perlekar-2020}
{\sc \au{Pandey, V.}, \au{Ramadugu, R.} \& \au{Perlekar, P.}} \yr{2020}
  \at{Liquid velocity fluctuations and energy spectra in three-dimensional
  buoyancy-driven bubbly flows}.  \jt{J. Fluid Mech.}  \bvol{884},  \pg{R6}.

\bibitem[Petersen {\em et~al.\/}(2019)Petersen, Baker \&
  Coletti]{petersen-etal-2019}
{\sc \au{Petersen, A.~J.}, \au{Baker, L.} \& \au{Coletti, F.}} \yr{2019}
  \at{Experimental study of inertial particles clustering and settling in
  homogeneous turbulence}.  \jt{J. Fluid Mech.}  \bvol{864},  \pg{925--970}.

\bibitem[Podvigina \& Pouquet(1994)]{podvigina-pouquet-1994}
{\sc \au{Podvigina, O.} \& \au{Pouquet, A.}} \yr{1994}  \at{On the non-linear
  stability of the 1:1:1 {{ABC}} flow}.  \jt{Phys. D: Nonlinear Phenom.}
  \bvol{75}~(4),  \pg{471--508}.

\bibitem[Poelma {\em et~al.\/}(2007)Poelma, Westerweel \&
  Ooms]{poelma-etal-2007}
{\sc \au{Poelma, C.}, \au{Westerweel, J.} \& \au{Ooms, G}} \yr{2007}
  \at{Particle-fluid-interactions in grid generated turbulence}.  \jt{J. Fluid
  Mech.}  \bvol{589},  \pg{315--351}.

\bibitem[Pope(2000)]{pope-2000}
{\sc \au{Pope, S.B.}} \yr{2000} {\em Turbulent {{Flows}}\/}.  \publ{{Cambridge
  University Press, Cambridge}}.

\bibitem[Prakash {\em et~al.\/}(2016)Prakash, Martínez~Mercado, van
  Wijngaarden, Mancilla, Tagawa, Lohse \& Sun]{prakash-etal-2016}
{\sc \au{Prakash, V.~N.}, \au{Martínez~Mercado, J.}, \au{van Wijngaarden, L.},
  \au{Mancilla, E.}, \au{Tagawa, Y.}, \au{Lohse, D.} \& \au{Sun, C.}} \yr{2016}
   \at{Energy spectra in turbulent bubbly flows}.  \jt{J. Fluid Mech.}
  \bvol{791},  \pg{174–190}.

\bibitem[Prasath {\em et~al.\/}(2019)Prasath, Vasan \&
  Govindarajan]{prasath-etal-2019}
{\sc \au{Prasath, S.~Ganga}, \au{Vasan, Vishal} \& \au{Govindarajan, Rama}}
  \yr{2019}  \at{Accurate solution method for the maxey–riley equation, and
  the effects of basset history}.  \jt{J. Fluid Mech.}  \bvol{868},
  \pg{428–460}.

\bibitem[Prosperetti \& Oguz(2001)]{prosperetti-oguz-2001}
{\sc \au{Prosperetti, A.} \& \au{Oguz, H.N.}} \yr{2001}  \at{Physalis: a new
  o(n) method for the numerical simulation of disperse systems: a potential
  flow of spheres}.  \jt{J. Comput. Phys.}  \bvol{167},  \pg{196--216}.

\bibitem[Qureshi {\em et~al.\/}(2007)Qureshi, Bourgoin, Baudet, Cartellier \&
  Gagne]{qureshi-etal-2007}
{\sc \au{Qureshi, N.M.}, \au{Bourgoin, M.}, \au{Baudet, C.}, \au{Cartellier,
  A.} \& \au{Gagne, Y.}} \yr{2007}  \at{Turbulent {{Transport}} of {{Material
  Particles}}: {{An Experimental Study}} of {{Finite Size Effects}}}.
  \jt{Phys. Rev. Lett.}  \bvol{99}~(18),  \pg{184502}.

\bibitem[Ramirez {\em et~al.\/}(2024)Ramirez, Burlot, Zamansky, Bois \&
  Risso]{ramirez-etal-2024}
{\sc \au{Ramirez, G.}, \au{Burlot, A.}, \au{Zamansky, R.}, \au{Bois, G.} \&
  \au{Risso, F.}} \yr{2024}  \at{Spectral analysis of dispersed multiphase
  flows in the presence of fluid interfaces}.  \jt{Int. J. Multiph. Flow}
  \bvol{177},  \pg{104860}.

\bibitem[Riboux {\em et~al.\/}(2010)Riboux, Risso \&
  Legendre]{riboux-risso-legendre-2010}
{\sc \au{Riboux, G.}, \au{Risso, F.} \& \au{Legendre, D.}} \yr{2010}
  \at{Experimental characterization of the agitation generated by bubbles
  rising at high reynolds number}.  \jt{J. Fluid Mech.}  \bvol{643},
  \pg{509–539}.

\bibitem[Risso(2018)]{risso-2018}
{\sc \au{Risso, F.}} \yr{2018}  \at{Agitation, mixing, and transfers induced by
  bubbles}.  \jt{Ann. Rev. Fluid Mech.}  \bvol{50},  \pg{25--48}.

\bibitem[Saffman(1965)]{saffman-1965}
{\sc \au{Saffman, P.~G.}} \yr{1965}  \at{The lift on a small sphere in a slow
  shear flow}.  \jt{J. Fluid Mech.}  \bvol{22}~(2),  \pg{385–400}.

\bibitem[Salazar {\em et~al.\/}(2008)Salazar, Jong, Cao, Woodward, Meng \&
  Collins]{salazar-etal-2008}
{\sc \au{Salazar, J. P. L.~C.}, \au{Jong, J.~D.}, \au{Cao, L.}, \au{Woodward,
  S.~H.}, \au{Meng, H.} \& \au{Collins, L.~R.}} \yr{2008}  \at{Experimental and
  numerical investigation of inertial particle clustering in isotropic
  turbulence}.  \jt{J. Fluid Mech.}  \bvol{600},  \pg{245--256}.

\bibitem[Saw {\em et~al.\/}(2008)Saw, Shaw, Ayyalasomayajula, Chuang \&
  Gylfason]{saw-etal-2008}
{\sc \au{Saw, E.~W.}, \au{Shaw, R.~A.}, \au{Ayyalasomayajula, S.}, \au{Chuang,
  P.~Y.} \& \au{Gylfason, {\'A}.}} \yr{2008}  \at{Inertial {{Clustering}} of
  {{Particles}} in {{High-Reynolds-Number Turbulence}}}.  \jt{Phys. Rev. Lett.}
   \bvol{100}~(21),  \pg{214501}.

\bibitem[Schneiders {\em et~al.\/}(2017)Schneiders, Meinke \&
  Schr{\"o}der]{schneiders-etal-2017}
{\sc \au{Schneiders, L.}, \au{Meinke, M.} \& \au{Schr{\"o}der, W.}} \yr{2017}
  \at{Direct particle\textendash fluid simulation of
  {{Kolmogorov-length-scale}} size particles in decaying isotropic turbulence}.
   \jt{J. Fluid Mech.}  \bvol{819},  \pg{188--227}.

\bibitem[Schumacher {\em et~al.\/}(2007)Schumacher, Sreenivasan \&
  Yakhot]{schumacher-etal-2007}
{\sc \au{Schumacher, J.}, \au{Sreenivasan, K.R.} \& \au{Yakhot, V.}} \yr{2007}
  \at{Asymptotic exponents from low-{Reynolds}-number flows}.  \jt{New J.
  Phys.}  \bvol{9}~(4),  \pg{89}.

\bibitem[Sengupta {\em et~al.\/}(2017)Sengupta, Carrara \&
  Stocker]{sengupta-etal-2017}
{\sc \au{Sengupta, A.}, \au{Carrara, F.} \& \au{Stocker, R.}} \yr{2017}
  \at{Phytoplankton can actively diversify their migration strategy in response
  to turbulent cues}.  \jt{Nature}  \bvol{543}~(7646),  \pg{555--558}.

\bibitem[Soria {\em et~al.\/}(1994)Soria, Sondergaard, Cantwell, Chong \&
  Perry]{soria-etal-1994}
{\sc \au{Soria, J.}, \au{Sondergaard, R.}, \au{Cantwell, B.~J.}, \au{Chong,
  M.~S.} \& \au{Perry, A.~E.}} \yr{1994}  \at{A study of the fine‐scale
  motions of incompressible time‐developing mixing layers}.  \jt{Phys.
  Fluids}  \bvol{6}~(2),  \pg{871--884}.

\bibitem[Squires \& Eaton(1990)]{squires-eaton-1990}
{\sc \au{Squires, K.D.} \& \au{Eaton, J.K.}} \yr{1990}  \at{Particle response
  and turbulence modification in isotropic turbulence}.  \jt{Phys. Fluids A}
  \bvol{2}~(7),  \pg{1191--1203}.

\bibitem[Sumbekova {\em et~al.\/}(2017)Sumbekova, Cartellier, Aliseda \&
  Burgoin]{sumbekova-etal-2017}
{\sc \au{Sumbekova, S.}, \au{Cartellier, A.}, \au{Aliseda, A.} \& \au{Burgoin,
  M.}} \yr{2017}  \at{Preferential concentrationof sub-kolmogorov particles:
  The roles of mass loading of particles, stokes numbers, and reynolds
  numbers}.  \jt{Phys. Rev. Fluids}  \bvol{2},  \pg{024302}.

\bibitem[Tanaka \& Eaton(2010)]{tanaka-eaton-2010}
{\sc \au{Tanaka, T.} \& \au{Eaton, J.~K.}} \yr{2010}  \at{Sub-{Kolmogorov}
  resolution partical image velocimetry measurements of particle-laden forced
  turbulence}.  \jt{J. Fluid Mech.}  \bvol{643},  \pg{177--206}.

\bibitem[Ten~Cate {\em et~al.\/}(2004)Ten~Cate, Derksen, Portela \& van
  Den~Akker]{tenCate-etal-2004}
{\sc \au{Ten~Cate, A.}, \au{Derksen, J.~J.}, \au{Portela, L.~M.} \& \au{van
  Den~Akker, H. E.~a.}} \yr{2004}  \at{Fully resolved simulations of colliding
  monodisperse spheres in forced isotropic turbulence}.  \jt{J. Fluid Mech.}
  \bvol{519},  \pg{233--271}.

\bibitem[Truesdell(1954)]{truesdell-1954}
{\sc \au{Truesdell, C.}} \yr{1954} {\em The Kinematics of Vorticity\/}.
  \publ{Indiana University Press}.

\bibitem[Tsuji {\em et~al.\/}(1993)Tsuji, Kawaguchi \& Tanaka]{tsuji-etal-1993}
{\sc \au{Tsuji, Y.}, \au{Kawaguchi, T.} \& \au{Tanaka, T.}} \yr{1993}
  \at{Discrete particle simulation of two-dimensional fluidized bed}.
  \jt{Powder Technol.}  \bvol{77}~(1),  \pg{79--87}.

\bibitem[Uhlmann(2005)]{uhlmann-2005}
{\sc \au{Uhlmann, M.}} \yr{2005}  \at{An immersed boundary method with direct
  forcing for the simulation of particulate flows}.  \jt{J. Comput. Phys.}
  \bvol{209}~(2),  \pg{448--476}.

\bibitem[Uhlmann \& Chouippe(2017)]{uhlmann-chouippe-2017}
{\sc \au{Uhlmann, M.} \& \au{Chouippe, A.}} \yr{2017}  \at{Clustering and
  preferential concentration of finite-size particles in forced
  homogeneous-isotropic turbulence}.  \jt{J. Fluid Mech.}  \bvol{812},
  \pg{991--1023}.

\bibitem[Vreman(2016{\natexlab{{\em a\/}}})]{vreman-2017}
{\sc \au{Vreman, A.W.}} \yr{2016{\natexlab{{\em a\/}}}}  \at{A staggered
  overset grid method for resolved simulation of incompressible flow around
  moving spheres}.  \jt{J. Comput. Phys.}  \bvol{333},  \pg{269--296}.

\bibitem[Vreman(2016{\natexlab{{\em b\/}}})]{vreman-2016}
{\sc \au{Vreman, A.~W.}} \yr{2016{\natexlab{{\em b\/}}}}  \at{Particle-resolved
  direct numerical simulation of homogeneous isotropic turbulence modified by
  small fixed spheres}.  \jt{J. Fluid Mech.}  \bvol{796},  \pg{40–85}.

\bibitem[Wang {\em et~al.\/}(2023)Wang, Jiang \& Sun]{wang-etal-2023}
{\sc \au{Wang, C.}, \au{Jiang, L.} \& \au{Sun, C.}} \yr{2023}  \at{Numerical
  study on turbulence modulation of finite-size particles in plane-couette
  flow}.  \jt{J. Fluid Mech.}  \bvol{970},  \pg{A7}.

\bibitem[Wang \& Maxey(1993)]{wang-maxey-1993}
{\sc \au{Wang, L.-P.} \& \au{Maxey, M.R.}} \yr{1993}  \at{Settling velocity and
  concentration distribution of heavy particles in homogeneous isotropic
  turbulence}.  \jt{J. Fluid Mech.}  \bvol{256},  \pg{27–68}.

\bibitem[Yang \& Shy(2005)]{yang-shy-2005}
{\sc \au{Yang, T.~S.} \& \au{Shy, S.~S.}} \yr{2005}  \at{Two-way interaction
  between solid particles and homogeneous air turbulence: Particle settling
  rate and turbulence modification measurements}.  \jt{J. Fluids Eng.}
  \bvol{526},  \pg{171--216}.

\bibitem[Yeo {\em et~al.\/}(2010)Yeo, Dong, Climent \& Maxey]{yeo-etal-2010}
{\sc \au{Yeo, K.}, \au{Dong, S.}, \au{Climent, E.} \& \au{Maxey, M.~R.}}
  \yr{2010}  \at{Modulation of homogeneous turbulence seeded with finite size
  bubbles or particles}.  \jt{Int. J. Multiph. Flow}  \bvol{36}~(3),
  \pg{221--233}.

\bibitem[Zamansky {\em et~al.\/}(2024)Zamansky, De~Bonneville \&
  Risso]{zamansky-etal-2024}
{\sc \au{Zamansky, R.}, \au{De~Bonneville, F.L.R.} \& \au{Risso, F.}} \yr{2024}
   \at{Turbulence induced by a swarm of rising bubbles from coarse-grained
  simulations}.  \jt{J. Fluid Mech}  \bvol{984},  \pg{A68}.

\end{thebibliography}

\end{document}